\documentclass{aa}  
\usepackage{natbib}
\usepackage{graphicx}
\usepackage{footmisc}
\usepackage{float}
\usepackage{caption}
\usepackage[caption=false]{subfig}
\usepackage[colorlinks=true,citecolor=blue]{hyperref}
\usepackage{multirow}
\usepackage{amsmath}
\usepackage{nccmath}
\usepackage{color}
\usepackage{array} 
\usepackage{ulem}
\usepackage[flushleft]{threeparttable}
\DeclareTextFontCommand{\textroman}{\fontlibertine}
\usepackage{txfonts}
\begin{document}

\title{
The atmosphere of HD 209458b seen with ESPRESSO\thanks{Based on Guaranteed Time Observations collected at the European Southern Observatory under ESO programme 1102.C-0744 by the ESPRESSO Consortium.} }

\subtitle{No detectable planetary absorptions at high resolution}

   \author{N.~Casasayas-Barris\inst{1,2}
   \and
    E. Palle\inst{1,2}
    \and
    M. Stangret\inst{1,2}
    \and
    V Bourrier\inst{3}
    \and
    H. M. Tabernero\inst{4}
    \and
    F. Yan\inst{5}
    \and
    F. Borsa\inst{6}
    \and
    R. Allart\inst{3}
    \and
    M.R. Zapatero Osorio\inst{7}
    \and
    C. Lovis\inst{3}
    \and
    S. G. Sousa\inst{4}
    \and
    G. Chen\inst{8}
    \and
    M. Oshagh\inst{1,2}
    \and
    N. C. Santos\inst{4,9}
    \and
    F. Pepe\inst{3}
    \and
    R. Rebolo\inst{1,2,13}
    \and
    P. Molaro\inst{11,12}
    \and
    S. Cristiani\inst{11}
    \and
    V. Adibekyan\inst{4,9}
    \and
    Y. Alibert\inst{10}
    \and
    C. Allende Prieto\inst{1,2}
    \and
    F. Bouchy\inst{3}
    \and
    O. D. S. Demangeon\inst{4,9,16}
    \and
    P. Di Marcantonio\inst{11}
    \and
    V. D'Odorico\inst{11,12}
    \and
    D. Ehrenreich\inst{3}
    \and
    P. Figueira\inst{4,14}
    \and
    R. G\'enova Santos\inst{1,2}
    \and
    J. I. Gonz\'alez Hern\'andez\inst{1,2}
    \and
    B. Lavie\inst{3}
    \and
    J. Lillo-Box\inst{14}
    \and
    G. Lo Curto\inst{15}
    \and
    C. J. A. P. Martins\inst{4,16}
    \and
    A. Mehner\inst{17}
    \and
    G. Micela\inst{18}
    \and
    N. J. Nunes\inst{19}
    \and
    E. Poretti\inst{6,20}
    \and
    A. Sozzetti\inst{21}
    \and
    A. Su\'arez Mascare\~no\inst{1,2}
    \and
    S. Udry\inst{3}}


   \institute{\label{inst:iac}Instituto de Astrofsica de Canarias, Vía Láctea s/n, 38205 La Laguna, Tenerife, Spain
              \\
              \email{nuriacb@iac.es}
         \and
             \label{inst:ull}Departamento de Astrofísica, Universidad de La Laguna, E-38206, La Laguna, Tenerife, Spain
        \and
            \label{inst:OG}Observatoire Astronomique de l'Universit\'e de Gen\`eve, Chemin Pegasi 51b, Sauverny, CH-1290, Switzerland
        \and
            \label{inst:IACE} Instituto de Astrofísica e Ciências do Espaço, Universidade do Porto, CAUP, Rua das Estrelas, 4150-762 Porto, Portugal
        \and
            \label{inst:Gott}Institut für Astrophysik, Georg-August-Universität, Friedrich-Hund-Platz 1, D-37077 Göttingen, Germany
        \and
            \label{inst:INAF}INAF - Osservatorio Astronomico di Brera, Via Bianchi 46, 23807 Merate, Italy
        \and
            \label{inst:calb}Centro de Astrobiología (CSIC-INTA), Carretera de Ajalvir, km 4. E-28850 Torrejón de Ardoz, Madrid, Spain
        \and
            \label{inst:PMO}Key Laboratory of Planetary Sciences, Purple Mountain Observatory, Chinese Academy of Sciences, Nanjing 210023, China
        \and
            Departamento de F\'{\i}sica e Astronomia, Faculdade de Ci\^encias, Universidade do Porto, Rua Campo Alegre, 4169-007, Porto, Portugal
        \and
            Physics Institute, University of Bern, Sidlerstrasse 5, 3012 Bern, Switzerland
        \and
            \label{inst:trie}INAF - Osservatorio Astronomico di Trieste, via G. B. Tiepolo 11, I-34143 Trieste, Italy
        \and
            Institute for Fundamental Physics of the Universe, Via Beirut 2, I-34151 Grignano, Trieste, Italy
        \and
            Consejo Superior de Investigaciones Cient\'{\i}ficas, Spain
        \and
            Centro de Astrobiolog\'{\i}a (CSIC-INTA), ESAC campus, E-28692 Villanueva de la Ca\~nada, Madrid, Spain
        \and
            European Southern Observatory, Karl-Schwarzschild-Strasse 2, 85748, Garching b. M\"unchen, Germany
        \and
            Centro de Astrof\'{\i}sica da Universidade do Porto, Rua das Estrelas, 4150-762 Porto, Portugal
        \and
            European Southern Observatory, Alonso de C\'ordova 3107, Vitacura, Regi\'on Metropolitana, Chile
        \and
            INAF - Osservatorio Astronomico di Palermo, Piazza del Parlamento 1, I-90134 Palermo, Italy
        \and
            Instituto de Astrof\'isica e Ci\^encias do Espa\c{c}o, Faculdade de Ci\^encias da Universidade de Lisboa, Campo Grande, PT1749-016 Lisboa, Portugal
        \and
            \label{inst:gal}INAF - Fundación Galileo Galilei, Rambla José Ana Fernandez Pérez 7, 38712 Breña Baja, Tenerife, Spain
        \and
            INAF - Osservatorio Astrofisico di Torino, via Osservatorio 20, 10025 Pino Torinese, Italy
        }


   \date{Received 28 September 2020 / Accepted 09 December 2020}


  \abstract{
  We observed two transits of the iconic gas giant HD~209458b between 380 and 780~nm, using the high-resolution ESPRESSO spectrograph. The derived planetary transmission spectrum exhibits features at all wavelengths where the parent star shows strong absorption lines, for example, \ion{Na}{i}, \ion{Mg}{i}, \ion{Fe}{i}, \ion{Fe}{ii}, \ion{Ca}{i}, \ion{V}{i}, H$\alpha$, and \ion{K}{i}. We interpreted these features as the signature of the deformation of the stellar line profiles due to the Rossiter-McLaughlin effect, combined with the centre-to-limb effects on the stellar surface, which is in agreement with similar reports recently presented in the literature. We also searched for species that might be present in the planetary atmosphere but not in the stellar spectra, such as TiO and VO, and obtained a negative result. Thus, we find no evidence of any planetary absorption, including previously reported \ion{Na}{i}, in the atmosphere of  HD 209458b.
  The high signal-to-noise ratio in the transmission spectrum (${\sim1700}$ at 590~nm) allows us to compare the modelled deformation of the stellar lines in assuming different one-dimensional stellar atmospheric models. We conclude that the differences among various models and observations remain within the precision limits of the data. However, the transmission light curves are better explained when the centre-to-limb variation is not included in the computation and only the Rossiter-McLaughlin deformation is considered. This demonstrates that ESPRESSO is currently the best facility for spatially resolving the stellar surface spectrum in the optical range using transit observations and carrying out empirical validations of stellar models.  
 }


   \maketitle
%
\section{Introduction}
\label{intro}

High-dispersion spectroscopy has become one of the most powerful tools for the atmospheric characterisation of exoplanets. The technique relies on the wavelength separation of the planetary, stellar, and telluric spectral lines that is due to their differential velocities \citep{Snellen2010}. So far, high-dispersion spectroscopy has been applied to both transiting (e.g. \citealt{SanchezLopez2019,Brogi2018}) and non-transiting planets (e.g. \citealt{Brogi2014,Guilluy2019}) and has led to the detection of several species in exoplanetary atmospheres, including: alkali lines and molecules in hot Jupiter atmospheres (\citealt{2008Redfield, Wytt2015,2017A&A...602A..36W,Chen2020,birkby2017}), tracers of evaporation and escape processes \citep{YanKELT9, Nortmann2018Science, Allart2018}, and ionospheric species in ultra hot Jupiter atmospheres \citep{Hoeijmakers2018, Casasayas2019,Seidel2019,Pino2020}.

With the initialisation of the operation of  the Echelle SPectrograph for Rocky Exoplanets and Stable Spectroscopic Observations (ESPRESSO; \citealt{Pepe2010ESP,Pepe2014ESP,Pepe2020}), our capabilities for this type of studies have been enhanced at optical wavelengths. ESPRESSO is already delivering ground-breaking measurements of time-resolved transmission spectra \citep{Ehrenreich2020, Borsa2020}, accurate Rossiter-McLaughlin (RM) measurements \citep{Santos2020} and high-precision characterisation measurements of multiple-planet systems (\citealt{Damasso2020, Alejandro2020}). Moreover, relatively small effects such as the impact of the stellar centre-to-limb variation (CLV) and the RM effect on the transmission spectra, which had fallen within the signal-to-noise ratio (S/N) of the results shown in previous studies, can now be measured and taken into account given the very high S/N and extreme quality of the ESPRESSO data. The impact of the RM effect in atmospheric studies was already described by \citet{LoudenW2015} and similarly, other studies \citep{cze15, Khalafinehad2017A&A...598A.131K} pointed out the importance of the CLV. The detailed understanding of these effects is crucial for the success of future observations of small rocky planet atmospheres with the extremely large telescopes (ELTs; \citealt{Snellen2013}).

Here, we revisit the atmosphere of the benchmark exoplanet HD~209458b using two transit observations with ESPRESSO. It was the first planet to be observed transiting its host star \citep{Charbonneau2000, Henry2000ApJ...529L..41H} and the first for which the detection of an atmosphere was claimed \citep{2002ApJ...568..377C} using \textit{Hubble Space Telescope} (HST) observations. HD~209458b is one of the most studied planets to date. This hot Jupiter orbits an F9-type star, and has a mass of $0.682~M_J$, radius of $1.39~R_J$, and equilibrium temperature of $1449$\,K (see more parameters in Table~\ref{tab:Param}). Over the years,  several studies of its atmosphere have been carried out using low- and high-resolution spectroscopy facilities. At low-resolution, several detections have been performed using the HST observations. For example, \citet{2002ApJ...568..377C} and \citet{Sing2008HD209ApJ...686..658S} reported \ion{Na}{i} absorption, \citet{Deming2013} found water in the atmosphere of the planet, and \citet{Desert2008} found tentative features of TiO and VO molecules.

High-resolution spectroscopy studies have demonstrated the detection of different atomic and molecular species. Using the High Dispersion Spectrograph (HDS) on the Subaru telescope, \citet{Narita2005HD209} reported upper limits for several lines, including the \ion{Na}{i} doublet. With the same data sets,  \citet{2008SnellenHD209} found absorption of \ion{Na}{i} in the transmission light curves, and \citet{Astudillo2013} detected calcium, and possibly scandium and hydrogen (H$\alpha$) in the atmosphere of the planet, and reconfirmed the \ion{Na}{i} detection. Similarly, \ion{Na}{i} absorption was reported by \citet{Albrecht2009} using the Ultraviolet and Visual Echelle Spectrograph (UVES) at the Very Large Telescope (VLT), and tentative features were shown by \citet{Langland2009} and \citet{jensen2011} using the High Resolution Echelle Spectrometer (HIRES) at Keck and the High-Resolution Spectrograph (HRS) at Hobby-Eberly Telescope (HET), respectively. On the other hand, \citet{Winn2004} were unable to detect H$\alpha$ absorption using observations with HDS at Subaru, while \citet{Jensen2012} found a broad feature correlated with the orbital phase centred at the H$\alpha$ position using observations with the HET.

More recently, \citet{Yan2017A&A...603A..73Y} studied the impact of the CLV when attempting to study the atmosphere of this planet using high-resolution spectroscopy. \citet{Casasayas2020}, on the other hand, analysed transit observations of HD~209458b with the HARPS-N (\citealt{HARPSN22012SPIE.8446E..1VC}) and CARMENES (\citealt{CARMENES,CARMENES18}) spectrographs, suggesting that the features observed in the high-resolution transmission spectrum could be explained by the combination of the RM effect and the CLV. With the current ESPRESSO observations, we aim at achieving a better characterisation of the signals in the transmission spectra and at exploring the presence of species in the planetary spectrum that are not present in the stellar spectrum.

This paper is organised as follows. In Sect.~\ref{sec:obs}, we present the observations. The methods for extracting the stellar parameters, the reloaded-RM technique, the atmospheric analysis, and the stellar contamination modelling are presented in Sect.~\ref{sec:analysis}. In Sect.~\ref{sec:results}, we present the results obtained in the atmospheric analysis of HD~209458b around individual lines and using cross-correlation techniques. In Sect.~\ref{sec:accuracy} we discuss the accuracy of the modelled stellar contamination. Our conclusions are presented in Sect.~\ref{sec:conc}.

\section{Observations and data reduction}
\label{sec:obs}

Two transits of HD~209458b were observed with ESPRESSO, a fiber-fed spectrograph located at the VLT that covers the optical wavelength range between 3800 and 7880$~{\rm\AA}$ on the nights of 2019-07-20 and 2019-09-11. Both observations were performed at the UT3 Melipal telescope, as part of the Guaranteed Time Observation under programme 1102.C-0744, using the HR21 observing mode, which considers 1-arcsec fiber, a binning of a factor of 2 along the spatial direction, achieving a resolving power of $\Re\sim{140~000}$ \citep{Pepe2020}.

The observations were performed following the typical observing strategy for transmission spectroscopy studies of exoplanets: monitoring the star with consecutive exposures before, during, and after the transit of the exoplanet. Over the course of the first night, one hour was dedicated to observing before and after the transit. On the second night, one hour was used before the transit and one and a half hours  after the transit. We used fiber A to observe the target and fiber B to monitor and subtract the sky signature. We used the same exposure time on both occasions ($175$\,s) obtaining a total of $89$ and $85$ exposures with an averaged S/N of $234$ and $193$ at 588~nm (physical order 104) on the two nights, respectively. The observations are summarised in Table~\ref{Tab:Obs}. Due to limitations in the atmospheric dispersion compensator, the exposures with an airmass larger than 2 are excluded from the analysis. This affects the six first exposures of the first night (2019-07-20). Therefore, a total of 83 spectra are used from that night, achieving an averaged S/N of $239$.

Here, we use the one-dimensional spectra (s1d sky subtracted products) extracted by the Data Reduction Software (DRS) pipeline version 2.0. When inspecting the sky spectra, we noticed that telluric \ion{Na}{i} emission is present in both nights, but corrected in the target spectra by the pipeline when the sky subtraction is applied. We also observe and correct for telluric \ion{Na}{i} absorption, which cannot be monitored by fiber B (see Sect.~\ref{sec:analysis}).

\begin{table*}[]
\centering
\caption{Observing log of the HD~209458b transit observations.}
\begin{tabular}{ccccccccc}
\hline\hline
Night & Telescope & Date of     & Start & End &  Airmass\tablefootmark{a} &$T_\mathrm{exp}$ & $N_\mathrm{obs}$ & S/N\tablefootmark{b} \\
         & & observation & [UT] & [UT] & change & [s] &         & \ion{Na}{i} order   \\ \hline
\\[-1em]
1 & VLT-UT3 & 2019-07-20 & 03:43 & 09:30 & 2.23--1.38--1.89 &175 & 89 & 122-261 \\ 
\\[-1em]
2 & VLT-UT3 & 2019-09-11 & 00:35 & 06:06 & 2.00--1.38--1.93 &175 & 85 & 132-228  \\ 
\\[-1em]
\hline
\end{tabular}\\
\tablefoot{\tablefootmark{a}{Airmass change during the observation.} \tablefoottext{b}{Minimum and maximum S/N for each night, calculated in the \ion{Na}{i} echelle order.}}
\label{Tab:Obs}
\end{table*}

\section{Analysis}
\label{sec:analysis}

\subsection{Stellar parameters}
\label{subsec:stellar_params}

Based on the ESPRESSO spectra, we derive the stellar atmospheric parameters of HD~209458 using the Equivalent Width method by means of the {\sc StePar}\footnote{\url{https://github.com/hmtabernero/StePar}} code \citep{StePar2019}, following the same methodology applied in recent ESPRESSO observations \citep{Tabernero2020,Ehrenreich2020}. In summary, {\sc StePar} relies on the 2017 version of the MOOG code \citep{Sneden1973} and a grid of MARCS stellar atmospheric models \citep{Gustafsson2008}. We use the \ion{Fe}{i} and \ion{Fe}{ii} line list from \citep{StePar2019} for metal-rich main sequence stars.

For comparison, we also used ARES+MOOG (\citealt{Sousa-14, Santos-13}) to derive the stellar atmospheric parameters and respective uncertainties. We used the usual line list from \citet{Sousa2008}, where the equivalent widths were measured with the ARES code\footnote{The last version of ARES code (ARES v2 - \url{http://www.astro.up.pt/~sousasag/ares}; \url{https://github.com/sousasag/ARES}} \citep{Sousa-07, Sousa-15}. Through the minimisation process, we find the ionisation and excitation equilibrium to converge in the best set of spectroscopic parameters. This process uses a grid of Kurucz model atmospheres \citep{Kurucz1993} and the radiative transfer code MOOG \citep{Sneden1973}.

Using {\sc StePar}, we measure an effective temperature ($T_{\rm eff}$) of $6069\pm54~{\rm K}$, gravity ($\log g$) of $4.41\pm0.13~{\rm cgs}$, metallicity ([Fe/H]) of $0.02\pm0.04$, and microturbulence velocity ($\xi$) of $1.03\pm0.08$\,km s$^{-1}$. On the other hand, using ARES+MOOG, $T_{\rm eff}=6139\pm62~{\rm K}$, $\log g=4.46\pm0.10~{\rm cgs}$, [Fe/H]$=0.05\pm0.04$, and $\xi=1.221\pm0.025$\,km s$^{-1}$. Using these previous spectroscopic values and PARAM1.3\footnote{\url{http://stev.oapd.inaf.it/cgi-bin/param_1.3}} \citep{daSilva2006}, we derive a stellar radius and mass of $R_{\star}= 1.136\pm0.027~R_{\odot}$ and $M_{\star}= 1.153\pm0.029~M_{\odot}$ for {\sc StePar} parameters, and $R_{\star}= 1.160\pm0.027~R_{\odot}$ and $M_{\star}= 1.116\pm0.029~M_{\odot}$, using ARES+MOOG results. These values are consistent with most of the previous studies (\citealt{Torres2008ApJ...677.1324T, Bonomo2017, Stassun2017HD209,Sousa2008,delBurgo2016}, among others). The stellar parameters derived here are summarised in Table~\ref{tab:Param}.

\begin{table}[]
\centering
\caption{Physical and orbital parameters of the HD~209458 system.}
\resizebox{\columnwidth}{!}{\begin{tabular}{lcr}
\\ \hline \hline
\\[-1em]
 Parameter  & Value & Reference\\ \hline
 \\[-1em]
 \multicolumn{3}{c}{\dotfill\it Stellar parameters\dotfill}\\\noalign{\smallskip}
   \\[-1em]
\quad  $T_{\rm eff}$ [K] & $6069\pm 54$ & This work ({\sc StePar}) \\
  \\[-1em]
\quad  & $6139\pm 62$ & This work (ARES+MOOG) \\
  \\[-1em]
\quad  $\xi$ [km\,s$^{-1}$]& $1.03\pm0.08$ &This work ({\sc StePar})\\
  \\[-1em]
\quad   & $1.221\pm0.025$ &This work (ARES+MOOG)\\
  \\[-1em]
\quad  $\log g$ [cgs]& $4.41\pm 0.13$ & This work ({\sc StePar})\\
  \\[-1em]
\quad  & $4.45\pm 0.10$ & This work (ARES+MOOG)\\
  \\[-1em]
\quad  [Fe/H] & $0.02\pm 0.04$ & This work ({\sc StePar})\\
  \\[-1em]
\quad   & $0.05\pm 0.04$ & This work (ARES+MOOG)\\
  \\[-1em]
\quad $M_{\star}$ [$\rm{M_{\odot}}$]& $1.116\pm0.029$ & This work ({\sc StePar})\\
  \\[-1em]
\quad & $1.153\pm0.029$ & This work (ARES+MOOG)\\
  \\[-1em]
\quad $R_{\star}$ [$\rm{R_{\odot}}$]& $1.160\pm0.027$ & This work ({\sc StePar})\\
  \\[-1em]
\quad & $1.136\pm0.027$ & This work (ARES+MOOG)\\
  \\[-1em]
 \quad  $v\sin i_{\star}$ [km\,s$^{-1}$]& $4.228\pm0.007$ & This work (reloaded-RM)\\
  \\[-1em]
 \multicolumn{3}{c}{\dotfill\it Planet parameters\dotfill}\\\noalign{\smallskip}
   \\[-1em]
 \quad $M_{\rm p}$ [$\rm{M_{Jup}}$]&  $0.682^{+0.015}_{-0.014}$ & \citet{Torres2008ApJ...677.1324T}\\
  \\[-1em]
 \quad $R_{\rm p}/R_{\rm \star}$ & $0.12086\pm0.00010$ & \citet{Torres2008ApJ...677.1324T}\\
 \\[-1em]
\quad $K_{p}$ [km\,s$^{-1}$]& $145.0\pm1.6$ & This work\tablefootmark{a} \\
  \\[-1em]
 \multicolumn{3}{c}{\dotfill\it Transit parameters\dotfill}\\\noalign{\smallskip}
   \\[-1em]
 \quad $T_{\rm 0}$ [BJD$_{\rm TDB}$] & $2454560.80588\pm0.00008$ & \citet{Evans2015MNRAS.451..680E} \\
  \\[-1em]
 \quad $P$ [day] & $3.52474859\pm0.00000038$ & \citet{Bonomo2017} \\
  \\[-1em]
 \quad $T_{14}$ [h]& $2.978\pm0.051$ & \citet{HD209Richardson_2006}\\
  \\[-1em]
 \quad $T_{23}$ [h]& $2.254\pm0.058$ & \citet{HD209Richardson_2006}\\
 \\[-1em]
 \multicolumn{3}{c}{\dotfill\it System parameters\dotfill}\\\noalign{\smallskip}
   \\[-1em]
  \quad $a/R_{\star}$& $8.87\pm0.05$ & \citet{Evans2015MNRAS.451..680E}\\
 \\[-1em]
 \quad $i$ [deg]& $86.78\pm0.07$&\citet{Evans2015MNRAS.451..680E}\\
    \\[-1em]
  \quad $a$ [au]& $0.04707^{+0.00045}_{-0.00047}$ & \citet{Bonomo2017}\\
 \\[-1em]
 \quad $e$& 0 & \citet{Bonomo2017}\\
\\[-1em]
 \quad $\omega$ [deg]& $90$ & \citet{Bonomo2017}\\
\\[-1em]
\quad $K_{\star}$ [m\,s$^{-1}$]& $84.27^{+0.69}_{-0.70}$ &  \citet{Bonomo2017}\\
  \\[-1em]
\quad $\gamma$ [km\,s$^{-1}$]& $-14.741\pm0.002$ & \citet{HD2092004Naef} \\
  \\[-1em]
 \quad $\lambda$ [deg]& $1.58\pm0.08$ & This work (reloaded-RM)\\
\\[-1em]
 \quad $b$& $0.511$ & This work (reloaded-RM)\\
\\[-1em]
\lasthline
\end{tabular}}
\tablefoot{\tablefoottext{a}{Derived assuming zero eccentricity ($e=0$): $K_p=2\pi a\sin(i)/P$}}
\label{tab:Param}
\end{table}

\subsection{Reloaded Rossiter-McLaughlin technique}
\label{subsec:relRM}

\begin{center}
\begin{figure}
\includegraphics[trim=0cm 0cm 0cm 0cm,clip=true,width=\columnwidth]{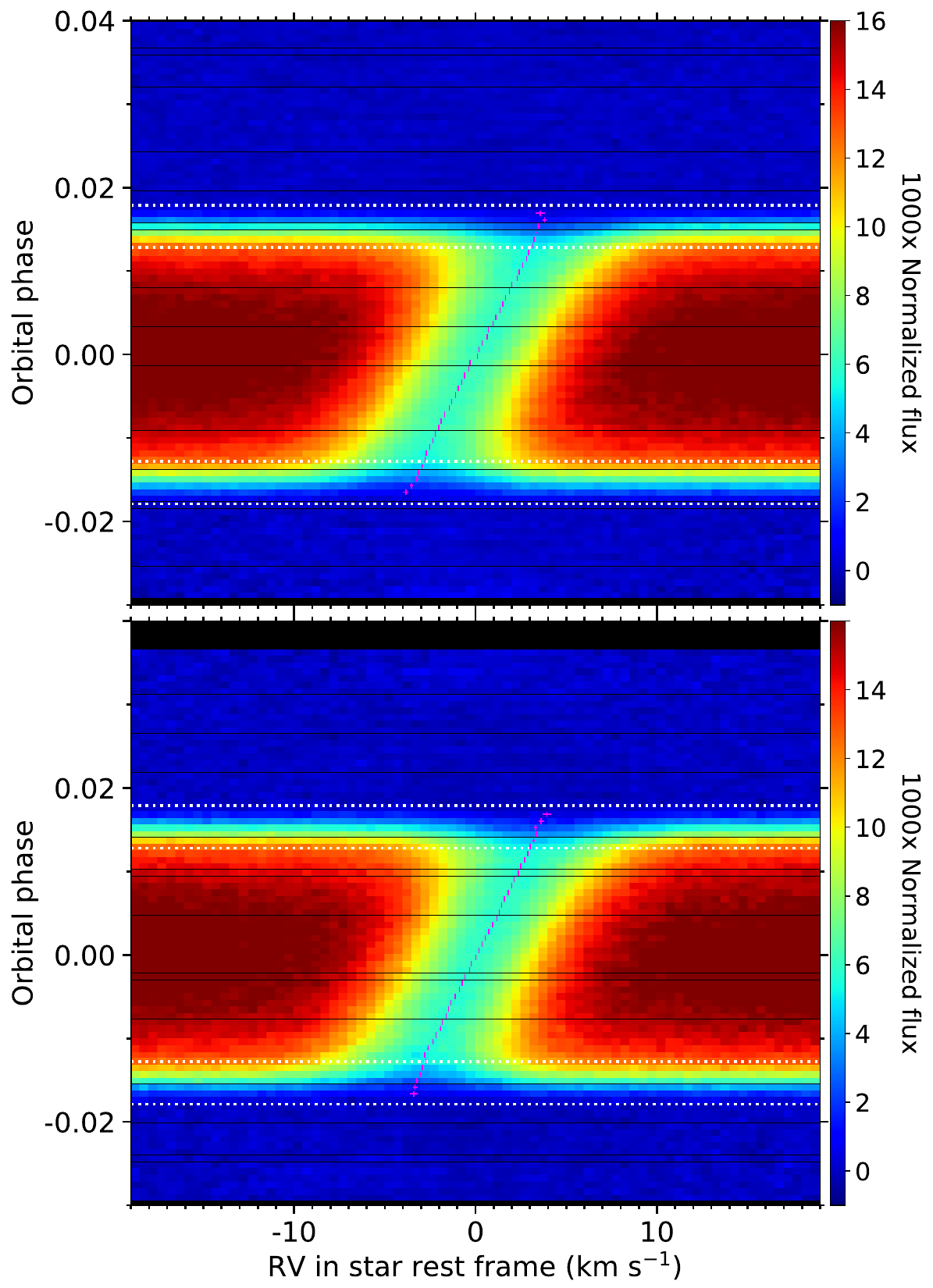}
\centering
\caption[]{Map of the CCF$_\mathrm{loc}$ series in the first (top panel) and second (bottom panel) night, as a function of orbital phase (in ordinate) and radial velocity in the stellar rest frame (in abscissa). Maps were obtained for the adjusted transit parameters (see text). Colours indicate the flux values. The four horizontal dashed white lines show the times of transit contacts. In-transit CCF$_\mathrm{loc}$ show the average stellar line profiles from the regions occulted by HD\,209458 b across the stellar disk. The magenta crosses are the measured centroids of the average stellar line profiles, corresponding to the local RVs of the planet-occulted regions. }
\label{fig:Local_maps}
\end{figure}
\end{center}

The CCFs generated by the ESPRESSO DRS (here after CCF$_\mathrm{DI}$, for disk-integrated) originate from starlight integrated over the disk of HD\,209458. We used the reloaded RM technique (\citealt{Cegla2016}, see also \citealt{Bourrier2017_WASP8,Bourrier_2018_Nat,Bourrier2020_HEARTSIII,Ehrenreich2020}) to isolate the local CCFs (heareafter, CCF$_\mathrm{loc}$) from the regions of the photosphere that are occulted by HD\,209458 b during its transit. The CCF$_\mathrm{DI}$ were first aligned by removing the Doppler-reflex motion of the star induced by the planet, calculated with the orbital properties in Table~\ref{tab:Param}. Since the ESPRESSO observations are not calibrated photometrically, each CCF$_\mathrm{DI}$ has to be continuum-scaled to reflect the planetary disk absorption. This was done using a light curve computed with the batman package (\citealt{Kreidberg2015}) and the properties shown in Table~\ref{tab:Param}. The CCF$_\mathrm{DI}$ outside of the transit were co-added to build a master-out CCF$_\mathrm{DI}$ representative of the unocculted star. We then aligned all CCF$_\mathrm{DI}$ in the star rest frame using the systemic velocity measured via a Gaussian fit to the master-out in each visit to account for possible nightly offsets in the instrumental, atmospheric, and astrophysical noise. Residual CCF$_\mathrm{loc}$ were then obtained by subtracting the scaled CCF$_\mathrm{DI}$ from the master-out in each visit (Fig.~\ref{fig:Local_maps}). Errors were propagated at each step from the CCF$_\mathrm{DI}$ to the CCF$_\mathrm{loc}$. \\

\begin{center}
\begin{figure}
\includegraphics[trim=0cm 0cm 0cm 0cm,clip=true,width=\columnwidth]{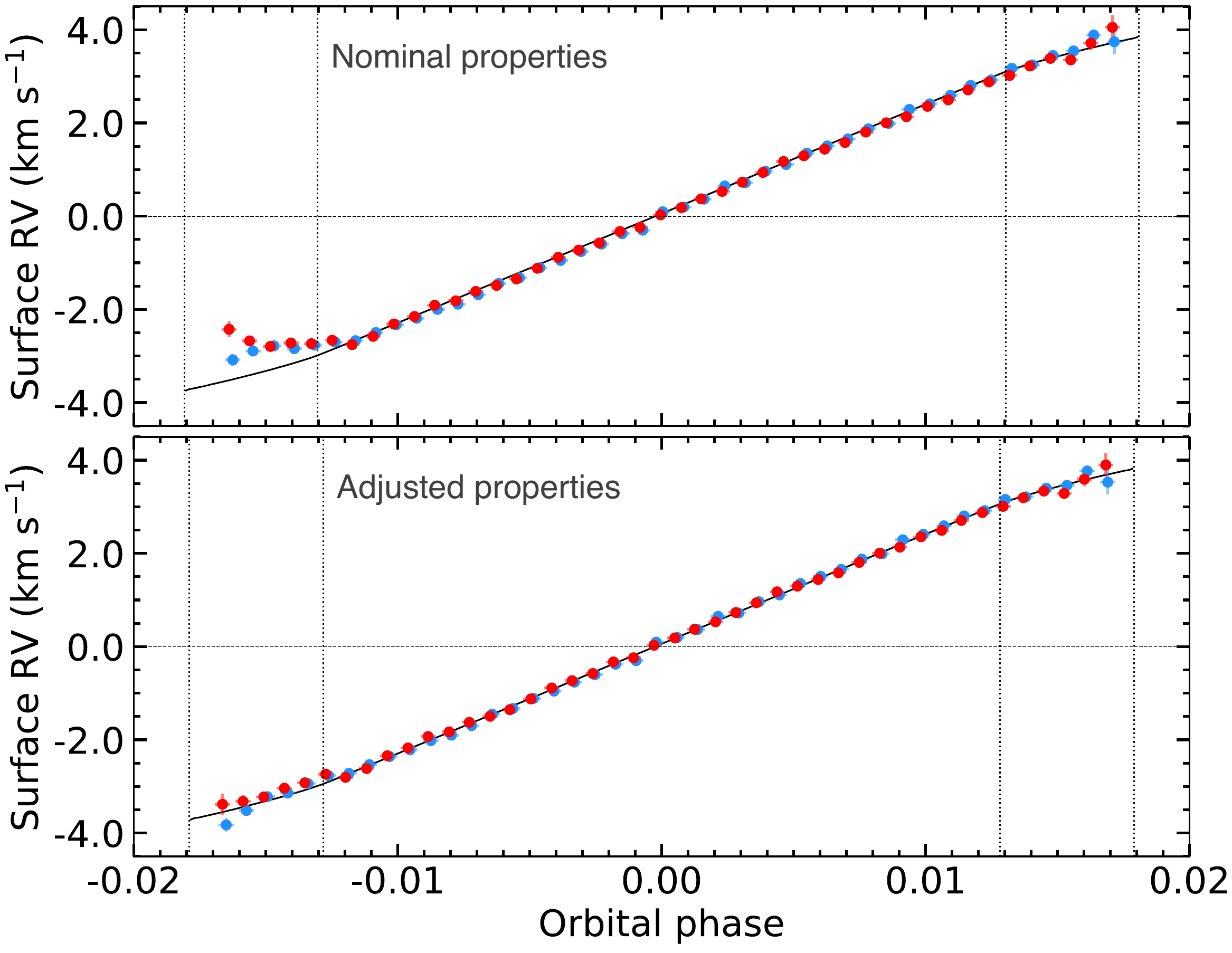}
\centering
\caption[]{RVs of the stellar surface regions occulted by HD\,209458 b in the first (blue points) and second (red points) night. Their extraction was performed for the nominal transit properties from \citet{Evans2015MNRAS.451..680E} in the top panel and for the adjusted properties in the bottom panel. The solid black line is the best-fit solid-body model to the RVs obtained for the adjusted properties. Vertical dotted lines show the times of transit contacts.}
\label{fig:Local_RVs}
\end{figure}
\end{center}

The CCF$_\mathrm{loc}$ spectrally and spatially resolve the photosphere of the star along the transit chord. The average stellar lines from the planet-occulted regions were fitted with independent Gaussian profiles and Levenberg-Marquardt least-squares minimisation to derive the local RVs of the stellar surface. The average local stellar lines are well fitted with Gaussian profiles and are detected in all exposures, except the first and last ones of ingress and egress on each night (the detection criterion is that the amplitude of the model CCF$_\mathrm{loc}$ is four times larger than the dispersion in the measured CCF$_\mathrm{loc}$ continuum). Thus, ESPRESSO allows us to sample, with a high temporal resolution, the full transit window, which translates into a fine spatial sampling of the full transit chord out to the limbs of the star (Fig.~\ref{fig:Local_RVs}). This has led us to identify an abnormal deviation of the local RVs during ingress. We found that the local RV series, particularly at the limbs, is highly sensitive to minute variations in the assumed mid-transit time and impact parameter. The efficiency of the reloaded RM technique relies on the possibility of analyzing spectrocopic data using photometry of a similar quality, so that the shape of the transit light curve and the orbital phase of the exposures are known to a sufficient level of precision. Deviations from the true photometric scaling and phasing of the CCF$_\mathrm{DI}$ may otherwise bias the extracted CCF$_\mathrm{loc}$ and their measured centroid. Here, the high quality of HD\,209458~b observations with ESPRESSO is not matched by our knowledge of its transit properties, which were derived from photometry obtained several years ago, resulting in a present-day uncertainty on T$_\mathrm{0}$ of about 40\,s.\\    

In order to estimate the precision required on the transit properties and their impact on the derived obliquity $\lambda$ and projected stellar rotational velocity, $v\sin i_{\star}$, we assumed that HD\,209458 rotates as a solid body and we performed the extraction of the local RVs over a grid of T$_\mathrm{0}$ and $b$ values. Different values for these properties can affect which exposures are considered in-transit, and which ingress or egress exposures yield a detection for the average local stellar line. Thus, different extractions can yield different series of local RVs, preventing us from using $\chi^2$ minimisation. We searched instead for the approximate T$_\mathrm{0}$ and $b$ values that minimise the dispersion of the residuals between a given local RV series and its best-fit solid-body model (described in \citealt{Cegla2016}, \citealt{Bourrier2017_WASP8}). We found that the local RVs best agree with solid-body rotation when the mid-transit time is shifted by 76\,s (2454560.80676\,BJD) and the impact parameter equals 0.511 (Fig.~\ref{fig:Local_RVs}). Provided that those properties are correct and HD\,209458 rotates as a solid-body, we then derive $\lambda$ = 1.58$\pm$0.08$^{\circ}$ and $v\sin i_{\star}$ = 4.228$\pm$0.007\,km\,s$^{-1}$. We caution that these uncertainties are underestimated, as they do not account for the additional uncertainty on the transit depth (Table
~\ref{tab:Param}) and limb-darkening coefficients (set using the code provided by \citealt{Espinoza2015}). Nonetheless, varying those properties changes the derived $v\sin i_{\star}$ by less than 50\,m\,s$^{-1}$ and has a negligible impact on $\lambda$. Even fitting the biased local RVs associated with \citet{Evans2015MNRAS.451..680E} transit properties only changes $\lambda$ by $\sim$1$^{\circ}$ (0.48$^{\circ}$) and $v\sin i_{\star}$ by 44\,m\,s$^{-1}$ (4.184\,km\,s$^{-1}$). Therefore, while our adjusted values for T$_\mathrm{0}$ and $b$ should be considered with caution, the derived obliquity is likely precise within a degree and the projected stellar rotational velocity within 100\,m\,s$^{-1}$. We note that the analysis of the HD\,209458b data by \citet{Santos2020}, based on the classical RM approach, similarly yields a mid-transit time shifted to later values. We also note the high repeatability of the local RV series, despite them being separated by about two months -- highlighting the stability of ESPRESSO.

Despite the strong influence of T$_\mathrm{0}$ and $b$ on the RM analysis of  HD\,209458~b, our adjusted values differ by less than 2$\sigma$ from the nominal values of \citet{Evans2015MNRAS.451..680E}. Reaching a high accuracy on $\lambda$ and $v\sin i_{\star}$ in the HD\,209458 system, and searching for fine RV variations associated with the stellar surface motion (e.g. differential rotation, convective blueshift), will require an extreme precision of the transit properties that only a space-based facility such as CHEOPS or TESS can provide.

\subsection{Telluric correction}
\label{subsec:tell}

As in other recent studies using ESPRESSO observations \citep{Chen2020,Tabernero2020,Allart2020}, we use {\tt molecfit} (\citealt{Molecfit1} and \citealt{Molecfit2}) to correct for the telluric absorption contamination (O$_2$ and H$_2$O) from the Earth's atmosphere as presented in \citet{Allart2017}. In this first correction, other telluric contamination such as emission or absorption of \ion{Na}{i}, for example, are not considered. 

In the ESPRESSO observations used here, we observe both telluric \ion{Na}{i} emission and absorption contamination. The telluric emission is monitored with fiber B and is already corrected by the DRS during the sky subtraction, but this is not the case of the telluric absorption. The mean Earth radial velocity is $17.6$\,km\,s$^{-1}$ and $-3.8$\,km\,s$^{-1}$ for the first and second night, respectively. Considering the systemic velocity of HD~209458 ($\gamma=-14.7$\,km\,s$^{-1}$), the telluric absorption is located at $32.3$\,km s$^{-1}$ ($0.6~{\rm \AA}$) and $10.9$\,km s$^{-1}$ ($0.2~{\rm \AA}$) from the \ion{Na}{i} lines core. Although its presence does not impact the results of the first night because of the distance with respect the stellar \ion{Na}{i} lines core, it is indeed important for the second night and needs to be corrected.

\begin{figure}[]
\centering
\includegraphics[width=0.45\textwidth]{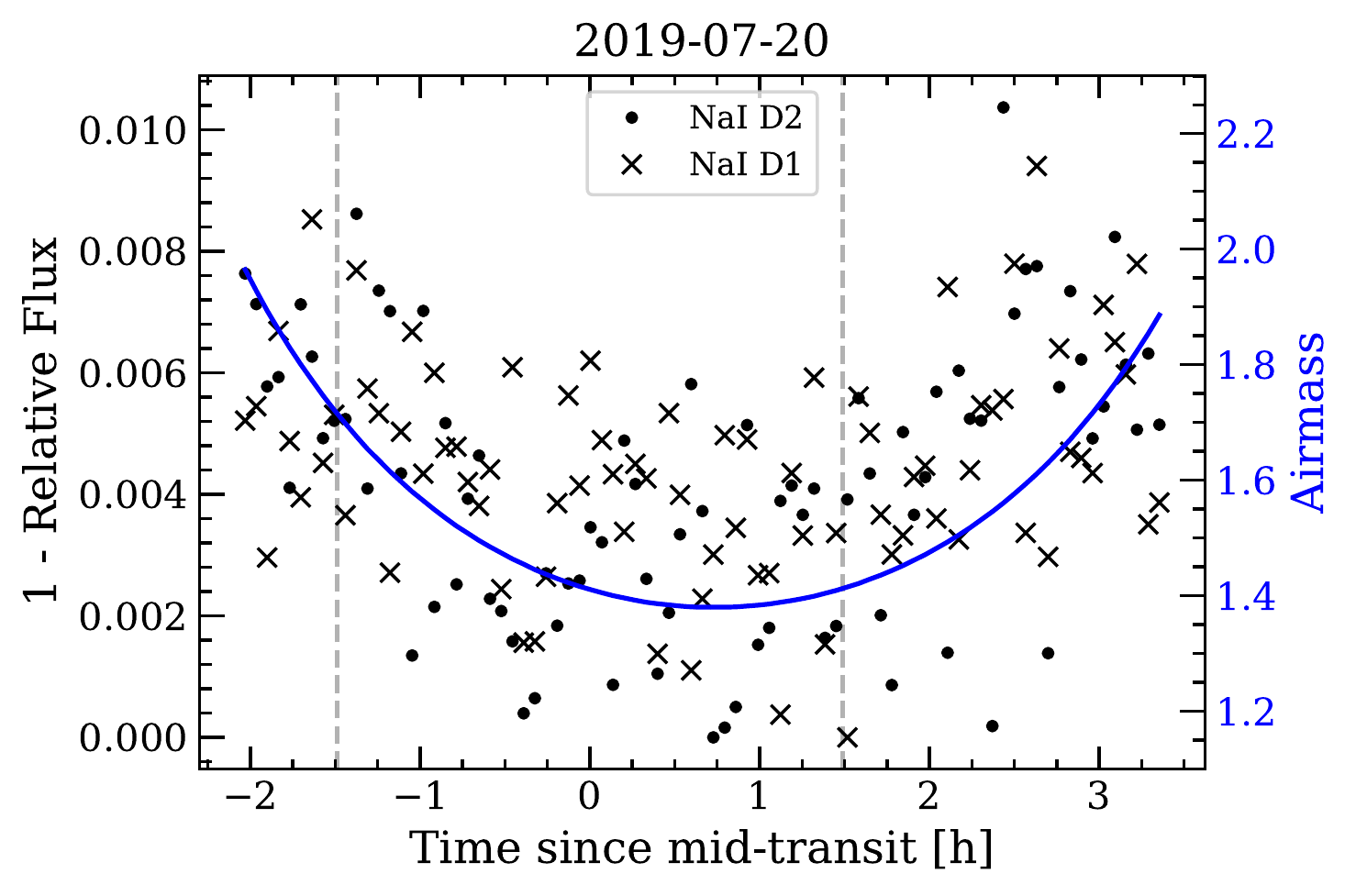}
\includegraphics[width=0.45\textwidth]{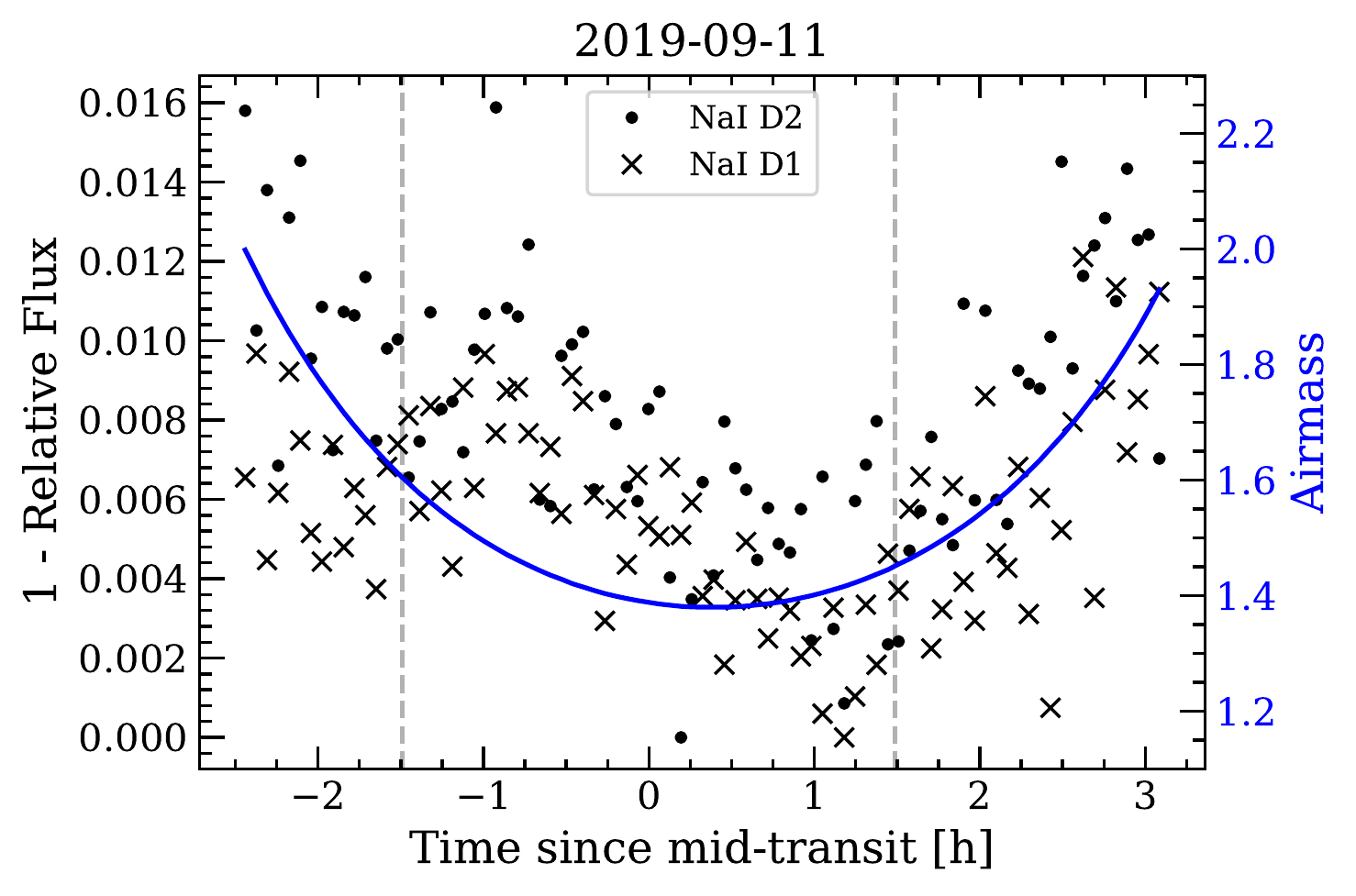}
\caption{Integrated flux in a $\pm5$\,km s$^{-1}$ passband centred on the telluric \ion{Na}{i} absorption lines of the first night (top panel) and second night (bottom panel). The integrated flux is normalised to its maximum value. In dots, we show the results for the  \ion{Na}{i} D2 line and in crosses, the results of the \ion{Na}{i} D1 line. The vertical dashed lines mark the origin and end of the transit.}
\label{fig:Na_sky}
\end{figure}

It is known that the telluric \ion{Na}{i} absorption can show strong seasonable variability \citep{2008SnellenHD209}. However, in order to correct for this contamination we assume that, within a night, the telluric \ion{Na}{i} can be corrected as presented, for example, in \citet{Borsa2020}, \citet{Wytt2015}, \citet{Zhou2012}, and \citet{VidalMadjar2010}. That is, assuming that the contrast variation of these lines during the observations is correlated with the airmass. This correction is only applied in a small region of $\pm5$\,km s$^{-1}$ centred at the position of the telluric \ion{Na}{i} absorption lines in order to not influence other regions. In Figure~\ref{fig:Na_sky}, we show the flux variation of this region with time. For the first night, the absorption lines follow the airmass change for the full observation while for the second night this only happens in the out-of-transit exposures. This is because the telluric \ion{Na}{i} of the second night is very close to the lines core and thus influenced by the line profile changes produced by the planet during the transit. For the first night, telluric \ion{Na}{i} is far from the lines core. 

\subsection{Transmission spectrum and light curves extraction}
\label{subsec:TS_extract}

After correcting for the telluric contamination, we extract the transmission spectrum as presented in previous studies (such as \citealt{Wytt2015,2017CasasayasB,YanKELT9,Borsa2020,Allart2017}). First, the spectra are normalised and then moved to the stellar rest frame. To do so, we use the Keplerian model computed with the parameters presented in Table~\ref{tab:Param} using the {\tt SinRadVel} model from {\tt PyAstronomy} \citep{PyAstronomy2019ascl.soft06010C}. Then we compute the master stellar spectrum resulting from the combination of all out-of-transit spectra and divide each individual spectrum by this master spectrum.

After computing the ratio of each stellar spectrum by the master stellar spectrum, we could notice a clear wiggle (sinusoidal) pattern already observed in recent ESPRESSO observations \citep{Tabernero2020,Borsa2020}, with amplitudes around $1~\%$ and periods $\sim30-40~{\rm \AA}$. In order to correct for this pattern, we follow the methodology presented by \citet{Borsa2020}: we fit a sinusoidal curve with varying period, amplitude, and phase at each individual spectrum after being divided by the master spectrum. The data are normalised using the resulting best fit model. We then move all the residuals to the planet rest frame by using the planet radial-velocity semi-amplitude $K_p = 145.0\pm1.6$\,km s$^{-1}$, calculated assuming zero eccentricity ($K_p=2\pi a\sin(i)/P$). Finally, the in-transit residuals are combined to compute the transmission spectrum. The combination of the in-transit residuals is performed using the simple average, as the residuals change at different orbital phases. The weighted average gives more importance to the orbital phases with higher S/N and, consequently, modifies the shape of the transmission spectrum features.

\begin{figure}[]
\centering
\includegraphics[width=0.5\textwidth]{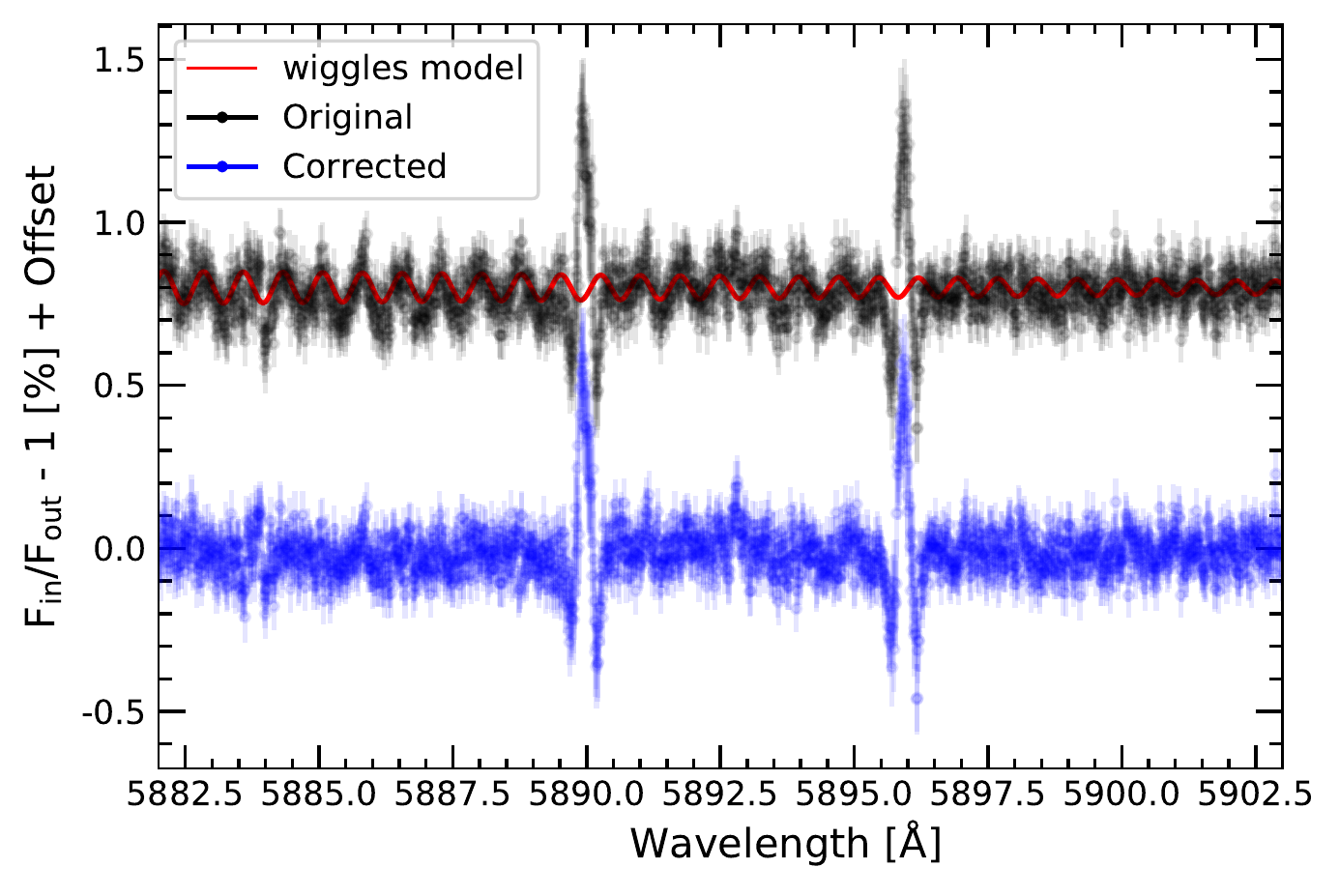}
\caption{Correction of the interference pattern in the final mid-transit transmission spectrum of the first night (2019 July 20). The original transmission spectrum in shown in black and the corrected transmission spectrum is shown in blue. In red and on-top of the uncorrected spectrum, we show the best fit sinusoidal curve. An offset between the two spectra is added for a better visualisation. }
\label{fig:wigg}
\end{figure}

In the final transmission spectra we can observe a second sinusoidal pattern, which is particularly clear in the pseudo-continuum (see Figure~\ref{fig:wigg}) due to the high S/N of the data. The origin of these patterns lies in non-flat-fielded interference patterns that occur in some optical elements of the Coudé Train (CT), that is, in the system that routes the light from the telescope to the instrument. This system is not spectrally flat-fielded by the calibration sources, which are injected downstream. Due to the (small) changes of the light path through the optical elements during the observations, the fringe pattern will evolve spectrally and eventually pop up in transmission spectroscopy when dividing by each other spectra taken at different moments of the night \footnote{A forthcoming paper will explain in details the origin of those wiggles.}. In order to remove the pattern from the final transmission spectrum, we fit a sinusoidal curve to the data, masking the strong features observed in the \ion{Na}{i} position and we then use the solution to normalise the full wavelength range of the transmission spectrum. As we can see in the top spectrum of Figure~\ref{fig:wigg}, although the period of the oscillations is almost constant, their amplitude changes in wavelength ($\sim \pm0.1~\%$ at bluer wavelengths for the first night). For this reason, we let the amplitude of the sinusoidal change linearly in wavelength, while the remaining free parameters (period and phase of the origin) only have one value in a particular wavelength range.

On the other hand, the period of the oscillations is first estimated by applying a Fourier transform on the combination of all in- and out-of-transit residuals and it is then fitted. The period we derive in both Fourier transform and the sinusoidal fit is $0.75~{\rm \AA}$ around the \ion{Na}{i} and it slightly varies in wavelength. This correction is applied to all transmission spectra presented here, fitting the pattern in each case in a wavelength region centred around the line of interest.

In order to calculate the transmission light curves, once the residuals are moved to the planet rest frame, we fix a passband of a given width centred on the specific line we want to analyse. Then, we integrate the flux inside this passband using trapezoidal integration. The method is the same as presented in \citet{Casasayas2020}. The transmission light curves are presented in Section~\ref{sec:results}, in particular, in Figures~\ref{fig:tlc_Na} and \ref{fig:oth_lines1}.

\subsection{Cross-correlation}
\label{subsec:crosscorr}

In addition to the analysis of individual lines, we apply the cross-correlation technique in order to search for atoms and molecules that may originate from hundreds to thousands of individual absorption lines in the transmission spectrum, if present in the atmosphere of the planet. Using the cross-correlation method, the contribution of all these lines is combined, reducing the photon noise and reaching the detection of particular atoms and molecules hidden in the noise when analysed individually \citep{Snellen2010}.

Here, we use the one-dimensional {\tt molecfit}-corrected spectra, which are then normalised and moved to the stellar rest frame. We divide each spectrum by the master stellar spectrum (computed using only the out-of-transit data), remove the wiggle pattern as presented in Section~\ref{subsec:TS_extract}, and cross-correlate the result with atmospheric models, as presented in \citet{SanchezLopez2019} and \citet{Stangret2020}. We discard the strong telluric contaminated regions around 690 nm and 760 nm where the telluric correction is not accurate, as well as the first few blue bins from 380 to 450\,nm due to their lower S/N. With this, all the spectra for the two nights have a S/N larger than 90 in each order.

The atmospheric models used to cross-correlate with the residuals are computed using the {\tt petitRADTRANS} code \citep{petitRADTRANS2019} which allows the generation of high-resolution models of atoms and molecules for exoplanet atmospheres. Here, we generate \ion{Fe}{i}, \ion{Fe}{ii}, and \ion{Ca}{i} models assuming the parameters presented in Table~\ref{tab:Param}, isothermal temperature profile of $1459~{\rm K}$, a continuum level of $1~{\rm mbar}$, and the volume mixing ratio (VMR) set to solar abundance. For these atomic species, the Kurucz line lists are used \citep{Kurucz1993}. The resulting models are convolved to match the ESPRESSO spectral resolution. On the other hand, we generate the atmospheric models of TiO and VO molecules using two different methodologies. First, we use {\tt petitRADTRANS} as described above and in considering VMR$_{\rm TiO}=10^{-7}$ and VMR$_{\rm VO}=10^{-8}$ (solar abundance), and the line lists presented in \citet{Plez1998}. Second, as presented in \citet{Tabernero2020}, we use {\tt HELIOS} code\footnote{\url{https://github.com/exoclime/HELIOS}} \citep{Malik2017,Malik2019} to calculate the atmospheric structure of HD~209458b. Then, the radiative transfer problem is solved using {\tt turbospectrum}\footnote{\url{https://github.com/bertrandplez/Turbospectrum2019}} \citep{Plez2012}. We use the TiO line list from \citet{Plez1998} and the VO list from {\tt exomol} \citep{McKemmish2016}. The cross-correlation results are presented in Section~\ref{sec:results}.

\subsection{Transit effects on the stellar lines}
\label{subsec:CLVRMmodels}

Following \citet{Yan2017A&A...603A..73Y} and \citet{YanKELT9}, we model the CLV and RM effects in the stellar line profile. This method involves the {\tt Spectroscopy Made Easy Tool} (SME; \citealt{SME}) and the line list from {\tt VALD}\footnote{\url{http://vald.astro.uu.se}} database \citep{VALDPisk1995,VALDKup1999}. Using SME we are able to compute synthetic stellar spectra at different limb-darkening angles. With this information, it is possible to compute the disc integrated stellar spectrum considering the stellar rotation ($v\sin i_{\star}$), but also to exclude the spectrum of those regions blocked by the planet at each exposure when the integrated disc spectrum is built, producing the RM and CLV deformation. 

This has already been applied to different high resolution atmospheric studies such as \citet{Casasayas2019}, \citet{Yan2019}, \citet{Chen2020}, and \citet{Borsa2020}. For this computation we need accurate measurements of the system and stellar parameters. Here, we take advantage of the high precision achieved with the reloaded RM technique in the $v\sin i_{\star}$ and $\lambda$ measurements (see Sect.~\ref{subsec:relRM}). These parameters are presented in Table~\ref{tab:Param}. In Sect.~\ref{sec:accuracy}, we examine the resulting modelled effects depending on the stellar atmospheric models that are used for their computation. 

Although for some planets these effects are not important, it has been observed that they can become important for particular planets \citep{cze15,Khalafinehad2017A&A...598A.131K,LoudenW2015}, and specially when using very high S/N observations (see \citealt{Borsa2020}). In most cases, the RM effect can influence the final absorption from the planet atmosphere if both contributions (RM and atmosphere) overlap at some point during the transit \citep{YanKELT9,Hoeijmakers2018}. Here, as observed by \citet{Casasayas2020}, we are in an extreme case where the RM (almost) fully overlaps with the expected planetary atmosphere track. This deformation is not constant along the planet transit, hence, it is particularly important to use the same steps when processing the models and the data for useful comparisons.


\section{Results}
\label{sec:results}

In this section, we present the results obtained in the analysis of individual atomic lines (\ion{Na}{i}, \ion{Fe}{i}, \ion{Mg}{i}, H$\alpha$, and \ion{K}{i} D1) and forests of atomic lines (\ion{Fe}{i}, \ion{Fe}{ii}, \ion{Ca}{i}, and \ion{V}{i}) using the cross-correlation technique. For comparison, we show the results together with the RM and CLV models computed using MARCS \citep{Gustafsson2008} stellar atmospheric models and assuming LTE approximation. The models containing both CLV and RM effects combined and only RM are shown. Using the cross-correlation technique we search for relevant molecules, such as TiO and VO.

\subsection{\ion{Na}{i} doublet}
\label{subsec:Na}

The transmission spectrum and tomography maps around the \ion{Na}{i} of both nights combined are presented in Figure~\ref{fig:ts_Na}, and the results of the individual nights are shown in Figure~\ref{fig:ts_Na_indiv} in the appendix. In all cases, as presented in \citet{Casasayas2020}, at the \ion{Na}{i} line positions in the transmission spectrum we observe emission-like features consistent with the deformation of the stellar line profile during the transit of HD~209458b, which are due to the RM and CLV effects. The particularity of HD~209458 system is that the expected exoplanet atmosphere and the RM effect deformation (which is the main contribution of the final transmission spectrum shape) fall at almost the same radial velocities (see tomography maps of Figure~\ref{fig:ts_Na}). For this reason, when the individual transmission spectra are moved to the planet rest frame before being combined, the positive part of the RM effect is almost aligned with the \ion{Na}{i} position and creates an emission-like feature in the transmission spectrum. 

\begin{figure*}[h]
\centering
\includegraphics[width=1.0\textwidth]{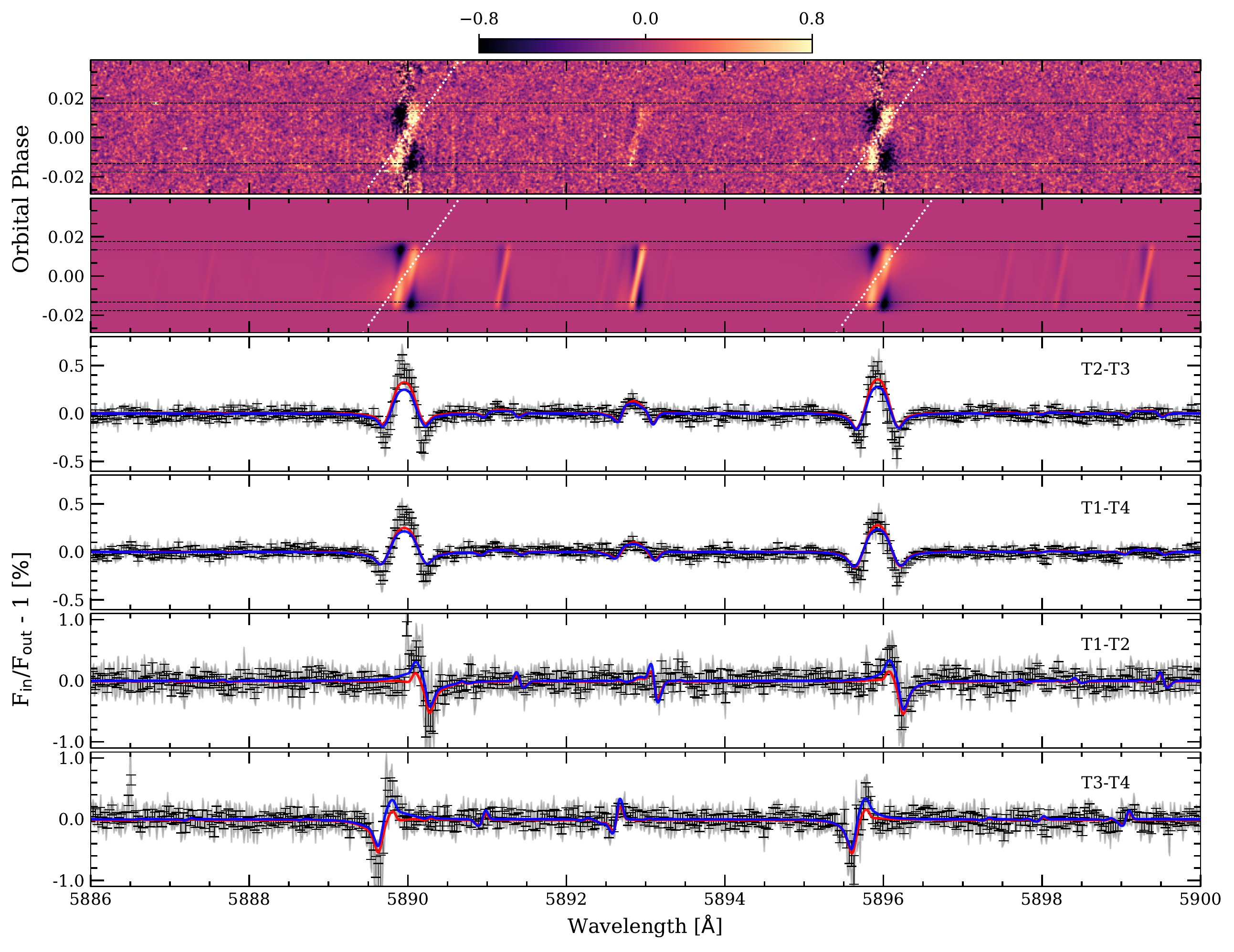}
\caption{\textit{First panel (top)}: Two-dimensional map of the individual transmission spectra around the \ion{Na}{i} doublet lines of HD~209458b. \textit{Second panel}: Two-dimensional map of the modelled CLV and RM effects around the \ion{Na}{i} doublet. In both panels, the results are presented in the stellar rest frame and the colour bar shows the relative flux (F$_{\rm in}$/F$_{\rm out}$-1) in \%. The black horizontal dashed lines indicate the four contacts of the transit and the dotted white line marks the expected position of the planet trail during the observations. \textit{Third panel}: Transmission spectrum computed combining the data between the second and third contacts of the transit. \textit{Fourth panel}: Combination of the in-transit exposures from the first to fourth contacts. \textit{Fifth panel}: Combination of ingress exposures. \textit{Sixth panel (bottom)}: Combination of egress exposures. The transmission spectra are computed combining the data in the planet rest frame. The original data error bars are shown in light grey and the data binned by $0.03~{\rm \AA}$ is shown in black. We note the different y-scale in the different panels. In red, we show the CLV and RM effects in the final transmission spectrum and in blue, the deformation due to the RM alone.} 
\label{fig:ts_Na}
\end{figure*}

Due to the system geometry, the stellar line deformation is asymmetric along the transit. With the high S/N of ESPRESSO data, these asymmetries can be explored by computing the transmission spectrum during different transit times. In Figure~\ref{fig:ts_Na}, we present the transmission spectrum computed using the spectra obtained during the ingress (T1-T2), the egress (T3-T4) between the first and last contacts of the transit (T1-T4) and between the second and third contacts (T2-T3). For the ingress and egress transmission spectra, we use around six spectra per night. The asymmetries of the ingress and egress regions can be clearly observed and are predicted by the CLV and RM modelled effects. In the time-regions of orbital phases $[-0.0175,-0.0100]$ and $[+0.0100,+0.0175],$ the planet radial-velocities do not fully overlap with the RM effect. However, we are not able to distinguish absorption due to the exoplanet atmosphere.

The asymmetries of the stellar line shapes during the transit are particularly important, especially if some orbital phases of the planet are not covered during the observations, as they could result in false absorption-like signals depending on the geometry of the system (see \citealt{Chen2020}). In Figure~\ref{fig:ts_Na}, we also see how the transmission spectra of the ingress and egress show absorption-like features shifted from the laboratory \ion{Na}{i} lines position, which result from the deformation of the lines and are not from atmospheric origin. Although the impact of the effects is smaller in the mid-transit times when the RM radial-velocities are used to move the spectra to the stellar rest frame, these asymmetries increase, especially during the ingress and egress.

In addition, we compute the transmission light curves for two different bandwidths ($0.4~{\rm \AA}$ and $0.75~{\rm \AA}$). The small passband size is selected to only include the positive RM effect ($\sim 10$\,km\,s$^{-1}$), while the larger passband includes the overall RM effect and has also been used in previous studies \citep{2008SnellenHD209, Albrecht2009}. In Figure~\ref{fig:tlc_Na} we present the results after combining both nights, and in Figure~\ref{fig:tlc_Na_indiv} of the Appendix we show the results of each individual night. In all cases both \ion{Na}{i} lines of the doublet are combined. We note that observations on the second night (2019 September 11) could be partially affected by telluric \ion{Na}{i} absorption residuals. As they are calculated in the planet rest frame, for the narrower passband we are mainly combining the positive part of the RM deformation, while with the wider passband both contributions are included, resulting in a compensation of the final effects. The transmission light curves follow the general shape predicted by the model containing only RM deformation; this is especially clear for the $0.4~{\rm \AA}$ passband, where the differences with the out-of-transit are more intense. In this case, in the centre of the transit, the transmission light curve clearly deviates from the model containing the CLV (see discussion in Sect.~\ref{sec:accuracy}). 

\begin{figure}[h]
\centering
\includegraphics[width=0.4\textwidth]{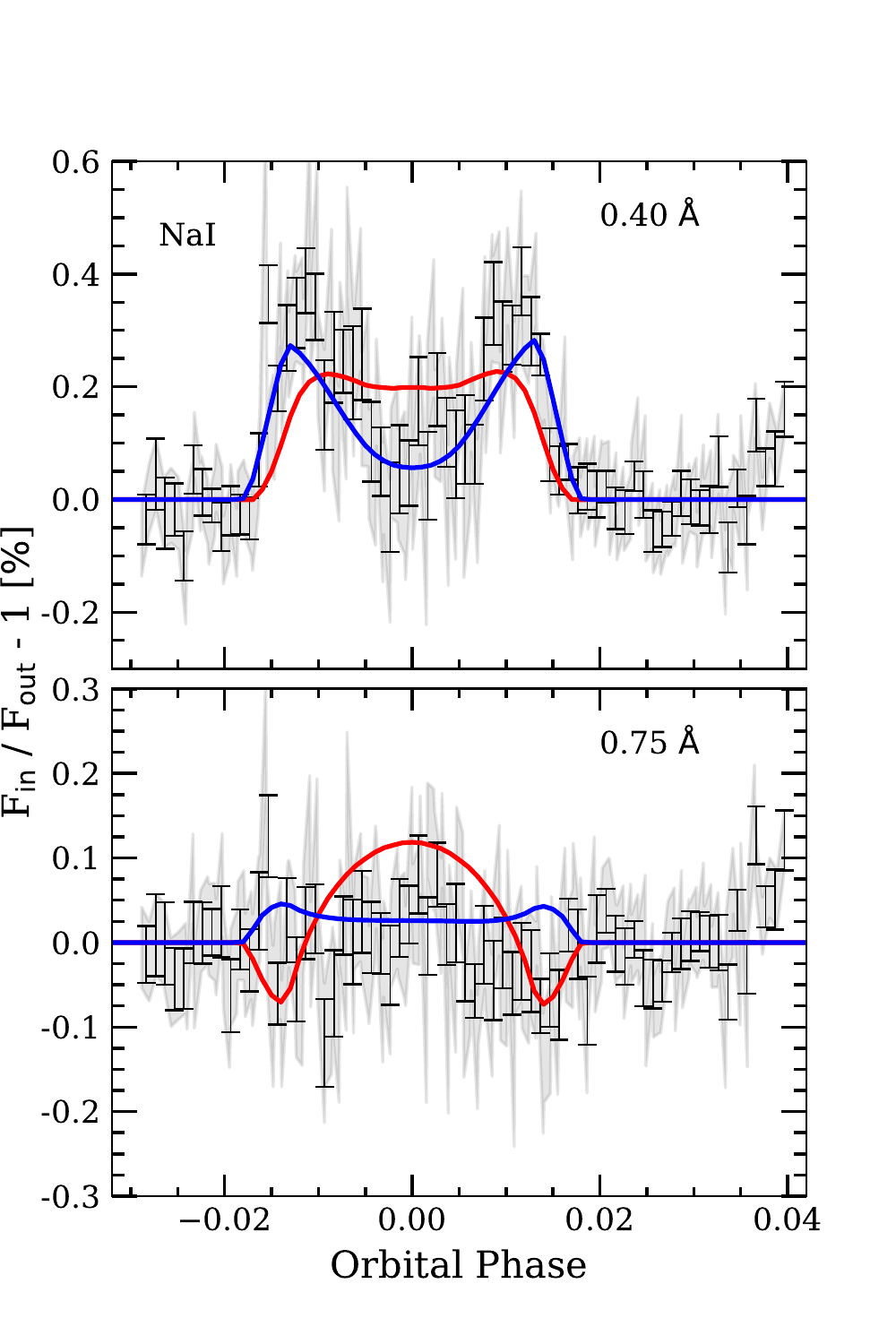}
\caption{Transmission light curves calculated in \ion{Na}{i} doublet after combining both nights, using a $0.4~{\rm \AA}$ (\textit{top}) and $0.75~{\rm \AA}$ (\textit{bottom}) passbands. In light-grey. we show the original data. In black, we show the data binned by $0.001$ in orbital phase. In red we show the impact of the CLV and RM effects in the transmission light curve. In blue, only the RM effect is considered.}
\label{fig:tlc_Na}
\end{figure}

\subsection{Other atomic lines}
\label{subsec:otherlines}

\begin{figure}[]
\centering
\includegraphics[width=0.5\textwidth]{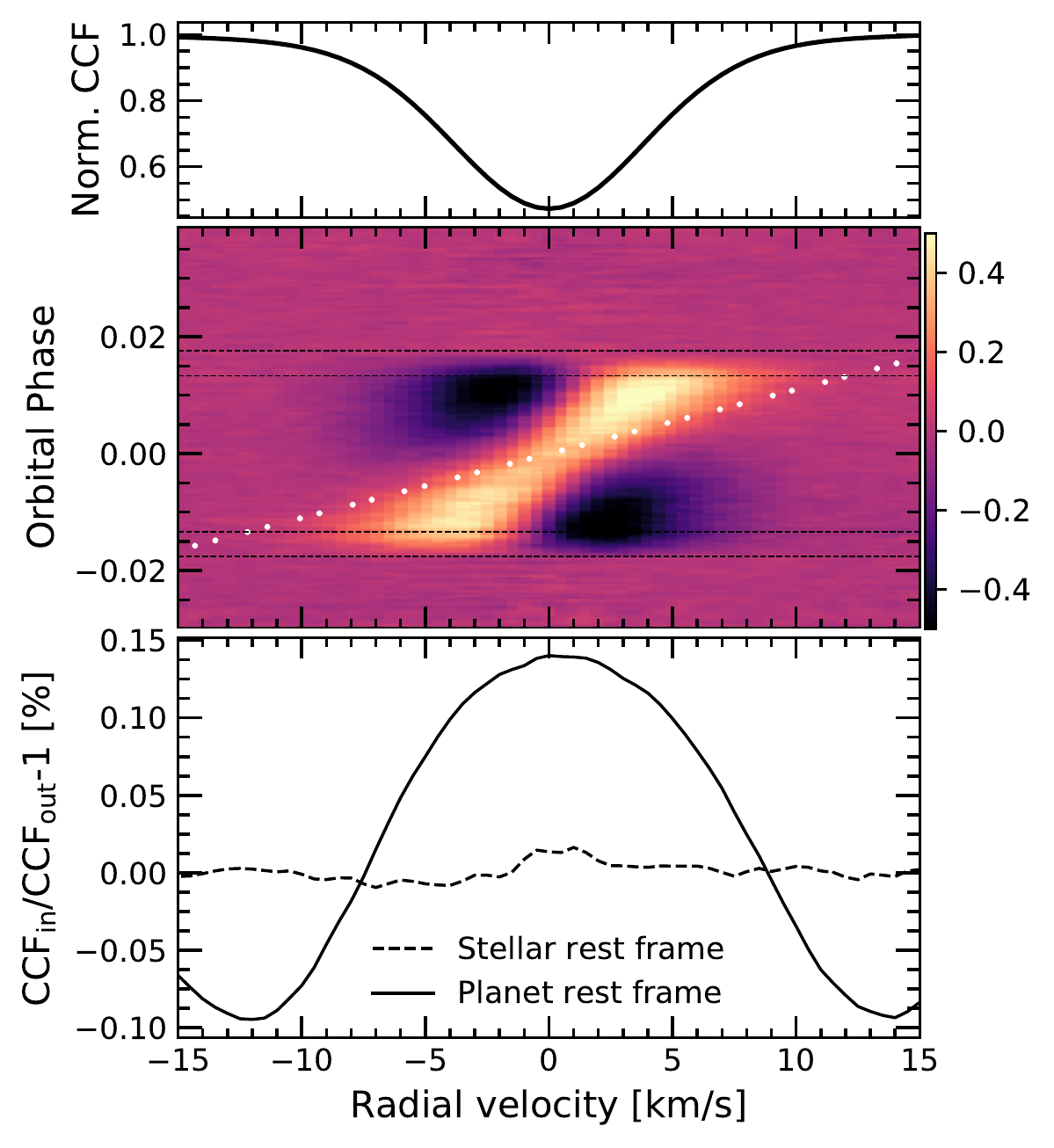}
\caption{\textit{Top panel:} Average out-of-transit stellar CCF. \textit{Middle panel:} Tomography of the CCFs deformation for the two transits combined in the stellar rest frame. The dotted white line indicates the planet radial velocities and the black horizontal dashed lines show the four contacts of the transit. The colour bar shows the relative variation of the CCFs with respect the combined out-of-transit CCF in $\%$. \textit{Bottom panel}: Combination of the CCF residuals from the \textit{middle panel} in the stellar rest frame (dashed line) and in the planet rest frame (solid line) between the first and fourth contact.}
\label{fig:ccf}
\end{figure}

The deformation of the lines profile caused by the RM, presented in Sect.~\ref{subsec:Na}, is expected to have an impact in all the stellar spectral lines, at different amplitudes, however, depending on the transition. In order to check this, we analyse the cross-correlation function (CCF) from the DRS as presented in \citet{Borsa2020}. In this case, the CCF values are generated with a F9 stellar mask using the DRS ESO version 2.0, with a step of 0.5~km s$^{-1}$. In order to extract the deformation of the lines profile propagated in the CCFs, we use the same process as for the transmission spectrum (see Sect.~\ref{subsec:TS_extract}): we move each CCF in the stellar rest frame and divide each of them by the averaged out-of-transit CCF. The result is presented in Figure~\ref{fig:ccf}, where the RM effect is clearly observed in the CCF values. As \ion{Fe}{i} lines are the major contributor to the stellar mask used to compute the CCFs \citep{Ehrenreich2020}, the RM deformation observed in the CCF tomography is dominated by the \ion{Fe}{i} lines. In contrast to \citet{Borsa2020}, where the atmospheric trail of the planet is clearly intercepted by the mask, we are not able to observe any feature with a possible atmospheric origin in the HD~209458b CCFs.

\begin{figure*}[]
\centering
\includegraphics[width=0.99\textwidth]{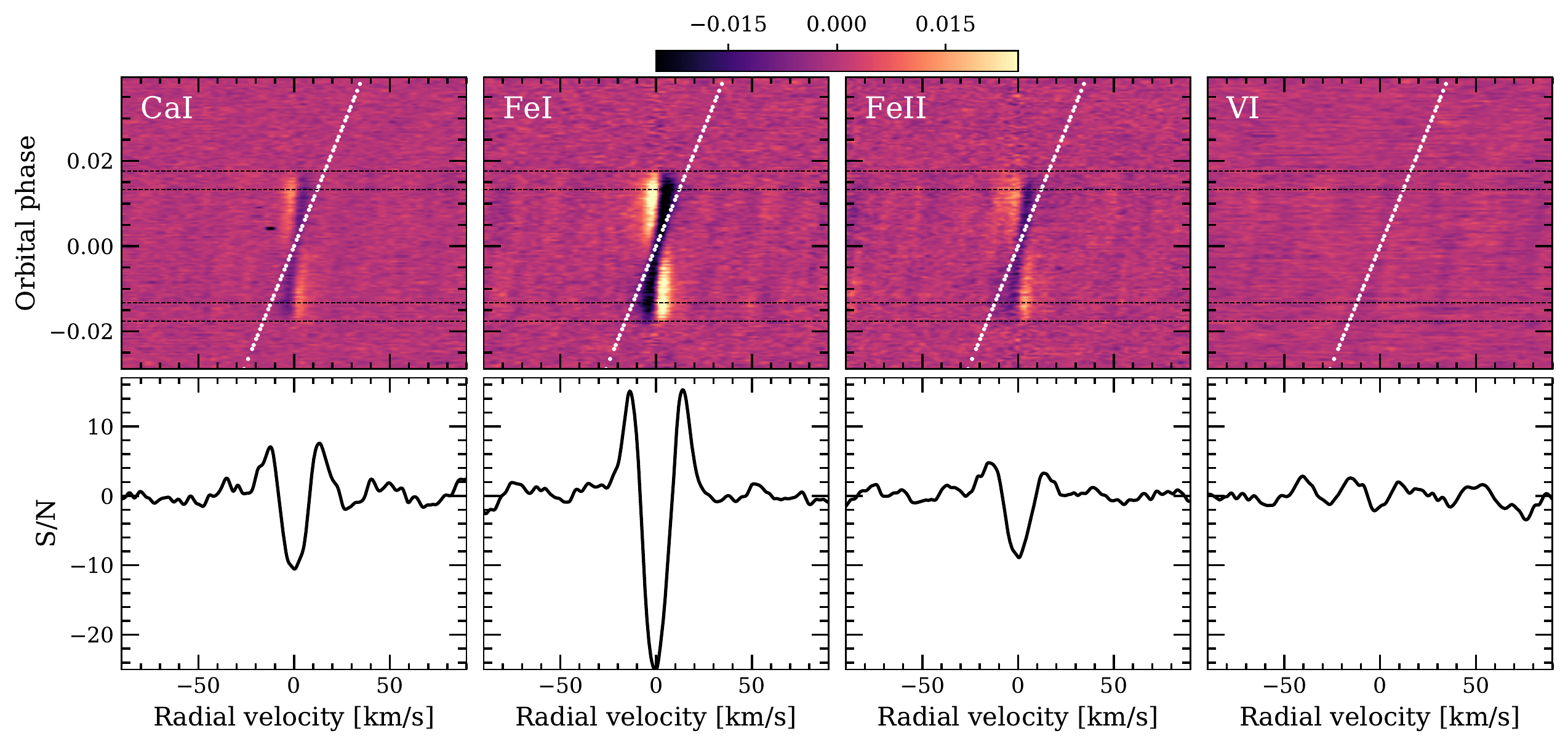}
\caption{Cross-correlation results of both nights combined for \ion{Ca}{i}, \ion{Fe}{i}, \ion{Fe}{ii}, and \ion{V}{i} (from left to right, respectively). \textit{Top panel}: Cross-correlation maps of the three different species. \textit{Bottom panel}: Average of the in-transit cross-correlation values in the planet rest frame between the first and fourth contacts. The y-axis is shown in S/N units, where negative values mean anti-correlation (see the cross-correlation coefficients in the colour bar). }
\label{fig:cc_atoms}
\end{figure*}

In addition, we search for \ion{Fe}{i}, \ion{Fe}{ii}, \ion{Ca}{i}, and \ion{V}{i} absorption cross-correlating the residuals spectra with the atmospheric templates described in Sect.~\ref{subsec:crosscorr}. The results can be observed in Figure~\ref{fig:cc_atoms}. For \ion{Fe}{i}, \ion{Fe}{ii}, and \ion{Ca}{i} the deformation of the stellar line profiles is visible at high S/N, without any clear feature of atmospheric origin in the ingress and egress regions where the planet atmosphere could be disentangled from the stellar contamination. For \ion{V}{i,} the strength of the deformation is fainter, probably indicating that this species is poorly (or not at all) present in the stellar spectrum, given the spectral type of the star. The S/N of the bottom row panels from Figure~\ref{fig:cc_atoms} is calculated by dividing the cross-correlation values by the standard deviation calculated far from the position of the signal, as presented in \citet{birkby2017}, \citet{Brogi2018}, and \citet{SanchezLopez2019}. Here, we use the ranges from $-150$ to $-50$\,km s$^{-1}$ and from $+50$ to $+150$\,km s$^{-1}$. In these units, a positive S/N means correlation with the atmospheric absorption template and negative S/N means anti-correlation. We note that an increase in the temperature of the isothermal models up to $2000$~K for \ion{Fe}{i}, \ion{Fe}{ii} and \ion{Ca}{i}, and to $3000$\,K for \ion{V}{i} does not significantly alter the results. 

\begin{figure*}[h]
\centering
\includegraphics[width=1.0\textwidth]{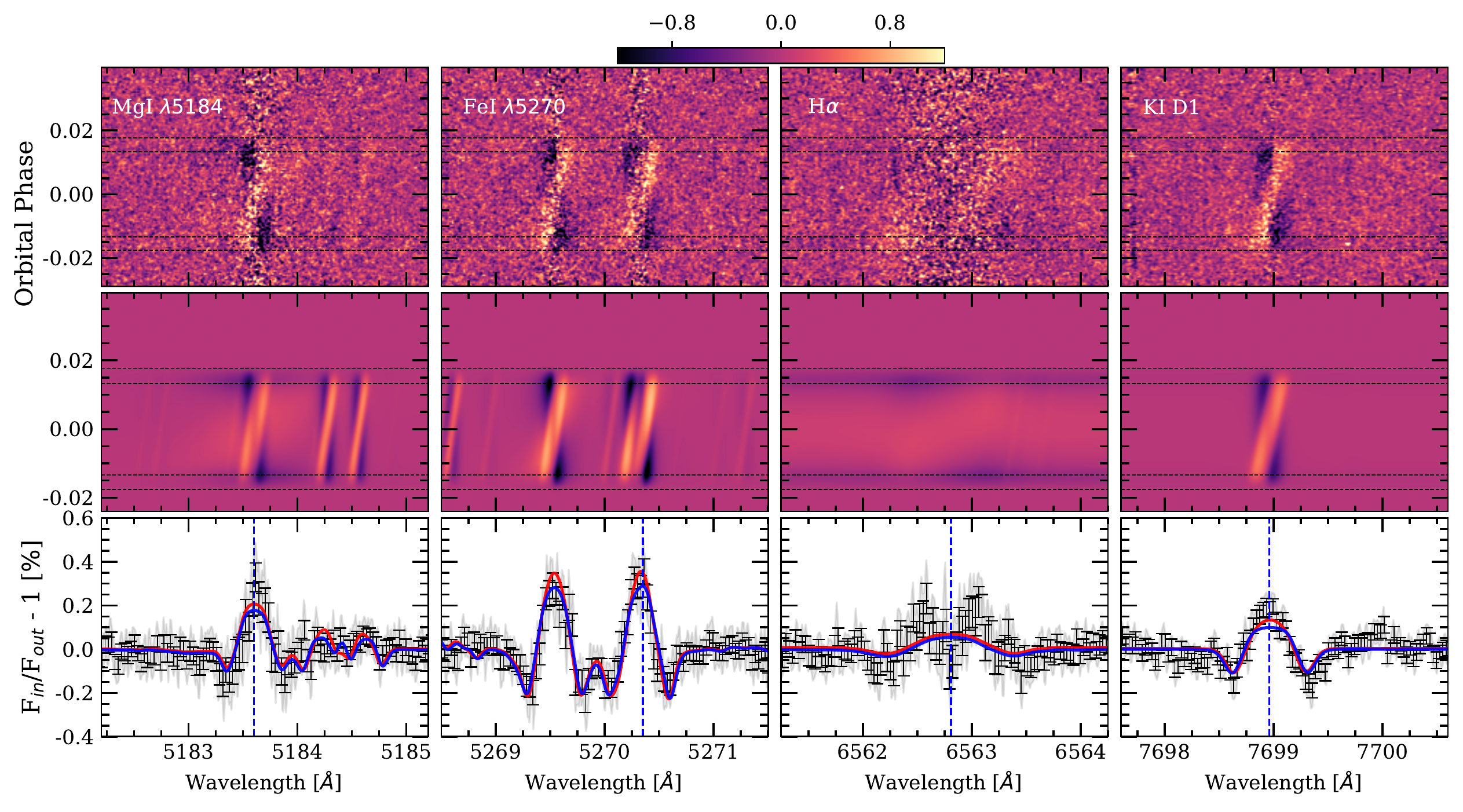}
\includegraphics[width=0.98\textwidth]{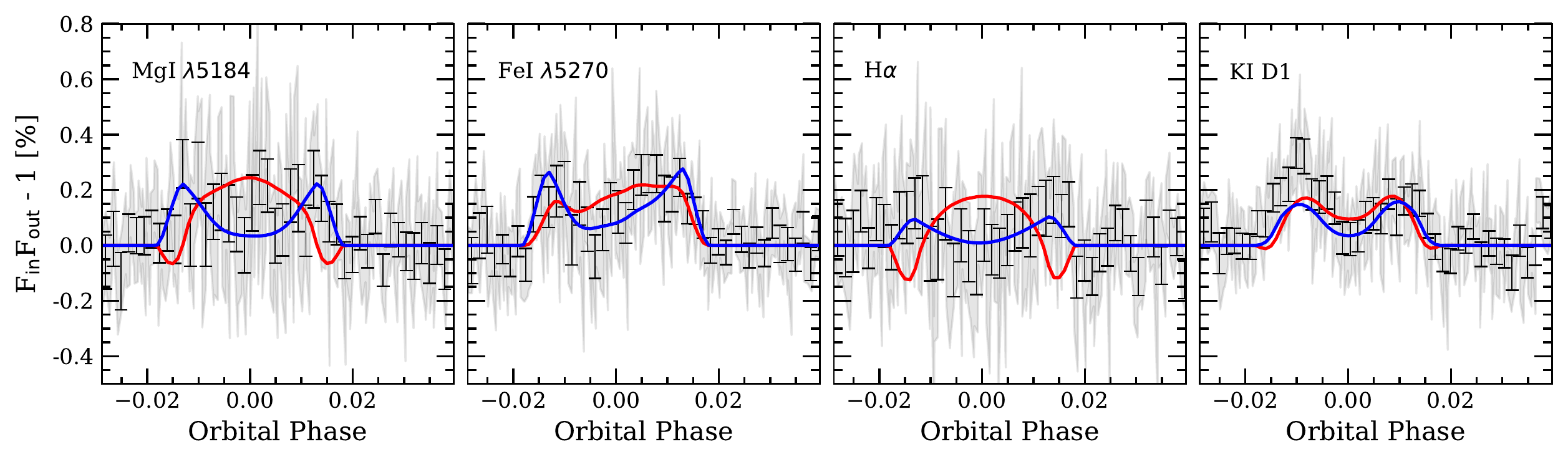}
\caption{Results around the \ion{Mg}{i} $\lambda 5184$ (first column), \ion{Fe}{i} $\lambda 5270$ (second column), H$\alpha$ (third column), and \ion{K}{i} D1 (fourth column) spectral lines. \textit{Top panel}, \textit{first row}: Two-dimensional map of the transmission residuals of HD~209458b around different spectral lines. \textit{Second row}: Two-dimensional map of the modelled CLV and RM effects. In both panels the residuals are presented in the stellar rest frame. The black horizontal dashed lines indicate the four contacts of the transit. \textit{Third row}: Transmission spectrum computed in the planet rest frame considering the data between the first and fourth contacts of the transit. The original data are shown in light gray and the data binned by $0.03~{\rm \AA}$ is shown in black. The blue-dashed vertical lines show the laboratory position of the lines. The colour bar shows the relative flux (F$_{\rm in}$/F$_{\rm out}$-1) in \%. \textit{Bottom panel:} Transmission light curves calculated for each line using $0.4~{\rm \AA}$ bandwidth and $0.5~{\rm \AA}$ for H$\alpha$. In light-grey, we show the original data and in black the data binned by $0.0015$ in orbital phase. In both panels, the red and blue lines correspond to the RM model with and without considering the CLV effect, respectively.}
\label{fig:oth_lines1}
\end{figure*}

The scenario observed in \ion{Na}{i}, \ion{Fe}{i}, \ion{Fe}{ii,} and \ion{Ca}{i} is repeated for the spectral lines observed in the full stellar spectrum when we attempt to detect the exoplanet atmosphere. As an example, we compute the transmission spectrum for particular individual lines such as \ion{Mg}{i} at $5183.60~{\rm \AA}$, \ion{Fe}{i} at $5270.36~{\rm \AA}$, \ion{H}{$\alpha$} at $6562.81~{\rm \AA}$, and \ion{K}{i} D1 line at $7698.96~{\rm \AA}$. The results are presented in Figure~\ref{fig:oth_lines1}. As in the case of \ion{Na}{i}, in all cases, the transmission spectra are consistent with the modelled RM effect, with and without considering the CLV effect due to the small (i.e. within the error bars) difference this last one introduces. However, the light curves seem to be benefit from a better explaination when only the RM effect is considered, even at lower S/N. The estimated effects on the H$\alpha$ transmission spectrum do not describe the observations as well as they do for the other lines. In Figure~\ref{fig:other_models}, the modelled stellar line profile, computed using the MARCS stellar models and LTE, are shown together with the observations. It is clear that the \ion{Mg}{i} and \ion{Fe}{i} lines, for example, are better reproduced by the stellar models than H$\alpha$. This is probably due to its chromospheric origin, as also observed for the \ion{Ca}{ii}~IRT in \citet{Casasayas2020}. These lines require more accurate stellar modelling.

\subsection{Analysis of systematic effects}
\label{subsec:syseff}

The error bars of the measurements come from the propagation of the random photon noise. Following \citet{2008Redfield}, we can statistically quantify the systematic effects of our data using the empirical Monte Carlo (EMC) analysis. This method has already been applied in several atmospheric studies, such as \citet{Wytt2015,2017A&A...602A..36W}, \citet{jensen2011, Jensen2012}, and \citet{Chen2020}. It consists of measuring the absorption depth of a transmission spectrum calculated using different in- and out-of-transit samples (scenarios), which do not need to match the real in- and out-of-transit data. With this, if a feature appears only during the transit, we would expect it to only be measured when the real in- and out-of-transit data are considered in the samples.

Here, we assume the four different scenarios described in \citet{Casasayas2020}, called 'out-out', 'in-in', 'mix-mix' and 'in-out' depending on the data considered  as in- and out-of-transit samples, respectively. For each scenario we run the EMC $20~000$ times, measuring the absorption depth in the expected position of the line using a $0.4~{\rm \AA}$ passband. In Figure~\ref{fig:EMC} of the appendix we show the absorption depth distributions. For the \ion{Na}{i}, it is clear that the 'in-out' samples of both nights are centred at a different position in comparison with the control distributions. Their positions are $0.19\%,$ on average, for the two nights. The control distributions are all centred at $0\%$, and the standard deviation of the Gaussian profiles from the 'out-out' distribution are $0.04\%$ and $0.05\%$ (for the first and second night, in the same order), which gives us an idea of the noise level of the data.

This exercise is also applied to the other lines analysed here (\ion{Mg}{i} $\lambda5184$, \ion{Fe}{i} $\lambda5270$, H$\alpha$, and \ion{K}{i} D1). As for the \ion{Na}{i} case, in all these lines, the absorption scenario shows distributions centred at positive absorption depth values ($0.12\%$, $0.15\%$ $0.03\%$, and $0.12\%$, respectively) while the control distributions remain centred at $0\%$. We note that for H$\alpha$, as the line is broader, we use a larger passband of $0.5~{\rm \AA}$, and the expected effects in this line are fainter (see Figure~\ref{fig:oth_lines1}). The EMC histograms of these lines are presented in Figure~\ref{fig:EMC} in the appendix.

\subsection{Searching for molecules}
\label{subsec:molecules}

In contrast with the atomic lines analysed in the previous sections, CLV and RM effects are not expected to impact the search for those atoms and molecules in the exoplanet atmosphere that are not present in the stellar spectrum in cases where there is no overlap with other stellar lines. For this reason, we attempted to search for the presence of such molecules as TiO and VO. The spectral features possibly associated to these same molecules were tentatively observed in HD~209458b atmosphere around $6~000-8 000~{\rm \AA}$ by \citet{Desert2008} using HST-STIS observations. However, other studies at high resolution spectroscopy such as  \citet{Hoeijmakers2015} found no evidence of TiO in this same planet.

Here, in the wavelength region covered by ESPRESSO and using the different TiO and VO models generated as described in Sect.~\ref{subsec:crosscorr}, we find no evidence of these species in the atmosphere of HD~209458b. In the final cross-correlation residuals (see Figure~\ref{fig:molec}), we observe several faint structures produced by the RM effect of closer stellar lines that are introduced in the radial-velocity space explored in the cross-correlation, but not at the expected planet position ($\sim0$\,km\,s$^{-1}$). It is known that the TiO and VO line lists are not very accurate at a high spectral resolution \citep{Merritt2020, McKemmish2019}. As was already studied by \citet{Hoeijmakers2015}, the lack of an accurate line list can become a critical aspect when attempting to detect atmospheric features using the cross-correlation technique.

\section{Accuracy of the modelled transit effects}
\label{sec:accuracy}

In Sects.~\ref{subsec:Na} and \ref{subsec:otherlines}, we observe that the transmission spectra can be explained by the combination of both CLV and RM effects, but also when only the dominant effect (the RM) is included in the calculations, considering the error bars of the data. The differences between both assumptions are small when the in-transit residuals are combined in the planet rest frame, as the CLV effect at the \ion{Na}{i} D2 line position is more than four times smaller than the RM when calculated using the MARCS LTE stellar models. However, in the transmission light curves (Figures~\ref{fig:tlc_Na} and \ref{fig:oth_lines1}), we are sensitive to time variations of a particular spectral region. In this case, we observe that the results are better described when only the RM effect in the stellar lines is considered, probably indicating an overestimation of the CLV effect.

\begin{figure*}[h]
\centering
\includegraphics[width=0.9\textwidth]{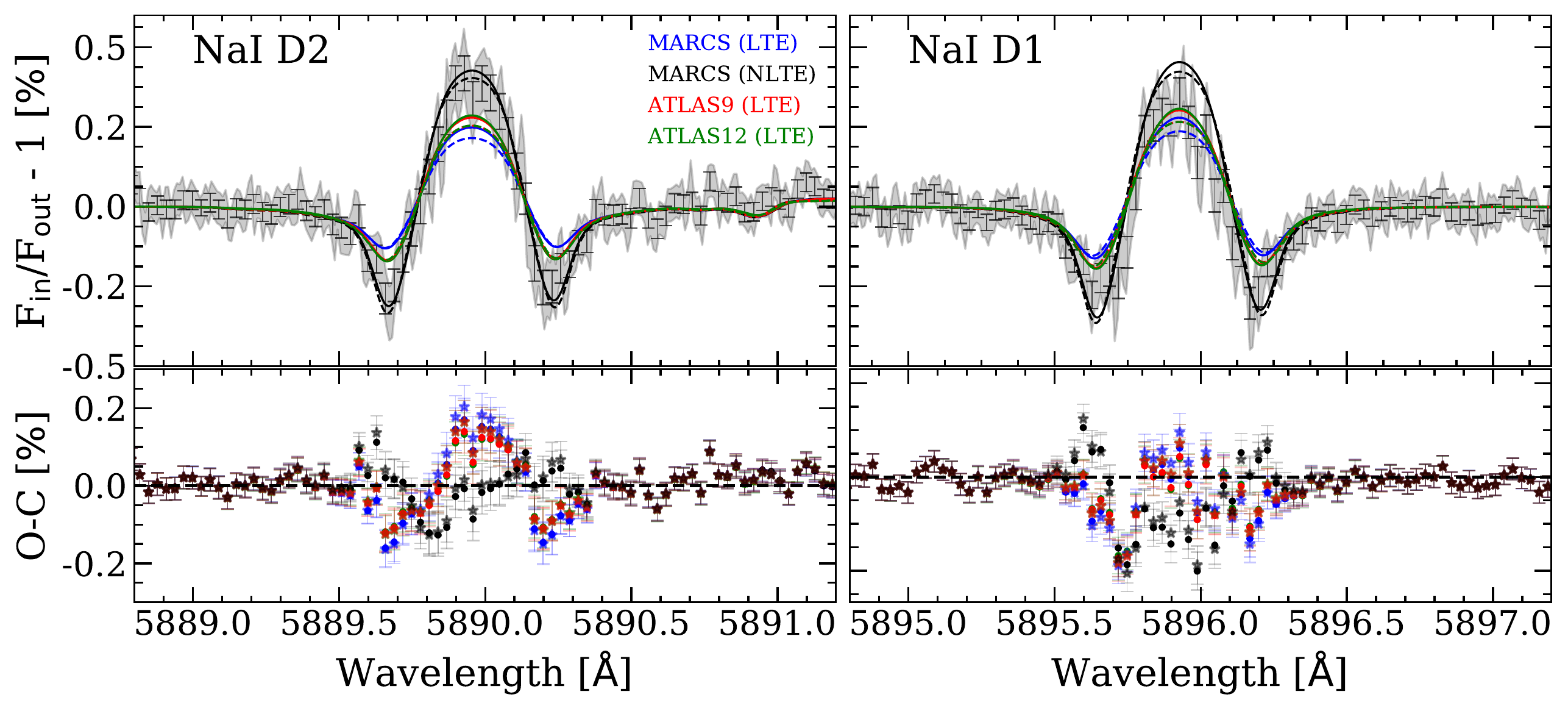}
\includegraphics[width=0.9\textwidth]{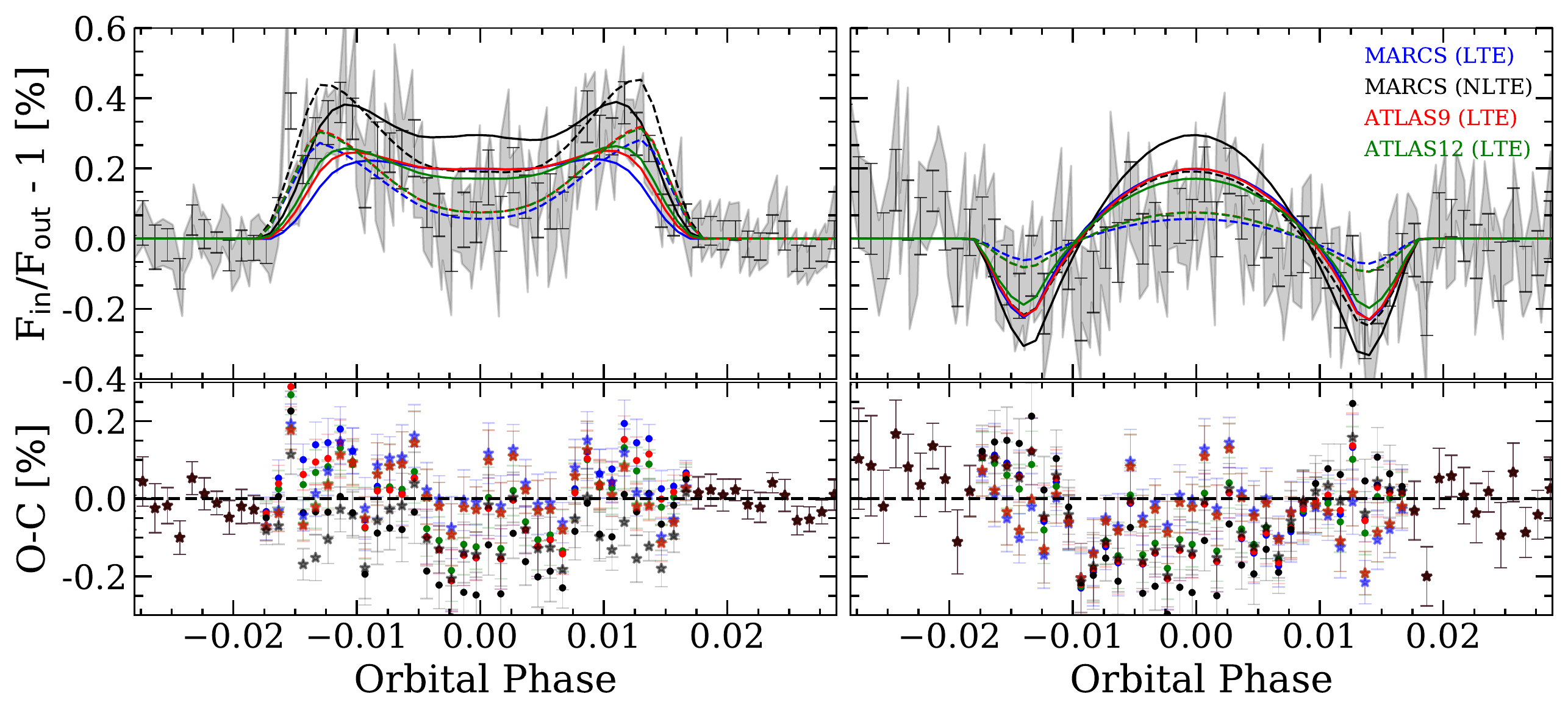}
\caption{Top panel, from left to right: Modelled CLV and RM effects of the \ion{Na}{i} doublet lines assuming different stellar atmospheric models compared with the transmission spectrum around \ion{Na}{i} D2 (left panel) and D1 (right panel). Bottom panel: Transmission light curve of both lines combined calculated in the planet rest frame (left panel) and the transmission light curve in the stellar rest frame (right panel). In solid lines, we show the models containing the CLV and RM deformation. In dashed lines, it is only the RM that is considered. In light-grey, we show the original data and in black error bars the data binned by $0.03~{\rm \AA}$ (top panels) and $0.001$ in the orbital phase (bottom panels). The different stellar models are shown in different colours and indicated in the legends. Bottom row: Residuals between the data and the models. The colours indicate the residuals between the data and the model shown with the same colour. We indicate the residuals with the models computed considering only the RM deformation with stars and the residuals with the models containing the CLV and RM deformation with dots.}
\label{fig:models_comp}
\end{figure*}

\citet{Casasayas2020}, as well as previous sections of this work, use the MARCS stellar atmospheric models \citep{Gustafsson2008}, assuming local thermodynamic equilibrium (LTE). We use solar abundances, based on the finding from \citet{Adibekyan2012} that HD~209458 has [Na/H]$= 0.010\pm0.057$. Here, thanks to the unprecedented S/N achieved by the ESPRESSO observations, we analyse the differences in the modelled effects when considering different stellar atmospheric models. In particular, we compare the results around the \ion{Na}{i} when assuming MARCS stellar models calculated with LTE and non-LTE approximations \citep{Mashonkina2008}. We also use ATLAS9 \citep{ATLAS92003} and ATLAS12 \citep{ATLAS122013K} models, assuming LTE. We note that non-LTE grids are only offered for MARCS models \citep{SMEEvolution2017}. The results are presented in Figure~\ref{fig:models_comp}. Furthermore, in Figure~\ref{fig:linevsmod}, the modelled stellar spectra in the \ion{Na}{i} region can be compared with the observed stellar spectrum.

In the transmission spectrum, we observe that the different models and assumptions produce different intensities of the line deformations (left panel of Figure~\ref{fig:models_comp}). The different models assuming LTE approximation produce very similar results, while the models assuming non-LTE differ by ${\sim 0.2\%}$ in relative flux at the \ion{Na}{i} lines centre. Around the \ion{Na}{i}, LTE models remain underestimated in comparison with the data and the non-LTE models seem to reproduce the results much better. We observe that the predicted CLV effect at the \ion{Na}{i} lines core of the transmission spectrum changes by a factor of two, depending on the stellar model selection. When we model the transmission light curves in the planetary rest frame (third panel of Figure~\ref{fig:models_comp}), we can clearly see that the observations are better reproduced when only the RM deformation is considered, especially at the centre of the transit. As presented in \citet{LoudenW2015} and \citet{Borsa2018}, in the planet rest frame the RM symmetry is broken, resulting in spurious residuals in the stellar line position. For the transmission light curve calculated in the stellar rest frame (right panel of Figure~\ref{fig:models_comp}), where we are actually tracking the stellar line deformations in their own frame, the RM symmetry is maintained and thus averaged out when integrating the full effect inside the $0.4~\rm{\AA}$ passband. In this frame, the CLV is the main contribution of the transmission light curve. As can be observed in the figure, the ingress and egress regions are better described considering the CLV contribution, as expected, but the overall effect in the mid-transit time region remains slightly overestimated.

When the residuals between the models and the data are computed (see bottom row of Figure~\ref{fig:models_comp}), we observe a large scatter expanding approximately $\pm0.2~\%$ depending on the model used. Therefore, attempting to measure features at these atmospheric absorption levels can become very challenging for HD~209458b. Using a $0.75~\rm{\AA}$ passband centred at the \ion{Na}{i} lines position in the residuals, we measure a mean excess ranging from $-0.05~\%$ to $-0.02~\%$, depending on the stellar model considered in the subtraction. In the transmission light curves, the in-transit residuals show an excess change between $-0.09~\%$ and $+0.04~\%$. 

\begin{figure*}[h]
\centering
\includegraphics[width=0.98\textwidth]{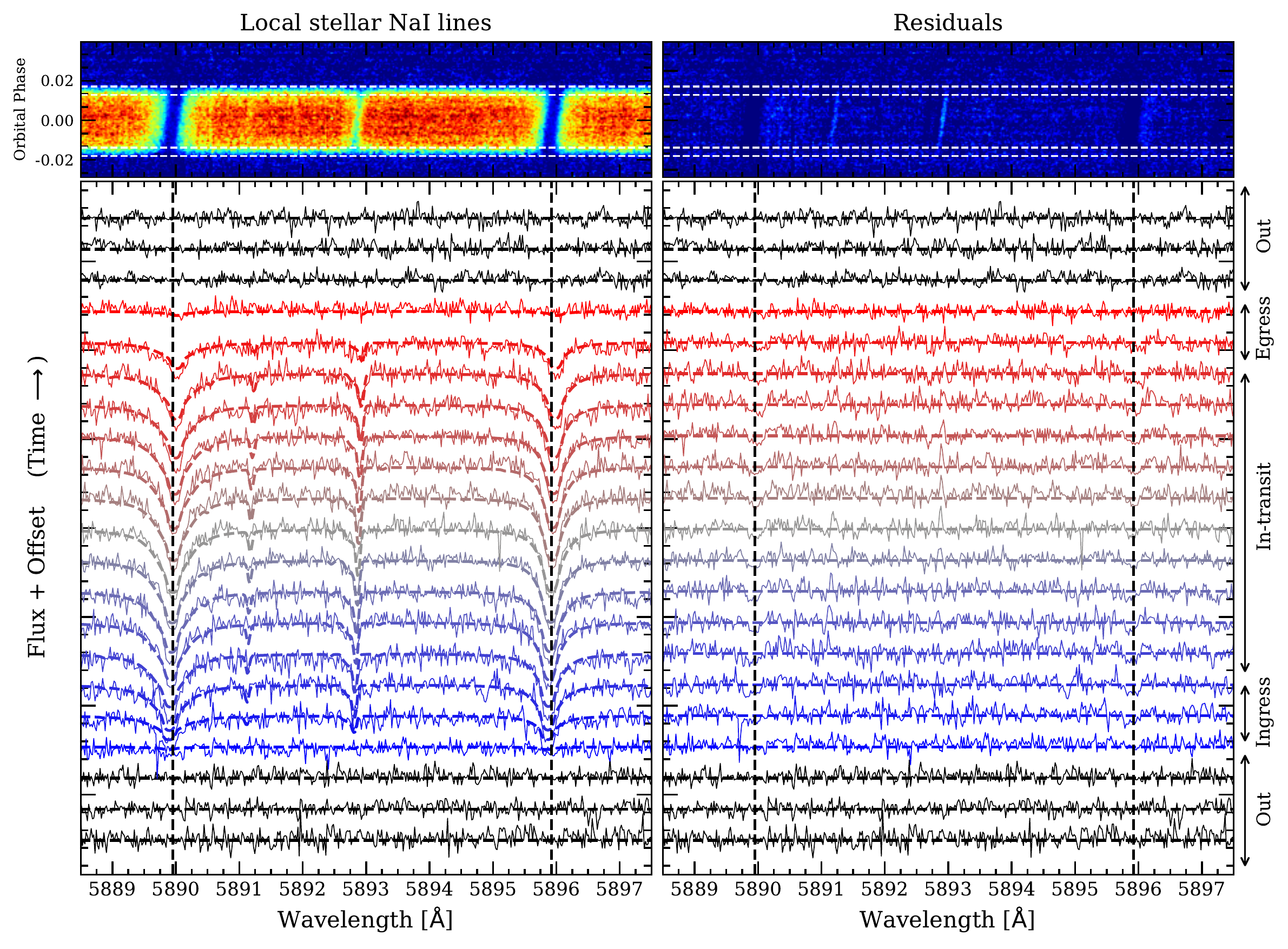}
\caption{\textit{Left panel}: Local stellar spectrum of HD~209458 around the \ion{Na}{i} doublet lines. In the \textit{top-left panel,} we show the local stellar spectrum following the same format presented in Figure~\ref{fig:Local_maps}, but around the \ion{Na}{i} doublet. \textit{Bottom-left panel:}  Time-evolution of the local stellar spectrum presented in the top panel but, in this case, the flux information is shown in the vertical axis. The spectra are not normalised to their own continuum level, so they still contain the flux information (the normalised profiles are shown in Figure~\ref{fig:local_Na_comb}). For a clear visualisation, we modified the continuum flux to include an offset so the spectra are shown ordered in orbital phase. The colours are indicative of the orbital phase of the planet and, therefore, the radial-velocity of the stellar disc surface. The observations are shown with solid lines and the stellar models with dashed lines of the same colour. The stellar models are computed using MARCS in LTE. The out-of-transit results are shown in black. The spectra are shown binned by $0.02~{\rm \AA}$ and $0.0025$ in orbital phase. The black-dashed vertical lines indicate the position of the \ion{Na}{i} doublet lines. \textit{Right panel:} Residuals between the observed and modelled local spectra presented in the left panel. }
\label{fig:LP_NaI}
\end{figure*}

With the S/N achieved with ESPRESSO observations, we are able to obtain the local stellar spectrum following the methodology presented in Section~\ref{subsec:relRM}. The local stellar spectrum is the spectrum of those regions in the stellar surface that are blocked by the planet at each exposure during the transit, and which are therefore missing in the integrated disc spectrum observations. In Figure~\ref{fig:LP_NaI} we show the local spectrum around the \ion{Na}{i} doublet lines at different orbital phases of the planet. A similar analysis is shown in \citet{Dravins2017} for this same planet using several \ion{Fe}{i} lines. Here, we observe that, during the transit, the exoplanet obscures blue-shifted regions of the stellar disc and continues to the red-shifted stellar surface, producing the RM effect that is propagated to the transmission spectrum. We also observe that the predicted models are able to reproduce the observations, although the lines core remain slightly underestimated. If the local stellar spectra are normalised to their continuum level and the two lines of the doublet are combined to get higher S/N profiles, we observe the expected wavelength shift of the lines produced by RM effect, but the CLV is not clearly seen in the data, although predicted by the stellar models (see Figure~\ref{fig:local_Na_comb}). This is one more indication of the CLV overestimation for this particular star.

The importance of the CLV for atmospheric characterisation of exoplanets has been pointed out in several previous studies (e.g. \citealt{cze15,Yan2017A&A...603A..73Y,Khalafinehad2017A&A...598A.131K,2017CasasayasB}). The CLV in the stellar atmosphere strongly depends on wavelength, and it is expected to be more intense for cooler stars. In the optical, it is strong in the deep stellar absorption lines profile such as \ion{Na}{i} D and \ion{Ca}{ii} H\&K lines \citep{cze15}. In the study of exoplanet transit light-curves, the CLV is one of the critical parameters \citep{MullerCLV2013}. Using HD~209458b transit light curves, \citet{Hayek2012} compared the CLV predictions resulting from MARCS one-dimensional (1D) atmospheric models and three-dimensional (3D) models, concluding that the 3D model provides a considerably better description of the atmospheric temperature structure of HD 209458. \citet{CiavassaBrogi2019} also show that, in general, 1D stellar models can differ significantly from 3D models.

On the other hand, the differences could be also produced by the combination of different stellar phenomena that are not contemplated in the stellar spectra or due to a misinterpretation of the stellar atmospheric structure. For example, the RM could be affected by convective shifts (\citealt{Cegla2016,Dravins2017}). The convective shifts are a consequence of the stellar granulation and occurs because the emerging granules are brighter than the surroundings and cover a greater fraction of the stellar surface \citep{Dravins1982}. It typically results in an additional blue shift between the position of the stellar lines and their laboratory position, with expected amplitudes of $0-500$\,m\,s$^{-1}$ for G-type stars \citep{Meunier2017} and it is less intense in those deep lines formed  higher in the stellar atmosphere \citep{Allende1998,Reiners2016}, such as \ion{Na}{i}. Moreover, it is known that granulation and star-spots can produce asymmetric line profiles on the stellar disc, which could be considered using 3D magneto-hydrodynamic solar simulations as presented in \citet{Cegla2016b}. On the other hand, solar-like differential rotation can become important in main sequence stars \citep{Karoff2018} and it is expected to be stronger in F-G stars \citep{Balona2016}. In our calculations, we assume solid-body rotation, meaning that we are not considering the amplitude variation of the RM signal that could be introduced by this effect \citep{WASP72020}.

Currently, ESPRESSO is the best facility with which we can spatially resolve the stellar surface by means of transit spectroscopy observations. This will help to empirically validate the stellar models, and constrain the 3D hydrodynamic models of the stellar photosphere \citep{Dravins2015CLV,Dravins2017}.

\section{Summary and conclusions}
\label{sec:conc}

Using two transit observations with ESPRESSO we revisit the transmission spectrum of HD~209458b and the system architecture using the reloaded-RM technique. We analyse individual atomic lines (\ion{Na}{i} doublet, \ion{Mg}{i} $\lambda5184$, \ion{Fe}{i} $\lambda5270$, H$\alpha$, and \ion{K}{i} D1) and we also use the cross-correlation technique to search for atomic lines (\ion{Fe}{i}, \ion{Ca}{i}, \ion{Fe}{ii}, and \ion{V}{i}) and molecules (TiO and VO).

The ESPRESSO observations confirm the results obtained in \citet{Casasayas2020}, where the \ion{Na}{i} features observed in the transmission spectrum can be explained by the deformation of the stellar line profile due to the transit of HD~209458b, mainly produced by the RM effect and without the need of absorption from the exoplanet atmosphere. This is also observed in the transmission spectra around other atomic lines and in the results obtained using cross-correlation techniques. Moreover, we examine the CCF values computed by the DRS as presented in \citet{Borsa2020}, where the deformation of the lines profile is clearly visible. This effect is present for all stellar lines and it is thus propagated through the method when attempting to extract any exoplanet atmospheric signals. Similarly, this change in the stellar line profile is imprinted in the CCF values when a stellar template is cross-correlated with the stellar spectra \citep{Borsa2018}. The radial-velocity change of the planet during the transit is very similar to the velocity range affected by the RM deformation and only a small time-region in the ingress and egress regions could be used to disentangle the exoplanet atmospheric signals by the stellar contamination.

Several atomic lines, including the \ion{Na}{i}, have been previously detected in the atmosphere of HD~209458b using high-resolution spectroscopy observations. This is the case of the $0.135\pm0.017~\%$ \ion{Na}{i} absorption detected in the transmission light curves by \citet{2008SnellenHD209} and \citet{Albrecht2009}, for example. The \ion{Na}{i} transmission light curve obtained with ESPRESSO observations and following the methodology presented in these previous studies (see Figure~\ref{fig:SnellNaI}) does not reproduce the absorption and, in contrast, shows a better agreement with the modelled RM light curve. On the other hand, \citet{Langland2009} found $>3\sigma$ \ion{Na}{i} absorption and also $>2\sigma$ absorption in \ion{Fe}{i} and \ion{Ni}{i} lines. \citet{jensen2011} tentatively detected \ion{Na}{i} absorption, and \citet{Astudillo2013} found significant absorption excess due to \ion{Ca}{i}, H$\alpha$, \ion{Na}{i}, and possibly \ion{Sc}{ii}. All these lines are present in the stellar atmosphere and we can detect the RM deformation at high S/N in our analysis, however, we do not find evidences of absorption due to the exoplanet atmosphere. Focusing on H$\alpha$, \citet{Jensen2012} detected a symmetric feature correlated in orbital phase and expanding $\sim15~{\rm \AA}$. This broad feature can not be reproduced with the methodology presented here.

The S/N of the observations used in these previous studies is similar to the observations used in this work. This is the case of the Subaru observations used by \citet{Narita2005HD209}, \citet{2008SnellenHD209} and \citet{Astudillo2013}, for example, which achieved a S/N$\sim350$ per pixel in the continuum using 500\,s exposures. The observations used here show a mean S/N of $\sim220$ in the \ion{Na}{i} order, but using exposures of 175\,s which results in a larger number of spectra. However, most of these studies used a different methodology. Thanks to the different high-resolution studies, the method to extract the transmission spectrum of the exoplanets has rapidly improved, including some steps that were not considered before, such as the planet radial-velocity correction. In the particular case of HD~209458b, using different methodology or partial transit observations leads to significantly different transmission spectra (see Figures~\ref{fig:ts_Na} and \ref{fig:models_comp}) due to the RM deformation, which could easily lead to a misinterpretation.

At high resolution, the RM contamination could be avoided by searching for molecules not present in the stellar atmosphere (if they fall far from other deep stellar lines). Here, we explore the presence of TiO and VO, as it was tentatively observed in \citet{Desert2008}. These molecules are expected at temperatures above $1800~{\rm K}$, in strongly irradiated hot Jupiters \citep{Seager1998,Fortney2008}. Indeed, TiO has already been tentatively detected in ultra hot planets (T$_{eq}>2000~{\rm K}$) such as WASP-19b \citep{Elyar2017} and WASP-121b \citep{Evans2016}. However, one of the main difficulties when attempting to detect these molecules using the cross-correlation techniques is the lack of accurate line lists \citep{Hoeijmakers2015}. HD~209458b has an equilibrium temperature (T$_{eq}$) around $1400~{\rm K}$ \citep{Torres2008ApJ...677.1324T} and it is probably too cold for these species to be present in its atmosphere. Despite the high S/N of our data, we are not able to identify  any clear feature in the residuals that could be generated by these molecules.

In the atmospheric analysis performed at a high resolution, a relatively small region around the lines core is probed and the continuum information is lost, while at low resolution, the wings of the atmospheric lines are probed, which may present widths as large as $100$\,nm (see e.g. \citealt{Sing2016}). The RM contamination seen in the \ion{Na}{i} lines occupies $\sim1~{\rm \AA}$ in the high-resolution transmission spectrum. At low resolution, this contamination would be completely included in the wavelength bin centred on the \ion{Na}{i} doublet lines. For this reason, with our ESPRESSO data we cannot exclude the atmospheric detections reported using broad-band transmission spectroscopy methods (e.g. \citealt{2002ApJ...568..377C}). However, this particular scenario, where the broad wings of \ion{Na}{i} are detected but the cores of the lines are not, is difficult to explain in terms of atmospheric physics.

Here, thanks to the S/N achieved with ESPRESSO observations, we are able to observe that the resulting transmission light curves can be better explained when only the RM contribution is considered, especially at the centre of the transit. However, during the ingress and egress, the contribution of the CLV is needed to fit the observations. Additionally, we explore the differences of the modelled RM and CLV effects when assuming different 1D stellar atmospheric models. It is clear, however, that the modelled CLV and RM effects do not fully reproduce the results around the \ion{Na}{i}; this is probably due to the approximations that their computation assumes and stellar phenomena that are not considered (1D, solid-body rotation, no convective blue shift, among others), although the non-LTE models describe the transmission spectrum better. Resolving the stellar surface using transit observations of the exoplanets \citep{Dravins2015CLV} would help to empirically validate and constrain the stellar atmospheric models and, consequently, to improve the characterisation of the exoplanet atmospheres. Currently, ESPRESSO provides the best opportunity for performing these observations.


\begin{acknowledgements}

 This work is partly financed by the Spanish Ministry of Science, Innovation and Universities through project ESP2016-80435-C2-2-R, ESP2017-87143-R, AYA2017-86389-P, AYA2016-79425-C3-2-P, and PID2019-109522GB-C51. 
 G.C. acknowledges the support by the National Natural Science Foundation of China (Grant No. 42075122) and the Natural Science Foundation of Jiangsu Province (Grant No. BK20190110). 
 F.Y. acknowledges the support of the DFG priority program SPP 1992 "Exploring the Diversity of Extrasolar Planets (RE 1664/16-1)". 
 This project has received funding from the European Research Council (ERC) under the European Union's Horizon 2020 research and innovation programme (project Four Aces grant agreement No 724427). 
 N.J.N acknowledges support from FCT through Investigador FCT contract and exploratory project IF/00852/2015, and project PTDC/FIS-OUT/29048/2017. 
 This work was supported by FCT - Funda\c{c}\~ao para a Ci\^encia e a Tecnologia through national funds and by FEDER through COMPETE2020 - Programa Operacional Competitividade e Internacionaliza\c{c}\~ao by these grants: UID/FIS/04434/2019; UIDB/04434/2020; UIDP/04434/2020; PTDC/FIS-AST/32113/2017 \& POCI-01-0145-FEDER-032113; PTDC/FIS-AST/28953/2017 \& POCI-01-0145-FEDER-028953; PTDC/FIS-AST/28987/2017 \& POCI-01-0145-FEDER-028987. 
 V.A. acknowledges the support from FCT through Investigador FCT contracts nr. IF/00650/2015/CP1273/CT0001. 
 S.G.S acknowledges the support from FCT through Investigador FCT contract nr. CEECIND/00826/2018 and POPH/FSE (EC). 
 O.D.S.D. is supported in the form of work contract (DL 57/2016/CP1364/CT0004) funded by FCT. 
 This research has been funded by the Spanish State Research Agency (AEI) Projects No.ESP2017-87676-C5-1-R and No. MDM-2017-0737 Unidad de Excelencia "Mar\'ia de Maeztu"-Centro de Astrobiolog\'ia (INTA-CSIC). 
 The INAF authors acknowledge financial support of the Italian Ministry of Education, University, and Research with PRIN 201278X4FL and the "Progetti Premiali" funding scheme. 
 FPE and CLO would like to acknowledge the Swiss National Science Foundation (SNSF) for supporting research with ESPRESSO through the SNSF grants nr. 140649, 152721, 166227 and 184618. 
 J.I.G.H. acknowledges financial support from Spanish Ministry of Science and Innovation (MICINN) under the 2013 Ram\'on y Cajal program RYC-2013-14875. A.S.M. acknowledges financial support from the Spanish MICINN under the 2019 Juan de la Cierva Programme.  C.A.P., J.I.G.H., R.R., and A.S.M.  acknowledge financial support from the Spanish MICINN AYA2017-86389-P. 
 This work has been carried out within the framework of the National Centre of Competence in Research Planets supported by the Swiss National Science Foundation. The authors acknowledge the financial support of the SNSF.
 The ESPRESSO Instrument Project was partially funded through SNSF’s FLARE Programme for large infrastructures. 
 This work made use of PyAstronomy and of the VALD database, operated at Uppsala University, the Institute of Astronomy RAS in Moscow, and the University of Vienna.

\end{acknowledgements}

%
%
\bibliographystyle{aa.bst} 
\bibliography{aa.bib} 

\begin{thebibliography}{118}
\expandafter\ifx\csname natexlab\endcsname\relax\def\natexlab#1{#1}\fi

\bibitem[{{Adibekyan} {et~al.}(2012){Adibekyan}, {Sousa}, {Santos}, {Delgado
  Mena}, {Gonz{\'a}lez Hern{\'a}ndez}, {Israelian}, {Mayor}, \&
  {Khachatryan}}]{Adibekyan2012}
{Adibekyan}, V.~Z., {Sousa}, S.~G., {Santos}, N.~C., {et~al.} 2012, \aap, 545,
  A32

\bibitem[{{Albrecht} {et~al.}(2009){Albrecht}, {Snellen}, {de Mooij}, \& {Le
  Poole}}]{Albrecht2009}
{Albrecht}, S., {Snellen}, I., {de Mooij}, E., \& {Le Poole}, R. 2009, in IAU
  Symposium, Vol. 253, Transiting Planets, ed. F.~{Pont}, D.~{Sasselov}, \&
  M.~J. {Holman}, 520--523

\bibitem[{{Allart} {et~al.}(2018){Allart}, {Bourrier}, {Lovis}, {Ehrenreich},
  {Spake}, {Wyttenbach}, {Pino}, {Pepe}, {Sing}, \& {Lecavelier des
  Etangs}}]{Allart2018}
{Allart}, R., {Bourrier}, V., {Lovis}, C., {et~al.} 2018, Science, 362, 1384

\bibitem[{{Allart} {et~al.}(2017){Allart}, {Lovis}, {Pino}, {Wyttenbach},
  {Ehrenreich}, \& {Pepe}}]{Allart2017}
{Allart}, R., {Lovis}, C., {Pino}, L., {et~al.} 2017, \aap, 606, A144

\bibitem[{{Allart} {et~al.}(2020){Allart}, {Pino}, {Lovis}, {Sousa},
  {Casasayas-Barris}, {Zapatero Osorio}, {Cretignier}, {Palle}, {Pepe},
  {Cristiani}, {Rebolo}, {Santos}, {Borsa}, {Bourrier}, {Demangeon},
  {Ehrenreich}, {Lavie}, {Lillo-Box}, {Micela}, {Oshagh}, {Sozzetti},
  {Tabernero}, {Adibekyan}, {Allende Prieto}, {Alibert}, {Amate}, {Benz},
  {Bouchy}, {Cabral}, {Dekker}, {D'Odorico}, {Di Marcantonio}, {Dumusque},
  {Figueira}, {Genova Santos}, {Gonz{\'a}lez Hern{\'a}ndez}, {Lo Curto},
  {Manescau}, {Martins}, {M{\'e}gevand}, {Mehner}, {Molaro}, {Nunes},
  {Poretti}, {Riva}, {Su{\'a}rez Mascare{\~n}o}, {Udry}, \&
  {Zerbi}}]{Allart2020}
{Allart}, R., {Pino}, L., {Lovis}, C., {et~al.} 2020, arXiv e-prints,
  arXiv:2010.15143

\bibitem[{{Allende Prieto} \& {Garcia Lopez}(1998)}]{Allende1998}
{Allende Prieto}, C. \& {Garcia Lopez}, R.~J. 1998, \aaps, 129, 41

\bibitem[{{Astudillo-Defru} \& {Rojo}(2013)}]{Astudillo2013}
{Astudillo-Defru}, N. \& {Rojo}, P. 2013, \aap, 557, A56

\bibitem[{{Balona} \& {Abedigamba}(2016)}]{Balona2016}
{Balona}, L.~A. \& {Abedigamba}, O.~P. 2016, \mnras, 461, 497

\bibitem[{{Birkby} {et~al.}(2017){Birkby}, {de Kok}, {Brogi}, {Schwarz}, \&
  {Snellen}}]{birkby2017}
{Birkby}, J.~L., {de Kok}, R.~J., {Brogi}, M., {Schwarz}, H., \& {Snellen},
  I.~A.~G. 2017, \aj, 153, 138

\bibitem[{{Bonomo} {et~al.}(2017){Bonomo}, {Desidera}, {Benatti}, {Borsa},
  {Crespi}, {Damasso}, {Lanza}, {Sozzetti}, {Lodato}, {Marzari}, {Boccato},
  {Claudi}, {Cosentino}, {Covino}, {Gratton}, {Maggio}, {Micela}, {Molinari},
  {Pagano}, {Piotto}, {Poretti}, {Smareglia}, {Affer}, {Biazzo}, {Bignamini},
  {Esposito}, {Giacobbe}, {H{\'e}brard}, {Malavolta}, {Maldonado}, {Mancini},
  {Martinez Fiorenzano}, {Masiero}, {Nascimbeni}, {Pedani}, {Rainer}, \& {Scand
  ariato}}]{Bonomo2017}
{Bonomo}, A.~S., {Desidera}, S., {Benatti}, S., {et~al.} 2017, \aap, 602, A107

\bibitem[{{Borsa} {et~al.}(2020){Borsa}, {Allart}, {Casasayas-Barris},
  {Tabernero}, {Zapatero Osorio}, {Cristiani}, {Pepe}, {Rebolo}, {Santos},
  {Adibekyan}, {Bourrier}, {Demangeon}, {Ehrenreich}, {Pall{\'e}}, {Sousa},
  {Lillo-Box}, {Lovis}, {Micela}, {Oshagh}, {Poretti}, {Sozzetti}, {Allende
  Prieto}, {Alibert}, {Amate}, {Benz}, {Bouchy}, {Cabral}, {Dekker},
  {D'Odorico}, {Di Marcantonio}, {Figueira}, {Genova Santos}, {Gonz{\'a}lez
  Hern{\'a}ndez}, {Lo Curto}, {Manescau}, {Martins}, {M{\'e}gevand}, {Mehner},
  {Molaro}, {Nunes}, {Riva}, {Su{\'a}rez Mascare{\~n}o}, {Udry}, \&
  {Zerbi}}]{Borsa2020}
{Borsa}, F., {Allart}, R., {Casasayas-Barris}, N., {et~al.} 2020, arXiv
  e-prints, arXiv:2011.01245

\bibitem[{{Borsa} \& {Zannoni}(2018)}]{Borsa2018}
{Borsa}, F. \& {Zannoni}, A. 2018, \aap, 617, A134

\bibitem[{{Bourrier} {et~al.}(2017){Bourrier}, {Cegla}, {Lovis}, \&
  {Wyttenbach}}]{Bourrier2017_WASP8}
{Bourrier}, V., {Cegla}, H.~M., {Lovis}, C., \& {Wyttenbach}, A. 2017, \aap,
  599, A33

\bibitem[{{Bourrier} {et~al.}(2020){Bourrier}, {Ehrenreich}, {Lendl},
  {Cretignier}, {Allart}, {Dumusque}, {Cegla}, {Su{\'a}rez-Mascare{\~n}o},
  {Wyttenbach}, {Hoeijmakers}, {Melo}, {Kuntzer}, {Astudillo-Defru}, {Giles},
  {Heng}, {Kitzmann}, {Lavie}, {Lovis}, {Murgas}, {Nascimbeni}, {Pepe}, {Pino},
  {Segransan}, \& {Udry}}]{Bourrier2020_HEARTSIII}
{Bourrier}, V., {Ehrenreich}, D., {Lendl}, M., {et~al.} 2020, \aap, 635, A205

\bibitem[{{Bourrier} {et~al.}(2018){Bourrier}, {Lovis}, {Beust}, {Ehrenreich},
  {Henry}, {Astudillo-Defru}, {Allart}, {Bonfils}, {S{\'e}gransan}, {Delfosse},
  {Cegla}, {Wyttenbach}, {Heng}, {Lavie}, \& {Pepe}}]{Bourrier_2018_Nat}
{Bourrier}, V., {Lovis}, C., {Beust}, H., {et~al.} 2018, \nat, 553, 477

\bibitem[{{Brogi} {et~al.}(2014){Brogi}, {de Kok}, {Birkby}, {Schwarz}, \&
  {Snellen}}]{Brogi2014}
{Brogi}, M., {de Kok}, R.~J., {Birkby}, J.~L., {Schwarz}, H., \& {Snellen},
  I.~A.~G. 2014, \aap, 565, A124

\bibitem[{{Brogi} {et~al.}(2018){Brogi}, {Giacobbe}, {Guilluy}, {de Kok},
  {Sozzetti}, {Mancini}, \& {Bonomo}}]{Brogi2018}
{Brogi}, M., {Giacobbe}, P., {Guilluy}, G., {et~al.} 2018, \aap, 615, A16

\bibitem[{{Casasayas-Barris} {et~al.}(2017){Casasayas-Barris}, {Palle},
  {Nowak}, {Yan}, {Nortmann}, \& {Murgas}}]{2017CasasayasB}
{Casasayas-Barris}, N., {Palle}, E., {Nowak}, G., {et~al.} 2017, \aap, 608,
  A135

\bibitem[{{Casasayas-Barris} {et~al.}(2019){Casasayas-Barris}, {Pall{\'e}},
  {Yan}, {Chen}, {Kohl}, {Stangret}, {Parviainen}, {Helling}, {Watanabe},
  {Czesla}, {Fukui}, {Monta{\~n}{\'e}s-Rodr{\'\i}guez}, {Nagel}, {Narita},
  {Nortmann}, {Nowak}, {Schmitt}, \& {Zapatero Osorio}}]{Casasayas2019}
{Casasayas-Barris}, N., {Pall{\'e}}, E., {Yan}, F., {et~al.} 2019, \aap, 628,
  A9

\bibitem[{{Casasayas-Barris} {et~al.}(2020){Casasayas-Barris}, {Pall{\'e}},
  {Yan}, {Chen}, {Luque}, {Stangret}, {Nagel}, {Zechmeister}, {Oshagh},
  {Sanz-Forcada}, {Nortmann}, {Alonso-Floriano}, {Amado}, {Caballero},
  {Czesla}, {Khalafinejad}, {L{\'o}pez-Puertas}, {L{\'o}pez-Santiago},
  {Molaverdikhani}, {Montes}, {Quirrenbach}, {Reiners}, {Ribas},
  {S{\'a}nchez-L{\'o}pez}, \& {Zapatero Osorio}}]{Casasayas2020}
{Casasayas-Barris}, N., {Pall{\'e}}, E., {Yan}, F., {et~al.} 2020, \aap, 635,
  A206

\bibitem[{{Castelli} \& {Kurucz}(2003)}]{ATLAS92003}
{Castelli}, F. \& {Kurucz}, R.~L. 2003, in IAU Symposium, Vol. 210, Modelling
  of Stellar Atmospheres, ed. N.~{Piskunov}, W.~W. {Weiss}, \& D.~F. {Gray},
  A20

\bibitem[{{Cegla} {et~al.}(2016{\natexlab{a}}){Cegla}, {Lovis}, {Bourrier},
  {Beeck}, {Watson}, \& {Pepe}}]{Cegla2016}
{Cegla}, H.~M., {Lovis}, C., {Bourrier}, V., {et~al.} 2016{\natexlab{a}}, \aap,
  588, A127

\bibitem[{{Cegla} {et~al.}(2016{\natexlab{b}}){Cegla}, {Oshagh}, {Watson},
  {Figueira}, {Santos}, \& {Shelyag}}]{Cegla2016b}
{Cegla}, H.~M., {Oshagh}, M., {Watson}, C.~A., {et~al.} 2016{\natexlab{b}},
  \apj, 819, 67

\bibitem[{{Charbonneau} {et~al.}(2000){Charbonneau}, {Brown}, {Latham}, \&
  {Mayor}}]{Charbonneau2000}
{Charbonneau}, D., {Brown}, T.~M., {Latham}, D.~W., \& {Mayor}, M. 2000, \apjl,
  529, L45

\bibitem[{{Charbonneau} {et~al.}(2002){Charbonneau}, {Brown}, {Noyes}, \&
  {Gilliland}}]{2002ApJ...568..377C}
{Charbonneau}, D., {Brown}, T.~M., {Noyes}, R.~W., \& {Gilliland}, R.~L. 2002,
  \apj, 568, 377

\bibitem[{{Chen} {et~al.}(2020){Chen}, {Casasayas-Barris}, {Pall{\'e}}, {Yan},
  {Stangret}, {Cegla}, {Allart}, \& {Lovis}}]{Chen2020}
{Chen}, G., {Casasayas-Barris}, N., {Pall{\'e}}, E., {et~al.} 2020, \aap, 635,
  A171

\bibitem[{{Chiavassa} \& {Brogi}(2019)}]{CiavassaBrogi2019}
{Chiavassa}, A. \& {Brogi}, M. 2019, \aap, 631, A100

\bibitem[{{Cosentino} {et~al.}(2012){Cosentino}, {Lovis}, {Pepe}, {Collier
  Cameron}, {Latham}, {Molinari}, {Udry}, {Bezawada}, {Black}, {Born},
  {Buchschacher}, {Charbonneau}, {Figueira}, {Fleury}, {Galli}, {Gallie},
  {Gao}, {Ghedina}, {Gonzalez}, {Gonzalez}, {Guerra}, {Henry}, {Horne},
  {Hughes}, {Kelly}, {Lodi}, {Lunney}, {Maire}, {Mayor}, {Micela}, {Ordway},
  {Peacock}, {Phillips}, {Piotto}, {Pollacco}, {Queloz}, {Rice}, {Riverol},
  {Riverol}, {San Juan}, {Sasselov}, {Segransan}, {Sozzetti}, {Sosnowska},
  {Stobie}, {Szentgyorgyi}, {Vick}, \& {Weber}}]{HARPSN22012SPIE.8446E..1VC}
{Cosentino}, R., {Lovis}, C., {Pepe}, F., {et~al.} 2012, in \procspie, Vol.
  8446, Ground-based and Airborne Instrumentation for Astronomy IV, 84461V

\bibitem[{{Czesla} {et~al.}(2015){Czesla}, {Klocov{\'a}}, {Khalafinejad},
  {Wolter}, \& {Schmitt}}]{cze15}
{Czesla}, S., {Klocov{\'a}}, T., {Khalafinejad}, S., {Wolter}, U., \&
  {Schmitt}, J.~H.~M.~M. 2015, \aap, 582, A51

\bibitem[{{Czesla} {et~al.}(2019){Czesla}, {Schr{\"o}ter}, {Schneider},
  {Huber}, {Pfeifer}, {Andreasen}, \&
  {Zechmeister}}]{PyAstronomy2019ascl.soft06010C}
{Czesla}, S., {Schr{\"o}ter}, S., {Schneider}, C.~P., {et~al.} 2019, {PyA:
  Python astronomy-related packages}

\bibitem[{{da Silva} {et~al.}(2006){da Silva}, {Girardi}, {Pasquini},
  {Setiawan}, {von der L{\"u}he}, {de Medeiros}, {Hatzes}, {D{\"o}llinger}, \&
  {Weiss}}]{daSilva2006}
{da Silva}, L., {Girardi}, L., {Pasquini}, L., {et~al.} 2006, \aap, 458, 609

\bibitem[{{Damasso} {et~al.}(2020){Damasso}, {Sozzetti}, {Lovis}, {Barros},
  {Sousa}, {Demangeon}, {Faria}, {Lillo-Box}, {Cristiani}, {Pepe}, {Rebolo},
  {Santos}, {Zapatero Osorio}, {Gonz{\'a}lez Hern{\'a}ndez}, {Amate},
  {Pasquini}, {Zerbi}, {Adibekyan}, {Abreu}, {Affolter}, {Alibert}, {Aliverti},
  {Allart}, {Allende Prieto}, {{\'A}lvarez}, {Alves}, {Avila}, {Baldini},
  {Bandy}, {Benz}, {Bianco}, {Borsa}, {Bossini}, {Bourrier}, {Bouchy}, {Broeg},
  {Cabral}, {Calderone}, {Cirami}, {Coelho}, {Conconi}, {Coretti}, {Cumani},
  {Cupani}, {D'Odorico}, {Deiries}, {Dekker}, {Delabre}, {Di Marcantonio},
  {Dumusque}, {Ehrenreich}, {Figueira}, {Fragoso}, {Genolet}, {Genoni},
  {G{\'e}nova Santos}, {Hughes}, {Iwert}, {Kerber}, {Knudstrup}, {Landoni},
  {Lavie}, {Lizon}, {Lo Curto}, {Maire}, {Martins}, {M{\'e}gevand}, {Mehner},
  {Micela}, {Modigliani}, {Molaro}, {Monteiro}, {Monteiro}, {Moschetti},
  {Mueller}, {Murphy}, {Nunes}, {Oggioni}, {Oliveira}, {Oshagh}, {Pall{\'e}},
  {Pariani}, {Poretti}, {Rasilla}, {Rebord{\~a}o}, {Redaelli}, {Riva}, {Santana
  Tschudi}, {Santin}, {Santos}, {S{\'e}gransan}, {Schmidt}, {Segovia},
  {Sosnowska}, {Span{\`o}}, {Su{\'a}rez Mascare{\~n}o}, {Tabernero}, {Tenegi},
  {Udry}, \& {Zanutta}}]{Damasso2020}
{Damasso}, M., {Sozzetti}, A., {Lovis}, C., {et~al.} 2020, arXiv e-prints,
  arXiv:2007.06410

\bibitem[{{del Burgo} \& {Allende Prieto}(2016)}]{delBurgo2016}
{del Burgo}, C. \& {Allende Prieto}, C. 2016, \mnras, 463, 1400

\bibitem[{{Deming} {et~al.}(2013){Deming}, {Wilkins}, {McCullough}, {Burrows},
  {Fortney}, {Agol}, {Dobbs-Dixon}, {Madhusudhan}, {Crouzet}, {Desert},
  {Gilliland}, {Haynes}, {Knutson}, {Line}, {Magic}, {Mandell}, {Ranjan},
  {Charbonneau}, {Clampin}, {Seager}, \& {Showman}}]{Deming2013}
{Deming}, D., {Wilkins}, A., {McCullough}, P., {et~al.} 2013, \apj, 774, 95

\bibitem[{{D{\'e}sert} {et~al.}(2008){D{\'e}sert}, {Vidal-Madjar}, {Lecavelier
  Des Etangs}, {Sing}, {Ehrenreich}, {H{\'e}brard}, \& {Ferlet}}]{Desert2008}
{D{\'e}sert}, J.~M., {Vidal-Madjar}, A., {Lecavelier Des Etangs}, A., {et~al.}
  2008, \aap, 492, 585

\bibitem[{{Dravins}(1982)}]{Dravins1982}
{Dravins}, D. 1982, \araa, 20, 61

\bibitem[{{Dravins} {et~al.}(2015){Dravins}, {Ludwig}, {Dahlen}, \&
  {Pazira}}]{Dravins2015CLV}
{Dravins}, D., {Ludwig}, H.-G., {Dahlen}, E., \& {Pazira}, H. 2015, in
  Cambridge Workshop on Cool Stars, Stellar Systems, and the Sun, Vol.~18, 18th
  Cambridge Workshop on Cool Stars, Stellar Systems, and the Sun, 853--868

\bibitem[{{Dravins} {et~al.}(2017){Dravins}, {Ludwig}, {Dahl{\'e}n}, \&
  {Pazira}}]{Dravins2017}
{Dravins}, D., {Ludwig}, H.-G., {Dahl{\'e}n}, E., \& {Pazira}, H. 2017, \aap,
  605, A90

\bibitem[{{Ehrenreich} {et~al.}(2020){Ehrenreich}, {Lovis}, {Allart}, {Zapatero
  Osorio}, {Pepe}, {Cristiani}, {Rebolo}, {Santos}, {Borsa}, {Demangeon},
  {Dumusque}, {Gonz{\'a}lez Hern{\'a}ndez}, {Casasayas-Barris},
  {S{\'e}gransan}, {Sousa}, {Abreu}, {Adibekyan}, {Affolter}, {Allende Prieto},
  {Alibert}, {Aliverti}, {Alves}, {Amate}, {Avila}, {Baldini}, {Bandy}, {Benz},
  {Bianco}, {Bolmont}, {Bouchy}, {Bourrier}, {Broeg}, {Cabral}, {Calderone},
  {Pall{\'e}}, {Cegla}, {Cirami}, {Coelho}, {Conconi}, {Coretti}, {Cumani},
  {Cupani}, {Dekker}, {Delabre}, {Deiries}, {D'Odorico}, {Di Marcantonio},
  {Figueira}, {Fragoso}, {Genolet}, {Genoni}, {G{\'e}nova Santos}, {Hara},
  {Hughes}, {Iwert}, {Kerber}, {Knudstrup}, {Land oni}, {Lavie}, {Lizon},
  {Lendl}, {Lo Curto}, {Maire}, {Manescau}, {Martins}, {M{\'e}gevand },
  {Mehner}, {Micela}, {Modigliani}, {Molaro}, {Monteiro}, {Monteiro},
  {Moschetti}, {M{\"u}ller}, {Nunes}, {Oggioni}, {Oliveira}, {Pariani},
  {Pasquini}, {Poretti}, {Rasilla}, {Redaelli}, {Riva}, {Santana Tschudi},
  {Santin}, {Santos}, {Segovia Milla}, {Seidel}, {Sosnowska}, {Sozzetti},
  {Span{\`o}}, {Su{\'a}rez Mascare{\~n}o}, {Tabernero}, {Tenegi}, {Udry},
  {Zanutta}, \& {Zerbi}}]{Ehrenreich2020}
{Ehrenreich}, D., {Lovis}, C., {Allart}, R., {et~al.} 2020, \nat, 580, 597

\bibitem[{{Espinoza} \& {Jord{\'a}n}(2015)}]{Espinoza2015}
{Espinoza}, N. \& {Jord{\'a}n}, A. 2015, \mnras, 450, 1879

\bibitem[{{Evans} {et~al.}(2015){Evans}, {Aigrain}, {Gibson}, {Barstow},
  {Amundsen}, {Tremblin}, \& {Mourier}}]{Evans2015MNRAS.451..680E}
{Evans}, T.~M., {Aigrain}, S., {Gibson}, N., {et~al.} 2015, \mnras, 451, 680

\bibitem[{{Evans} {et~al.}(2016){Evans}, {Sing}, {Wakeford}, {Nikolov},
  {Ballester}, {Drummond}, {Kataria}, {Gibson}, {Amundsen}, \&
  {Spake}}]{Evans2016}
{Evans}, T.~M., {Sing}, D.~K., {Wakeford}, H.~R., {et~al.} 2016, \apjl, 822, L4

\bibitem[{{Fortney} {et~al.}(2008){Fortney}, {Lodders}, {Marley}, \&
  {Freedman}}]{Fortney2008}
{Fortney}, J.~J., {Lodders}, K., {Marley}, M.~S., \& {Freedman}, R.~S. 2008,
  \apj, 678, 1419

\bibitem[{{Guilluy} {et~al.}(2019){Guilluy}, {Sozzetti}, {Brogi}, {Bonomo},
  {Giacobbe}, {Claudi}, \& {Benatti}}]{Guilluy2019}
{Guilluy}, G., {Sozzetti}, A., {Brogi}, M., {et~al.} 2019, \aap, 625, A107

\bibitem[{{Gustafsson} {et~al.}(2008){Gustafsson}, {Edvardsson}, {Eriksson},
  {J{\o}rgensen}, {Nordlund}, \& {Plez}}]{Gustafsson2008}
{Gustafsson}, B., {Edvardsson}, B., {Eriksson}, K., {et~al.} 2008, \aap, 486,
  951

\bibitem[{{Hayek} {et~al.}(2012){Hayek}, {Sing}, {Pont}, \&
  {Asplund}}]{Hayek2012}
{Hayek}, W., {Sing}, D., {Pont}, F., \& {Asplund}, M. 2012, \aap, 539, A102

\bibitem[{{Henry} {et~al.}(2000){Henry}, {Marcy}, {Butler}, \&
  {Vogt}}]{Henry2000ApJ...529L..41H}
{Henry}, G.~W., {Marcy}, G.~W., {Butler}, R.~P., \& {Vogt}, S.~S. 2000, \apjl,
  529, L41

\bibitem[{{Hoeijmakers} {et~al.}(2015){Hoeijmakers}, {de Kok}, {Snellen},
  {Brogi}, {Birkby}, \& {Schwarz}}]{Hoeijmakers2015}
{Hoeijmakers}, H.~J., {de Kok}, R.~J., {Snellen}, I.~A.~G., {et~al.} 2015,
  \aap, 575, A20

\bibitem[{{Hoeijmakers} {et~al.}(2018){Hoeijmakers}, {Ehrenreich}, {Heng},
  {Kitzmann}, {Grimm}, {Allart}, {Deitrick}, {Wyttenbach}, {Oreshenko}, {Pino},
  {Rimmer}, {Molinari}, \& {Di Fabrizio}}]{Hoeijmakers2018}
{Hoeijmakers}, H.~J., {Ehrenreich}, D., {Heng}, K., {et~al.} 2018, \nat, 560,
  453

\bibitem[{{Jensen} {et~al.}(2012){Jensen}, {Redfield}, {Endl}, {Cochran},
  {Koesterke}, \& {Barman}}]{Jensen2012}
{Jensen}, A.~G., {Redfield}, S., {Endl}, M., {et~al.} 2012, \apj, 751, 86

\bibitem[{{Jensen} {et~al.}(2011){Jensen}, {Redfield}, {Endl}, {Cochran},
  {Koesterke}, \& {Barman}}]{jensen2011}
{Jensen}, A.~G., {Redfield}, S., {Endl}, M., {et~al.} 2011, \apj, 743, 203

\bibitem[{{Karoff} {et~al.}(2018){Karoff}, {Metcalfe}, {Santos}, {Montet},
  {Isaacson}, {Witzke}, {Shapiro}, {Mathur}, {Davies}, {Lund}, {Garcia},
  {Brun}, {Salabert}, {Avelino}, {van Saders}, {Egeland}, {Cunha}, {Campante},
  {Chaplin}, {Krivova}, {Solanki}, {Stritzinger}, \& {Knudsen}}]{Karoff2018}
{Karoff}, C., {Metcalfe}, T.~S., {Santos}, {\^A}. R.~G., {et~al.} 2018, \apj,
  852, 46

\bibitem[{{Kausch} {et~al.}(2015){Kausch}, {Noll}, {Smette}, {Kimeswenger},
  {Barden}, {Szyszka}, {Jones}, {Sana}, {Horst}, \& {Kerber}}]{Molecfit2}
{Kausch}, W., {Noll}, S., {Smette}, A., {et~al.} 2015, \aap, 576, A78

\bibitem[{{Khalafinejad} {et~al.}(2017){Khalafinejad}, {von Essen},
  {Hoeijmakers}, {Zhou}, {Klocov{\'a}}, {Schmitt}, {Dreizler}, {Lopez-Morales},
  {Husser}, {Schmidt}, \& {Collet}}]{Khalafinehad2017A&A...598A.131K}
{Khalafinejad}, S., {von Essen}, C., {Hoeijmakers}, H.~J., {et~al.} 2017, \aap,
  598, A131

\bibitem[{{Kreidberg}(2015)}]{Kreidberg2015}
{Kreidberg}, L. 2015, \pasp, 127, 1161

\bibitem[{{Kupka} {et~al.}(1999){Kupka}, {Piskunov}, {Ryabchikova}, {Stempels},
  \& {Weiss}}]{VALDKup1999}
{Kupka}, F., {Piskunov}, N., {Ryabchikova}, T.~A., {Stempels}, H.~C., \&
  {Weiss}, W.~W. 1999, \aaps, 138, 119

\bibitem[{{Kurucz}(1993)}]{Kurucz1993}
{Kurucz}, R.~L. 1993, {SYNTHE spectrum synthesis programs and line data}

\bibitem[{{Kurucz}(2013)}]{ATLAS122013K}
{Kurucz}, R.~L. 2013, {ATLAS12: Opacity sampling model atmosphere program}

\bibitem[{{Langland-Shula} {et~al.}(2009){Langland-Shula}, {Vogt},
  {Charbonneau}, {Butler}, \& {Marcy}}]{Langland2009}
{Langland-Shula}, L.~E., {Vogt}, S.~S., {Charbonneau}, D., {Butler}, P., \&
  {Marcy}, G. 2009, \apj, 696, 1355

\bibitem[{{Louden} \& {Wheatley}(2015)}]{LoudenW2015}
{Louden}, T. \& {Wheatley}, P.~J. 2015, \apjl, 814, L24

\bibitem[{{Malik} {et~al.}(2017){Malik}, {Grosheintz}, {Mendon{\c{c}}a},
  {Grimm}, {Lavie}, {Kitzmann}, {Tsai}, {Burrows}, {Kreidberg}, {Bedell},
  {Bean}, {Stevenson}, \& {Heng}}]{Malik2017}
{Malik}, M., {Grosheintz}, L., {Mendon{\c{c}}a}, J.~M., {et~al.} 2017, \aj,
  153, 56

\bibitem[{{Malik} {et~al.}(2019){Malik}, {Kitzmann}, {Mendon{\c{c}}a}, {Grimm},
  {Marleau}, {Linder}, {Tsai}, \& {Heng}}]{Malik2019}
{Malik}, M., {Kitzmann}, D., {Mendon{\c{c}}a}, J.~M., {et~al.} 2019, \aj, 157,
  170

\bibitem[{{Mashonkina} {et~al.}(2008){Mashonkina}, {Zhao}, {Gehren}, {Aoki},
  {Bergemann}, {Noguchi}, {Shi}, {Takada-Hidai}, \& {Zhang}}]{Mashonkina2008}
{Mashonkina}, L., {Zhao}, G., {Gehren}, T., {et~al.} 2008, \aap, 478, 529

\bibitem[{{McKemmish} {et~al.}(2019){McKemmish}, {Masseron}, {Hoeijmakers},
  {P{\'e}rez-Mesa}, {Grimm}, {Yurchenko}, \& {Tennyson}}]{McKemmish2019}
{McKemmish}, L.~K., {Masseron}, T., {Hoeijmakers}, H.~J., {et~al.} 2019,
  \mnras, 488, 2836

\bibitem[{{McKemmish} {et~al.}(2016){McKemmish}, {Yurchenko}, \&
  {Tennyson}}]{McKemmish2016}
{McKemmish}, L.~K., {Yurchenko}, S.~N., \& {Tennyson}, J. 2016, \mnras, 463,
  771

\bibitem[{{Merritt} {et~al.}(2020){Merritt}, {Gibson}, {Nugroho}, {de Mooij},
  {Hooton}, {Matthews}, {McKemmish}, {Mikal-Evans}, {Nikolov}, {Sing}, {Spake},
  \& {Watson}}]{Merritt2020}
{Merritt}, S.~R., {Gibson}, N.~P., {Nugroho}, S.~K., {et~al.} 2020, \aap, 636,
  A117

\bibitem[{{Meunier} {et~al.}(2017){Meunier}, {Mignon}, \&
  {Lagrange}}]{Meunier2017}
{Meunier}, N., {Mignon}, L., \& {Lagrange}, A.~M. 2017, \aap, 607, A124

\bibitem[{{Molli{\`e}re} {et~al.}(2019){Molli{\`e}re}, {Wardenier}, {van
  Boekel}, {Henning}, {Molaverdikhani}, \& {Snellen}}]{petitRADTRANS2019}
{Molli{\`e}re}, P., {Wardenier}, J.~P., {van Boekel}, R., {et~al.} 2019, \aap,
  627, A67

\bibitem[{{M{\"u}ller} {et~al.}(2013){M{\"u}ller}, {Huber}, {Czesla}, {Wolter},
  \& {Schmitt}}]{MullerCLV2013}
{M{\"u}ller}, H.~M., {Huber}, K.~F., {Czesla}, S., {Wolter}, U., \& {Schmitt},
  J.~H.~M.~M. 2013, \aap, 560, A112

\bibitem[{{Naef} {et~al.}(2004){Naef}, {Mayor}, {Beuzit}, {Perrier}, {Queloz},
  {Sivan}, \& {Udry}}]{HD2092004Naef}
{Naef}, D., {Mayor}, M., {Beuzit}, J.~L., {et~al.} 2004, \aap, 414, 351

\bibitem[{{Narita} {et~al.}(2005){Narita}, {Suto}, {Winn}, {Turner}, {Aoki},
  {Leigh}, {Sato}, {Tamura}, \& {Yamada}}]{Narita2005HD209}
{Narita}, N., {Suto}, Y., {Winn}, J.~N., {et~al.} 2005, \pasj, 57, 471

\bibitem[{{Nortmann} {et~al.}(2018){Nortmann}, {Pall{\'e}}, {Salz},
  {Sanz-Forcada}, {Nagel}, {Alonso-Floriano}, {Czesla}, {Yan}, {Chen},
  {Snellen}, {Zechmeister}, {Schmitt}, {L{\'o}pez-Puertas}, {Casasayas-Barris},
  {Bauer}, {Amado}, {Caballero}, {Dreizler}, {Henning}, {Lamp{\'o}n}, {Montes},
  {Molaverdikhani}, {Quirrenbach}, {Reiners}, {Ribas}, {S{\'a}nchez-L{\'o}pez},
  {Schneider}, \& {Zapatero Osorio}}]{Nortmann2018Science}
{Nortmann}, L., {Pall{\'e}}, E., {Salz}, M., {et~al.} 2018, Science, 362, 1388

\bibitem[{{Pepe} {et~al.}(2020){Pepe}, {Damasso}, {Cristiani}, {Rebolo}, \&
  {Santos}}]{Pepe2020}
{Pepe}, F., {Damasso}, M., {Cristiani}, S., {Rebolo}, R., \& {Santos}, N. C.~a.
  2020, \AA

\bibitem[{{Pepe} {et~al.}(2014){Pepe}, {Molaro}, {Cristiani}, {Rebolo},
  {Santos}, {Dekker}, {M{\'e}gevand}, {Zerbi}, {Cabral}, {Di Marcantonio},
  {Abreu}, {Affolter}, {Aliverti}, {Allende Prieto}, {Amate}, {Avila},
  {Baldini}, {Bristow}, {Broeg}, {Cirami}, {Coelho}, {Conconi}, {Coretti},
  {Cupani}, {D'Odorico}, {De Caprio}, {Delabre}, {Dorn}, {Figueira}, {Fragoso},
  {Galeotta}, {Genolet}, {Gomes}, {Gonz{\'a}lez Hern{\'a}ndez}, {Hughes},
  {Iwert}, {Kerber}, {Landoni}, {Lizon}, {Lovis}, {Maire}, {Mannetta},
  {Martins}, {Monteiro}, {Oliveira}, {Poretti}, {Rasilla}, {Riva}, {Santana
  Tschudi}, {Santos}, {Sosnowska}, {Sousa}, {Span{\'o}}, {Tenegi}, {Toso},
  {Vanzella}, {Viel}, \& {Zapatero Osorio}}]{Pepe2014ESP}
{Pepe}, F., {Molaro}, P., {Cristiani}, S., {et~al.} 2014, Astronomische
  Nachrichten, 335, 8

\bibitem[{{Pepe} {et~al.}(2010){Pepe}, {Cristiani}, {Rebolo Lopez}, {Santos},
  {Amorim}, {Avila}, {Benz}, {Bonifacio}, {Cabral}, {Carvas}, {Cirami},
  {Coelho}, {Comari}, {Coretti}, {De Caprio}, {Dekker}, {Delabre}, {Di
  Marcantonio}, {D'Odorico}, {Fleury}, {Garc{\'\i}a}, {Herreros Linares},
  {Hughes}, {Iwert}, {Lima}, {Lizon}, {Lo Curto}, {Lovis}, {Manescau},
  {Martins}, {M{\'e}gevand}, {Moitinho}, {Molaro}, {Monteiro}, {Monteiro},
  {Pasquini}, {Mordasini}, {Queloz}, {Rasilla}, {Rebord{\~a}o}, {Santana
  Tschudi}, {Santin}, {Sosnowska}, {Span{\`o}}, {Tenegi}, {Udry}, {Vanzella},
  {Viel}, {Zapatero Osorio}, \& {Zerbi}}]{Pepe2010ESP}
{Pepe}, F.~A., {Cristiani}, S., {Rebolo Lopez}, R., {et~al.} 2010, Society of
  Photo-Optical Instrumentation Engineers (SPIE) Conference Series, Vol. 7735,
  {ESPRESSO: the Echelle spectrograph for rocky exoplanets and stable
  spectroscopic observations}, 77350F

\bibitem[{{Pino} {et~al.}(2020){Pino}, {D{\'e}sert}, {Brogi}, {Malavolta},
  {Wyttenbach}, {Line}, {Hoeijmakers}, {Fossati}, {Bonomo}, {Nascimbeni},
  {Panwar}, {Affer}, {Benatti}, {Biazzo}, {Bignamini}, {Borsa}, {Carleo},
  {Claudi}, {Cosentino}, {Covino}, {Damasso}, {Desidera}, {Giacobbe},
  {Harutyunyan}, {Lanza}, {Leto}, {Maggio}, {Maldonado}, {Mancini}, {Micela},
  {Molinari}, {Pagano}, {Piotto}, {Poretti}, {Rainer}, {Scandariato},
  {Sozzetti}, {Allart}, {Borsato}, {Bruno}, {Fabrizio}, {Ehrenreich},
  {Fiorenzano}, {Frustagli}, {Lavie}, {Lovis}, {Magazz{\`u}}, {Nardiello},
  {Pedani}, \& {Smareglia}}]{Pino2020}
{Pino}, L., {D{\'e}sert}, J.-M., {Brogi}, M., {et~al.} 2020, \apjl, 894, L27

\bibitem[{{Piskunov} \& {Valenti}(2017)}]{SMEEvolution2017}
{Piskunov}, N. \& {Valenti}, J.~A. 2017, \aap, 597, A16

\bibitem[{{Piskunov} {et~al.}(1995){Piskunov}, {Kupka}, {Ryabchikova}, {Weiss},
  \& {Jeffery}}]{VALDPisk1995}
{Piskunov}, N.~E., {Kupka}, F., {Ryabchikova}, T.~A., {Weiss}, W.~W., \&
  {Jeffery}, C.~S. 1995, \aaps, 112, 525

\bibitem[{{Plez}(1998)}]{Plez1998}
{Plez}, B. 1998, \aap, 337, 495

\bibitem[{{Plez}(2012)}]{Plez2012}
{Plez}, B. 2012, {Turbospectrum: Code for spectral synthesis}

\bibitem[{{Quirrenbach} {et~al.}(2014){Quirrenbach}, {Amado}, {Caballero},
  {Mundt}, {Reiners}, {Ribas}, {Seifert}, {Abril}, {Aceituno},
  {Alonso-Floriano}, {Ammler-von Eiff}, {Antona Jim{\'e}nez},
  {Anwand-Heerwart}, {Azzaro}, {Bauer}, {Barrado}, {Becerril}, {B{\'e}jar},
  {Ben{\'{\i}}tez}, {Berdi{\~n}as}, {C{\'a}rdenas}, {Casal}, {Claret},
  {Colom{\'e}}, {Cort{\'e}s-Contreras}, {Czesla}, {Doellinger}, {Dreizler},
  {Feiz}, {Fern{\'a}ndez}, {Galad{\'{\i}}}, {G{\'a}lvez-Ortiz},
  {Garc{\'{\i}}a-Piquer}, {Garc{\'{\i}}a-Vargas}, {Garrido}, {Gesa}, {G{\'o}mez
  Galera}, {Gonz{\'a}lez {\'A}lvarez}, {Gonz{\'a}lez Hern{\'a}ndez},
  {Gr{\"o}zinger}, {Gu{\`a}rdia}, {Guenther}, {de Guindos},
  {Guti{\'e}rrez-Soto}, {Hagen}, {Hatzes}, {Hauschildt}, {Helmling}, {Henning},
  {Hermann}, {Hern{\'a}ndez Casta{\~n}o}, {Herrero}, {Hidalgo}, {Holgado},
  {Huber}, {Huber}, {Jeffers}, {Joergens}, {de Juan}, {Kehr}, {Klein},
  {K{\"u}rster}, {Lamert}, {Lalitha}, {Laun}, {Lemke}, {Lenzen}, {L{\'o}pez del
  Fresno}, {L{\'o}pez Mart{\'{\i}}}, {L{\'o}pez-Santiago}, {Mall}, {Mandel},
  {Mart{\'{\i}}n}, {Mart{\'{\i}}n-Ruiz}, {Mart{\'{\i}}nez-Rodr{\'{\i}}guez},
  {Marvin}, {Mathar}, {Mirabet}, {Montes}, {Morales Mu{\~n}oz}, {Moya},
  {Naranjo}, {Ofir}, {Oreiro}, {Pall{\'e}}, {Panduro}, {Passegger},
  {P{\'e}rez-Calpena}, {P{\'e}rez Medialdea}, {Perger}, {Pluto}, {Ram{\'o}n},
  {Rebolo}, {Redondo}, {Reffert}, {Reinhardt}, {Rhode}, {Rix}, {Rodler},
  {Rodr{\'{\i}}guez}, {Rodr{\'{\i}}guez-L{\'o}pez},
  {Rodr{\'{\i}}guez-P{\'e}rez}, {Rohloff}, {Rosich}, {S{\'a}nchez-Blanco},
  {S{\'a}nchez Carrasco}, {Sanz-Forcada}, {Sarmiento}, {Sch{\"a}fer},
  {Schiller}, {Schmidt}, {Schmitt}, {Solano}, {Stahl}, {Storz}, {St{\"u}rmer},
  {Su{\'a}rez}, {Ulbrich}, {Veredas}, {Wagner}, {Winkler}, {Zapatero Osorio},
  {Zechmeister}, {Abell{\'a}n de Paco}, {Anglada-Escud{\'e}}, {del Burgo},
  {Klutsch}, {Lizon}, {L{\'o}pez-Morales}, {Morales}, {Perryman}, {Tulloch}, \&
  {Xu}}]{CARMENES}
{Quirrenbach}, A., {Amado}, P.~J., {Caballero}, J.~A., {et~al.} 2014, in
  \procspie, Vol. 9147, Ground-based and Airborne Instrumentation for Astronomy
  V, 91471F

\bibitem[{{Quirrenbach} {et~al.}(2018){Quirrenbach}, {Amado}, {Ribas},
  {Reiners}, {Caballero}, {Seifert}, {Aceituno}, {Azzaro}, {Baroch}, {Barrado},
  \& et~al.}]{CARMENES18}
{Quirrenbach}, A., {Amado}, P.~J., {Ribas}, I., {et~al.} 2018, in Society of
  Photo-Optical Instrumentation Engineers (SPIE) Conference Series, Vol. 10702,
  Ground-based and Airborne Instrumentation for Astronomy VII, 107020W

\bibitem[{{Redfield} {et~al.}(2008){Redfield}, {Endl}, {Cochran}, \&
  {Koesterke}}]{2008Redfield}
{Redfield}, S., {Endl}, M., {Cochran}, W.~D., \& {Koesterke}, L. 2008, \apjl,
  673, L87

\bibitem[{{Reiners} {et~al.}(2016){Reiners}, {Mrotzek}, {Lemke}, {Hinrichs}, \&
  {Reinsch}}]{Reiners2016}
{Reiners}, A., {Mrotzek}, N., {Lemke}, U., {Hinrichs}, J., \& {Reinsch}, K.
  2016, \aap, 587, A65

\bibitem[{Richardson {et~al.}(2006)Richardson, Harrington, Seager, \&
  Deming}]{HD209Richardson_2006}
Richardson, L.~J., Harrington, J., Seager, S., \& Deming, D. 2006, The
  Astrophysical Journal, 649, 1043

\bibitem[{{S{\'a}nchez-L{\'o}pez} {et~al.}(2019){S{\'a}nchez-L{\'o}pez},
  {Alonso-Floriano}, {L{\'o}pez-Puertas}, {Snellen}, {Funke}, {Nagel}, {Bauer},
  {Amado}, {Caballero}, {Czesla}, {Nortmann}, {Pall{\'e}}, {Salz}, {Reiners},
  {Ribas}, {Quirrenbach}, {Anglada-Escud{\'e}}, {B{\'e}jar},
  {Casasayas-Barris}, {Galad{\'\i}-Enr{\'\i}quez}, {Guenther}, {Henning},
  {Kaminski}, {K{\"u}rster}, {Lamp{\'o}n}, {Lara}, {Montes}, {Morales},
  {Stangret}, {Tal-Or}, {Sanz-Forcada}, {Schmitt}, {Zapatero Osorio}, \&
  {Zechmeister}}]{SanchezLopez2019}
{S{\'a}nchez-L{\'o}pez}, A., {Alonso-Floriano}, F.~J., {L{\'o}pez-Puertas}, M.,
  {et~al.} 2019, \aap, 630, A53

\bibitem[{{Santos} {et~al.}(2020){Santos}, {Cristo}, {Demangeon}, {Oshagh},
  {Allart}, {Barros}, {Borsa}, {Bourrier}, {Casasayas-Barris}, {Ehrenreich},
  {Faria}, {Figueira}, {Martins}, {Micela}, {Pall{\'e}}, {Sozzetti},
  {Tabernero}, {Zapatero Osorio}, {Pepe}, {Cristiani}, {Rebolo}, {Adibekyan},
  {Allende Prieto}, {Alibert}, {Bouchy}, {Cabral}, {Dekker}, {Di Marcantonio},
  {D'Odorico}, {Dumusque}, {Gonz{\'a}lez Hern{\'a}ndez}, {Lavie}, {Lo Curto},
  {Lovis}, {Manescau}, {Martins}, {M{\'e}gevand}, {Mehner}, {Molaro}, {Nunes},
  {Poretti}, {Riva}, {Sousa}, {Su{\'a}rez Mascare{\~n}o}, \&
  {Udry}}]{Santos2020}
{Santos}, N.~C., {Cristo}, E., {Demangeon}, O., {et~al.} 2020, \aap, 644, A51

\bibitem[{{Santos} {et~al.}(2013){Santos}, {Sousa}, {Mortier}, {Neves},
  {Adibekyan}, {Tsantaki}, {Delgado Mena}, {Bonfils}, {Israelian}, {Mayor}, \&
  {Udry}}]{Santos-13}
{Santos}, N.~C., {Sousa}, S.~G., {Mortier}, A., {et~al.} 2013, \aap, 556, A150

\bibitem[{{Seager} \& {Sasselov}(1998)}]{Seager1998}
{Seager}, S. \& {Sasselov}, D.~D. 1998, \apjl, 502, L157

\bibitem[{{Sedaghati} {et~al.}(2017){Sedaghati}, {Boffin}, {MacDonald},
  {Gandhi}, {Madhusudhan}, {Gibson}, {Oshagh}, {Claret}, \&
  {Rauer}}]{Elyar2017}
{Sedaghati}, E., {Boffin}, H. M.~J., {MacDonald}, R.~J., {et~al.} 2017, \nat,
  549, 238

\bibitem[{{Seidel} {et~al.}(2019){Seidel}, {Ehrenreich}, {Wyttenbach},
  {Allart}, {Lendl}, {Pino}, {Bourrier}, {Cegla}, {Lovis}, {Barrado},
  {Bayliss}, {Astudillo-Defru}, {Deline}, {Fisher}, {Heng}, {Joseph}, {Lavie},
  {Melo}, {Pepe}, {S{\'e}gransan}, \& {Udry}}]{Seidel2019}
{Seidel}, J.~V., {Ehrenreich}, D., {Wyttenbach}, A., {et~al.} 2019, \aap, 623,
  A166

\bibitem[{{Serrano} {et~al.}(2020){Serrano}, {Oshagh}, {Cegla}, {Barros},
  {Santos}, {Faria}, \& {Akinsanmi}}]{WASP72020}
{Serrano}, L.~M., {Oshagh}, M., {Cegla}, H.~M., {et~al.} 2020, \mnras, 493,
  5928

\bibitem[{{Sing} {et~al.}(2016){Sing}, {Fortney}, {Nikolov}, {Wakeford},
  {Kataria}, {Evans}, {Aigrain}, {Ballester}, {Burrows}, {Deming},
  {D{\'e}sert}, {Gibson}, {Henry}, {Huitson}, {Knutson}, {Lecavelier Des
  Etangs}, {Pont}, {Showman}, {Vidal-Madjar}, {Williamson}, \&
  {Wilson}}]{Sing2016}
{Sing}, D.~K., {Fortney}, J.~J., {Nikolov}, N., {et~al.} 2016, \nat, 529, 59

\bibitem[{{Sing} {et~al.}(2008){Sing}, {Vidal-Madjar}, {D{\'e}sert},
  {Lecavelier des Etangs}, \& {Ballester}}]{Sing2008HD209ApJ...686..658S}
{Sing}, D.~K., {Vidal-Madjar}, A., {D{\'e}sert}, J.~M., {Lecavelier des
  Etangs}, A., \& {Ballester}, G. 2008, \apj, 686, 658

\bibitem[{{Smette} {et~al.}(2015){Smette}, {Sana}, {Noll}, {Horst}, {Kausch},
  {Kimeswenger}, {Barden}, {Szyszka}, {Jones}, {Gallenne}, {Vinther},
  {Ballester}, \& {Taylor}}]{Molecfit1}
{Smette}, A., {Sana}, H., {Noll}, S., {et~al.} 2015, \aap, 576, A77

\bibitem[{{Sneden}(1973)}]{Sneden1973}
{Sneden}, C. 1973, \apj, 184, 839

\bibitem[{{Snellen}(2013)}]{Snellen2013}
{Snellen}, I. 2013, in European Physical Journal Web of Conferences, Vol.~47,
  European Physical Journal Web of Conferences, 11001

\bibitem[{{Snellen} {et~al.}(2008){Snellen}, {Albrecht}, {de Mooij}, \& {Le
  Poole}}]{2008SnellenHD209}
{Snellen}, I.~A.~G., {Albrecht}, S., {de Mooij}, E.~J.~W., \& {Le Poole}, R.~S.
  2008, \aap, 487, 357

\bibitem[{{Snellen} {et~al.}(2010){Snellen}, {de Kok}, {de Mooij}, \&
  {Albrecht}}]{Snellen2010}
{Snellen}, I.~A.~G., {de Kok}, R.~J., {de Mooij}, E.~J.~W., \& {Albrecht}, S.
  2010, \nat, 465, 1049

\bibitem[{{Sousa}(2014)}]{Sousa-14}
{Sousa}, S.~G. 2014, {ARES + MOOG: A Practical Overview of an Equivalent Width
  (EW) Method to Derive Stellar Parameters}, 297--310

\bibitem[{{Sousa} {et~al.}(2015){Sousa}, {Santos}, {Adibekyan}, {Delgado-Mena},
  \& {Israelian}}]{Sousa-15}
{Sousa}, S.~G., {Santos}, N.~C., {Adibekyan}, V., {Delgado-Mena}, E., \&
  {Israelian}, G. 2015, \aap, 577, A67

\bibitem[{{Sousa} {et~al.}(2007){Sousa}, {Santos}, {Israelian}, {Mayor}, \&
  {Monteiro}}]{Sousa-07}
{Sousa}, S.~G., {Santos}, N.~C., {Israelian}, G., {Mayor}, M., \& {Monteiro},
  M.~J.~P.~F.~G. 2007, \aap, 469, 783

\bibitem[{{Sousa} {et~al.}(2008){Sousa}, {Santos}, {Mayor}, {Udry},
  {Casagrande}, {Israelian}, {Pepe}, {Queloz}, \& {Monteiro}}]{Sousa2008}
{Sousa}, S.~G., {Santos}, N.~C., {Mayor}, M., {et~al.} 2008, \aap, 487, 373

\bibitem[{{Stangret} {et~al.}(2020){Stangret}, {Casasayas-Barris}, {Pall{\'e}},
  {Yan}, {S{\'a}nchez-L{\'o}pez}, \& {L{\'o}pez-Puertas}}]{Stangret2020}
{Stangret}, M., {Casasayas-Barris}, N., {Pall{\'e}}, E., {et~al.} 2020, \aap,
  638, A26

\bibitem[{{Stassun} {et~al.}(2017){Stassun}, {Collins}, \&
  {Gaudi}}]{Stassun2017HD209}
{Stassun}, K.~G., {Collins}, K.~A., \& {Gaudi}, B.~S. 2017, \aj, 153, 136

\bibitem[{{Su{\'a}rez Mascare{\~n}o} {et~al.}(2020){Su{\'a}rez Mascare{\~n}o},
  {Faria}, {Figueira}, {Lovis}, {Damasso}, {Gonz{\'a}lez Hern{\'a}ndez},
  {Rebolo}, {Cristiani}, {Pepe}, {Santos}, {Zapatero Osorio}, {Adibekyan},
  {Hojjatpanah}, {Sozzetti}, {Murgas}, {Abreu}, {Affolter}, {Alibert},
  {Aliverti}, {Allart}, {Allende Prieto}, {Alves}, {Amate}, {Avila}, {Baldini},
  {Bandi}, {Barros}, {Bianco}, {Benz}, {Bouchy}, {Broeng}, {Cabral},
  {Calderone}, {Cirami}, {Coelho}, {Conconi}, {Coretti}, {Cumani}, {Cupani},
  {D'Odorico}, {Deiries}, {Delabre}, {Di Marcantonio}, {Dumusque},
  {Ehrenreich}, {Fragoso}, {Genolet}, {Genoni}, {G{\'e}nova Santos}, {Hughes},
  {Iwert}, {Kerber}, {Knusdstrup}, {Landoni}, {Lavie}, {Lillo-Box}, {Lizon},
  {Lo Curto}, {Maire}, {Manescau}, {Martins}, {M{\'e}gevand}, {Mehner},
  {Micela}, {Modigliani}, {Molaro}, {Monteiro}, {Monteiro}, {Moschetti},
  {Mueller}, {Nunes}, {Oggioni}, {Oliveira}, {Pall{\'e}}, {Pariani},
  {Pasquini}, {Poretti}, {Rasilla}, {Redaelli}, {Riva}, {Santana Tschudi},
  {Santin}, {Santos}, {Segovia}, {Sosnowska}, {Sousa}, {Span{\`o}}, {Tenegi},
  {Udry}, {Zanutta}, \& {Zerbi}}]{Alejandro2020}
{Su{\'a}rez Mascare{\~n}o}, A., {Faria}, J.~P., {Figueira}, P., {et~al.} 2020,
  \aap, 639, A77

\bibitem[{{Tabernero} {et~al.}(2019){Tabernero}, {Marfil}, {Montes}, \&
  {Gonz{\'a}lez Hern{\'a}ndez}}]{StePar2019}
{Tabernero}, H.~M., {Marfil}, E., {Montes}, D., \& {Gonz{\'a}lez
  Hern{\'a}ndez}, J.~I. 2019, \aap, 628, A131

\bibitem[{{Tabernero} {et~al.}(2020){Tabernero}, {Zapatero Osorio}, {Allart},
  {Borsa}, {Casasayas-Barris}, {Demangeon}, {Ehrenreich}, {Lillo-Box}, {Lovis},
  {Pall{\'e}}, {Sousa}, {Rebolo}, {Santos}, {Pepe}, {Cristiani}, {Adibekyan},
  {Allende Prieto}, {Alibert}, {Barros}, {Bouchy}, {Bourrier}, {D'Odorico},
  {Dumusque}, {Faria}, {Figueira}, {G{\'e}nova Santos}, {Gonz{\'a}lez
  Hern{\'a}ndez}, {Hojjatpanah}, {Lo Curto}, {Lavie}, {Martins}, {Martins},
  {Mehner}, {Micela}, {Molaro}, {Nunes}, {Poretti}, {Seidel}, {Sozzetti},
  {Su{\'a}rez Mascare{\~n}o}, {Udry}, {Aliverti}, {Affolter}, {Alves}, {Amate},
  {Avila}, {Bandy}, {Benz}, {Bianco}, {Broeg}, {Cabral}, {Conconi}, {Coelho},
  {Cumani}, {Deiries}, {Dekker}, {Delabre}, {Fragoso}, {Genoni}, {Genolet},
  {Hughes}, {Knudstrup}, {Kerber}, {Landoni}, {Lizon}, {Maire}, {Manescau}, {Di
  Marcantonio}, {M{\'e}gevand}, {Monteiro}, {Monteiro}, {Moschetti}, {Mueller},
  {Modigliani}, {Oggioni}, {Oliveira}, {Pariani}, {Pasquini}, {Rasilla},
  {Redaelli}, {Riva}, {Santana-Tschudi}, {Santin}, {Santos}, {Segovia},
  {Sosnowska}, {Span{\`o}}, {Tenegi}, {Iwert}, {Zanutta}, \&
  {Zerbi}}]{Tabernero2020}
{Tabernero}, H.~M., {Zapatero Osorio}, M.~R., {Allart}, R., {et~al.} 2020,
  arXiv e-prints, arXiv:2011.12197

\bibitem[{{Torres} {et~al.}(2008){Torres}, {Winn}, \&
  {Holman}}]{Torres2008ApJ...677.1324T}
{Torres}, G., {Winn}, J.~N., \& {Holman}, M.~J. 2008, \apj, 677, 1324

\bibitem[{{Valenti} \& {Piskunov}(1996)}]{SME}
{Valenti}, J.~A. \& {Piskunov}, N. 1996, \aaps, 118, 595

\bibitem[{{Vidal-Madjar} {et~al.}(2010){Vidal-Madjar}, {Arnold}, {Ehrenreich},
  {Ferlet}, {Lecavelier Des Etangs}, {Bouchy}, {Segransan}, {Boisse},
  {H{\'e}brard}, {Moutou}, {D{\'e}sert}, {Sing}, {Cabanac}, {Nitschelm},
  {Bonfils}, {Delfosse}, {Desort}, {Diaz}, {Eggenberger}, {Forveille},
  {Lagrange}, {Lovis}, {Pepe}, {Perrier}, {Pont}, {Santos}, \&
  {Udry}}]{VidalMadjar2010}
{Vidal-Madjar}, A., {Arnold}, L., {Ehrenreich}, D., {et~al.} 2010, \aap, 523,
  A57

\bibitem[{{Winn} {et~al.}(2004){Winn}, {Suto}, {Turner}, {Narita}, {Frye},
  {Aoki}, {Sato}, \& {Yamada}}]{Winn2004}
{Winn}, J.~N., {Suto}, Y., {Turner}, E.~L., {et~al.} 2004, \pasj, 56, 655

\bibitem[{{Wyttenbach} {et~al.}(2015){Wyttenbach}, {Ehrenreich}, {Lovis},
  {Udry}, \& {Pepe}}]{Wytt2015}
{Wyttenbach}, A., {Ehrenreich}, D., {Lovis}, C., {Udry}, S., \& {Pepe}, F.
  2015, \aap, 577, A62

\bibitem[{{Wyttenbach} {et~al.}(2017){Wyttenbach}, {Lovis}, {Ehrenreich},
  {Bourrier}, {Pino}, {Allart}, {Astudillo-Defru}, {Cegla}, {Heng}, {Lavie},
  {Melo}, {Murgas}, {Santerne}, {S{\'e}gransan}, {Udry}, \&
  {Pepe}}]{2017A&A...602A..36W}
{Wyttenbach}, A., {Lovis}, C., {Ehrenreich}, D., {et~al.} 2017, \aap, 602, A36

\bibitem[{{Yan} {et~al.}(2019){Yan}, {Casasayas-Barris}, {Molaverdikhani},
  {Alonso-Floriano}, {Reiners}, {Pall{\'e}}, {Henning}, {Molli{\`e}re}, {Chen},
  {Nortmann}, {Snellen}, {Ribas}, {Quirrenbach}, {Caballero}, {Amado},
  {Azzaro}, {Bauer}, {Cort{\'e}s Contreras}, {Czesla}, {Khalafinejad}, {Lara},
  {L{\'o}pez-Puertas}, {Montes}, {Nagel}, {Oshagh}, {S{\'a}nchez-L{\'o}pez},
  {Stangret}, \& {Zechmeister}}]{Yan2019}
{Yan}, F., {Casasayas-Barris}, N., {Molaverdikhani}, K., {et~al.} 2019, \aap,
  632, A69

\bibitem[{{Yan} \& {Henning}(2018)}]{YanKELT9}
{Yan}, F. \& {Henning}, T. 2018, Nature Astronomy, 2, 714

\bibitem[{{Yan} {et~al.}(2017){Yan}, {Pall{\'e}}, {Fosbury}, {Petr-Gotzens}, \&
  {Henning}}]{Yan2017A&A...603A..73Y}
{Yan}, F., {Pall{\'e}}, E., {Fosbury}, R.~A.~E., {Petr-Gotzens}, M.~G., \&
  {Henning}, T. 2017, \aap, 603, A73

\bibitem[{{Zhou} \& {Bayliss}(2012)}]{Zhou2012}
{Zhou}, G. \& {Bayliss}, D.~D.~R. 2012, \mnras, 426, 2483

\end{thebibliography}





   
  



%

%
\onecolumn
\begin{appendix} 

\section{Additional results around the \ion{Na}{i}}
\label{sec:ap_TSNa}
\begin{figure*}[h]
\centering
\includegraphics[width=0.75\textwidth]{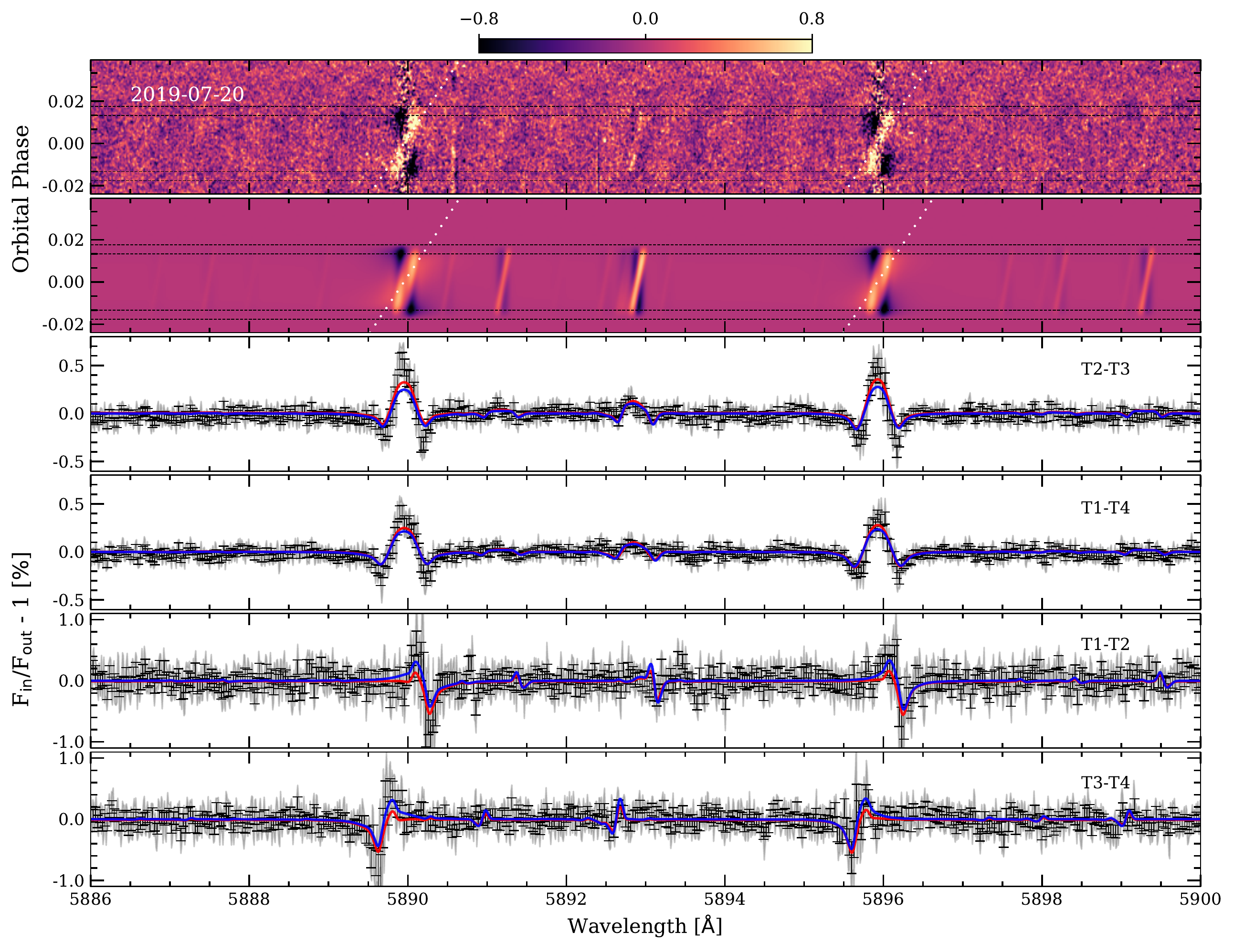}
\includegraphics[width=0.75\textwidth]{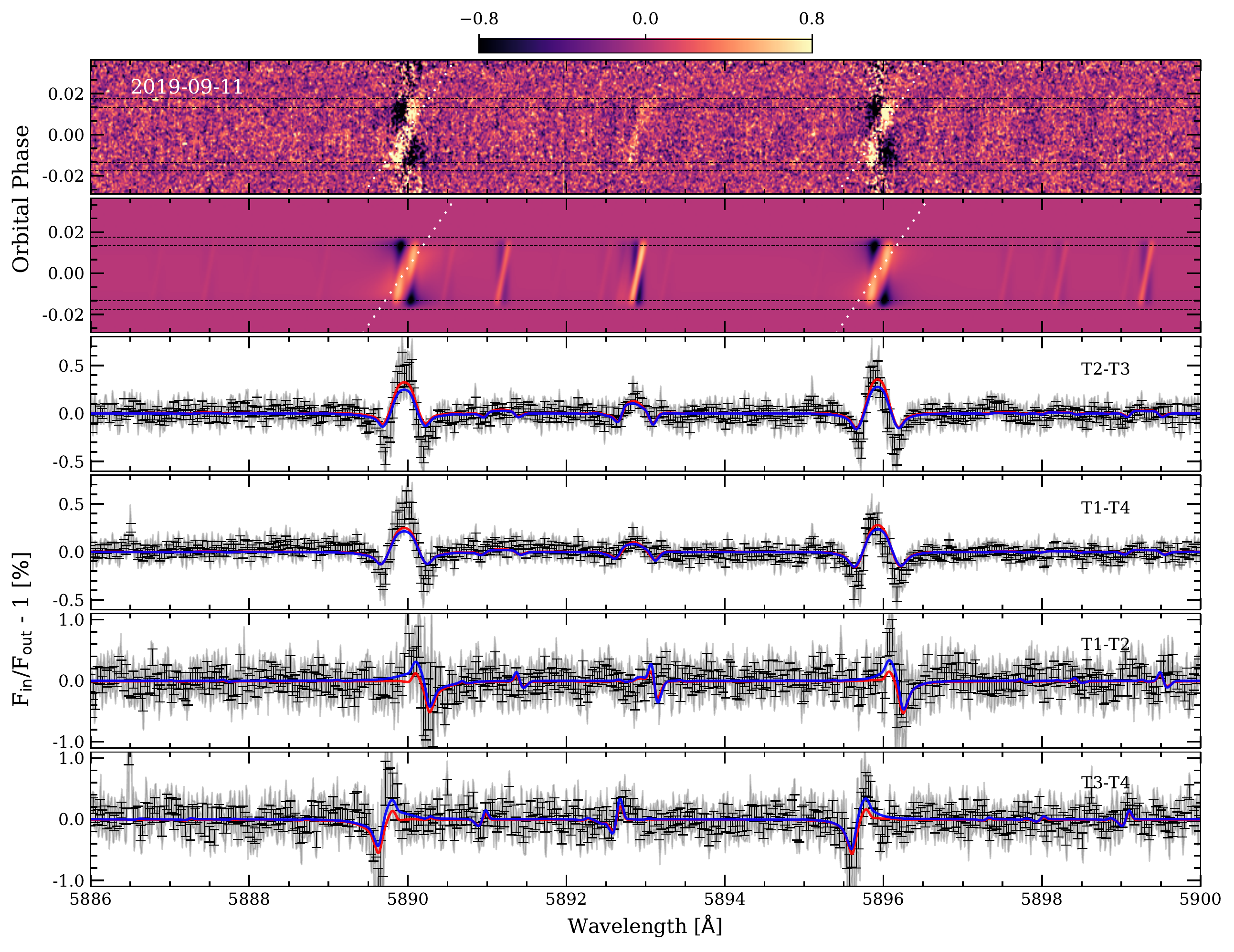}
\caption{Same as in Figure~\ref{fig:ts_Na}, but for the individual nights. The first night (2019-07-20) is presented in the top panel and the second night (2019-09-11) is shown in the bottom panel.}
\label{fig:ts_Na_indiv}
\end{figure*}

\begin{figure*}[h]
\centering
\includegraphics[width=0.8\textwidth]{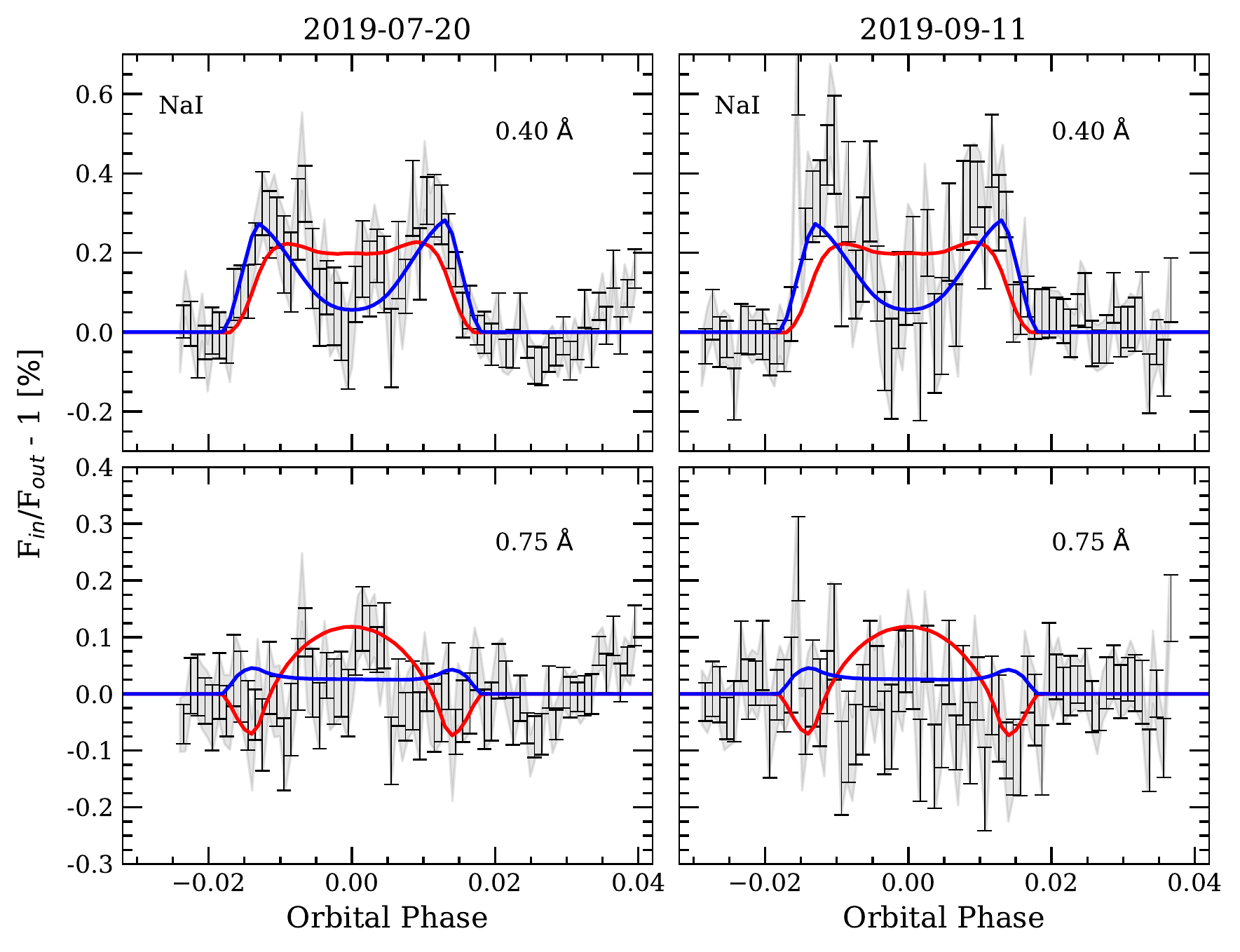}
\caption{Same as in Figure~\ref{fig:tlc_Na}, but for the individual nights: first night (2019-07-20) in the left column and second night (2019-09-11) in the right column. We note that the second night results could be partially affected by telluric \ion{Na}{i} absorption residuals.}
\label{fig:tlc_Na_indiv}
\end{figure*}

\begin{figure*}[h]
\centering
\includegraphics[width=0.7\textwidth]{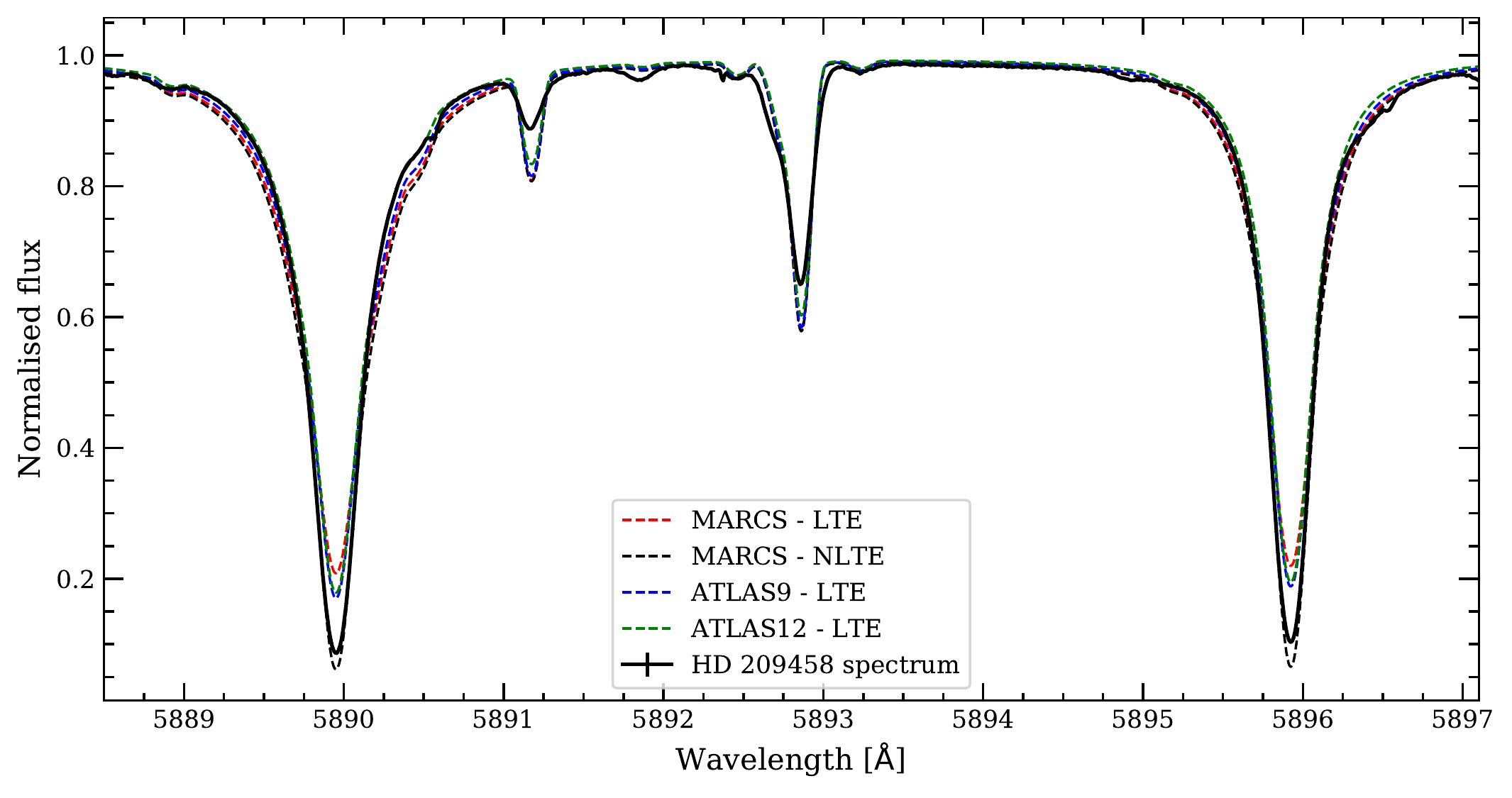}
\caption{Master stellar spectrum in the \ion{Na}{i} region of HD~209458 (black) computed as the combination of all out-of-transit spectra of the first night (2019-07-20). For comparison, in different colours we show the synthetic stellar spectrum computed with the different stellar models used in this work.}
\label{fig:linevsmod}
\end{figure*}

\begin{figure*}[h]
\centering
\includegraphics[width=0.8\textwidth]{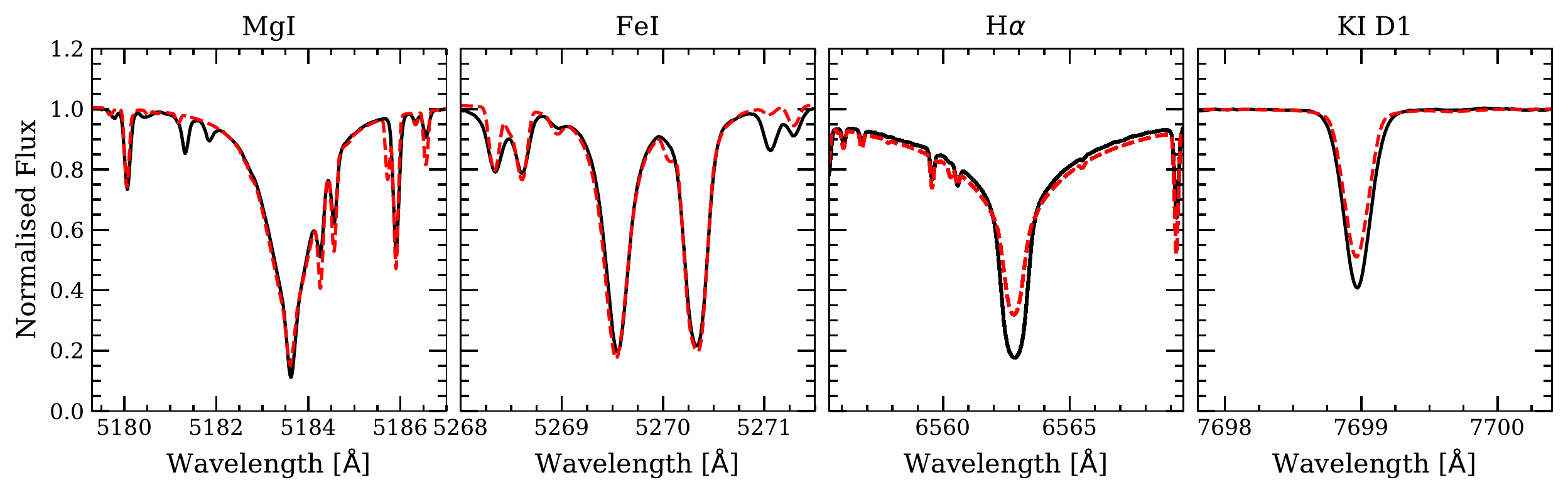}
\caption{Master stellar spectrum of HD~209458 (black) computed as the combination of all out-of-transit spectra of the first night (2019-07-20) in different wavelength regions. For comparison, we show in red colour the synthetic stellar spectrum computed with the MARCS stellar models assuming the stellar parameters from Table~\ref{tab:Param}, solar abundance, and LTE.}
\label{fig:other_models}
\end{figure*}

\begin{figure*}[h]
\centering
\includegraphics[width=0.55\textwidth]{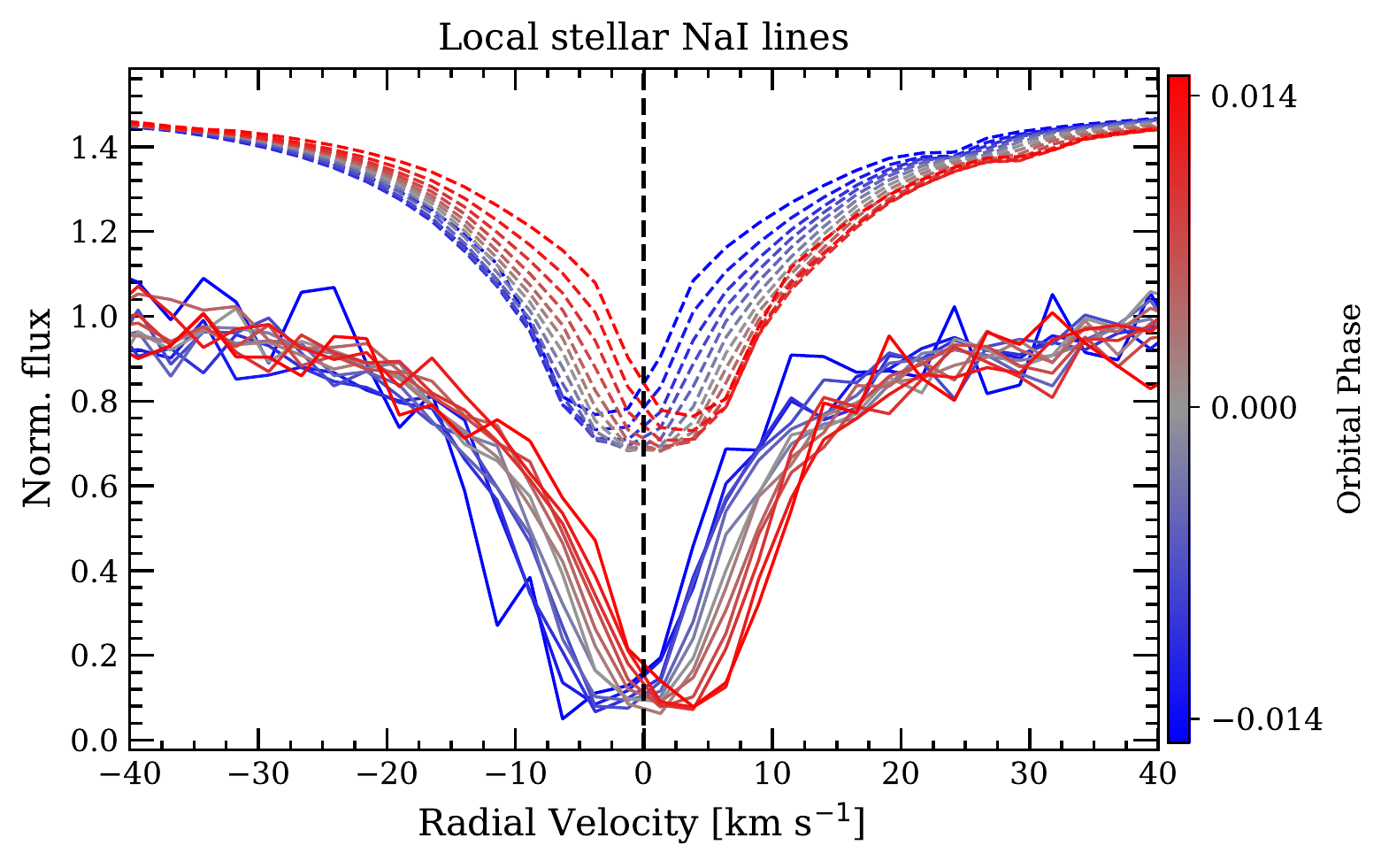}
\caption{Observed (solid lines) and modelled (dashed lines) local stellar sodium \ion{Na}{i} profiles at different orbital phases during the transit. The profiles are shown normalised to their continuum level, and result from the combination of both \ion{Na}{i} D lines of the two nights of observation. The data are shown binned by $2.5$\,km\,s$^{-1}$ and each profile is the result of combining the individual local stellar lines every $0.0025$ in orbital phase. The models are shown with an offset to the vertical axis for better visualisation, and are computed using the MARCS stellar models in LTE. The colours indicate the orbital phase of the planet (see colour bar). The vertical black-dashed line is the reference at 0\,km~s$^{-1}$.}
\label{fig:local_Na_comb}
\end{figure*}

\begin{figure*}[h]
\centering
\includegraphics[width=0.67\textwidth]{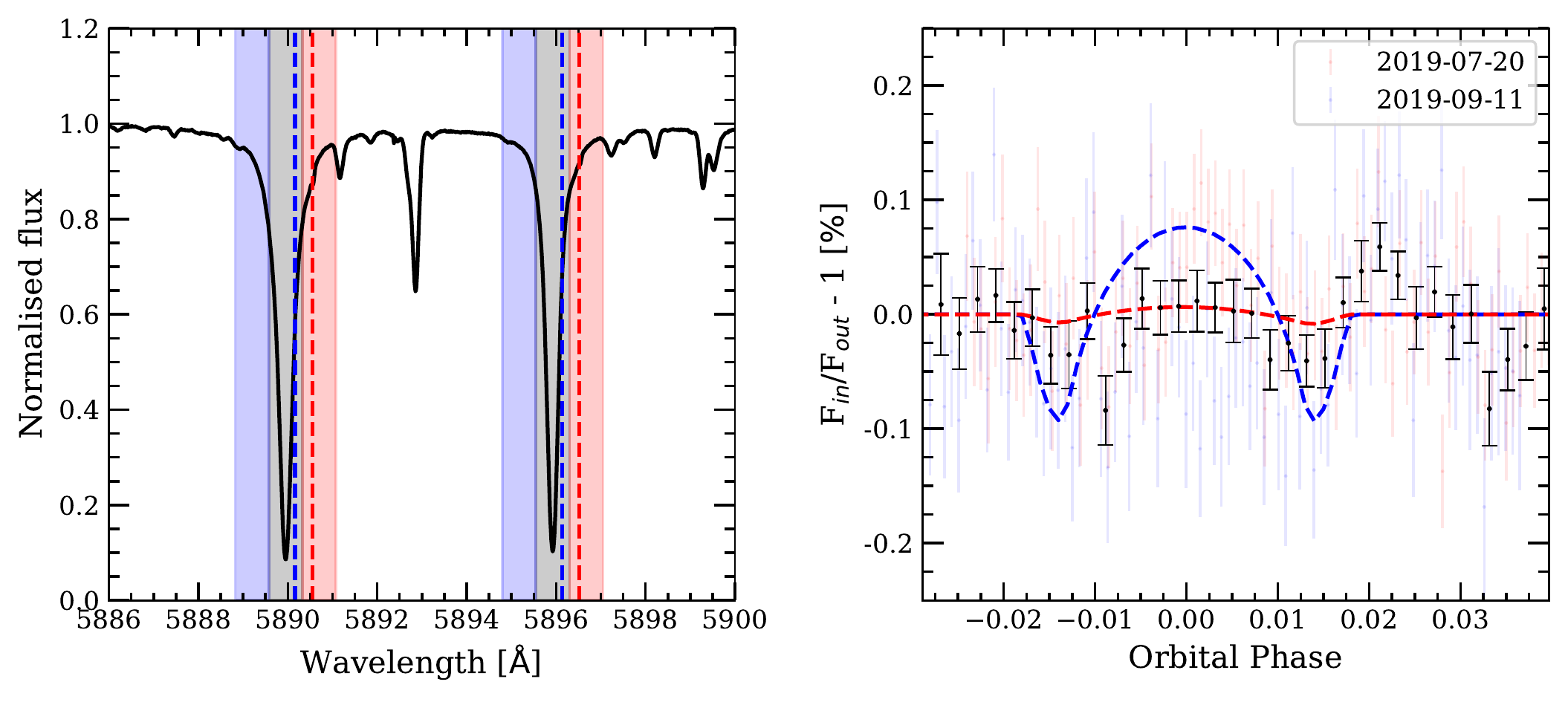}
\caption{Transmission light curves around the \ion{Na}{i} computed as presented in \citet{2008SnellenHD209} and \citet{Albrecht2009}. \textit{Left panel:} Stellar spectrum around the \ion{Na}{i} showing the central (grey) and reference (blue and red) passbands of each line of the doublet, all of them with a bandwidth of $0.75~{\rm \AA}$. The vertical dashed lines show the position of the telluric \ion{Na}{i} absorption residuals for the night of 2019-07-20 (red) and 2019-09-11 (blue). \textit{Right panel:} \ion{Na}{i} transmission light curve of each night (in colours) and the combination of the two nights (black dots). The combined result is binned by $0.002$ in orbital phase, similar to the results presented in the previous literature. The blue-dashed line corresponds to the modelled transmission light curve considering the RM and CLV effects in the stellar spectrum. The red-dashed line considers only the RM effect.}
\label{fig:SnellNaI}
\end{figure*}

\clearpage

\section{Empirical Monte Carlo distributions}
\label{sec:EMC_other}

\begin{figure*}[h]
\centering
\includegraphics[width=0.49\textwidth]{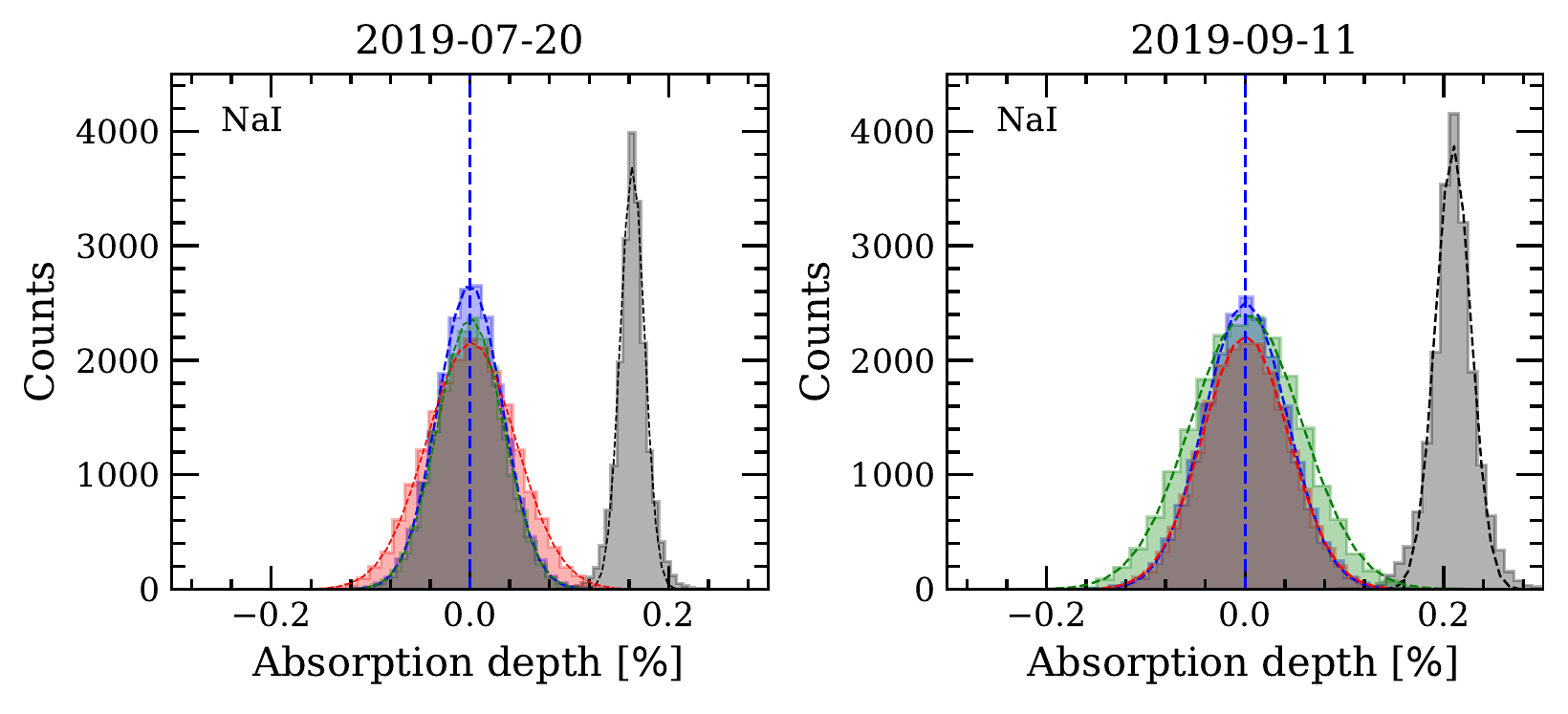}
\includegraphics[width=0.49\textwidth]{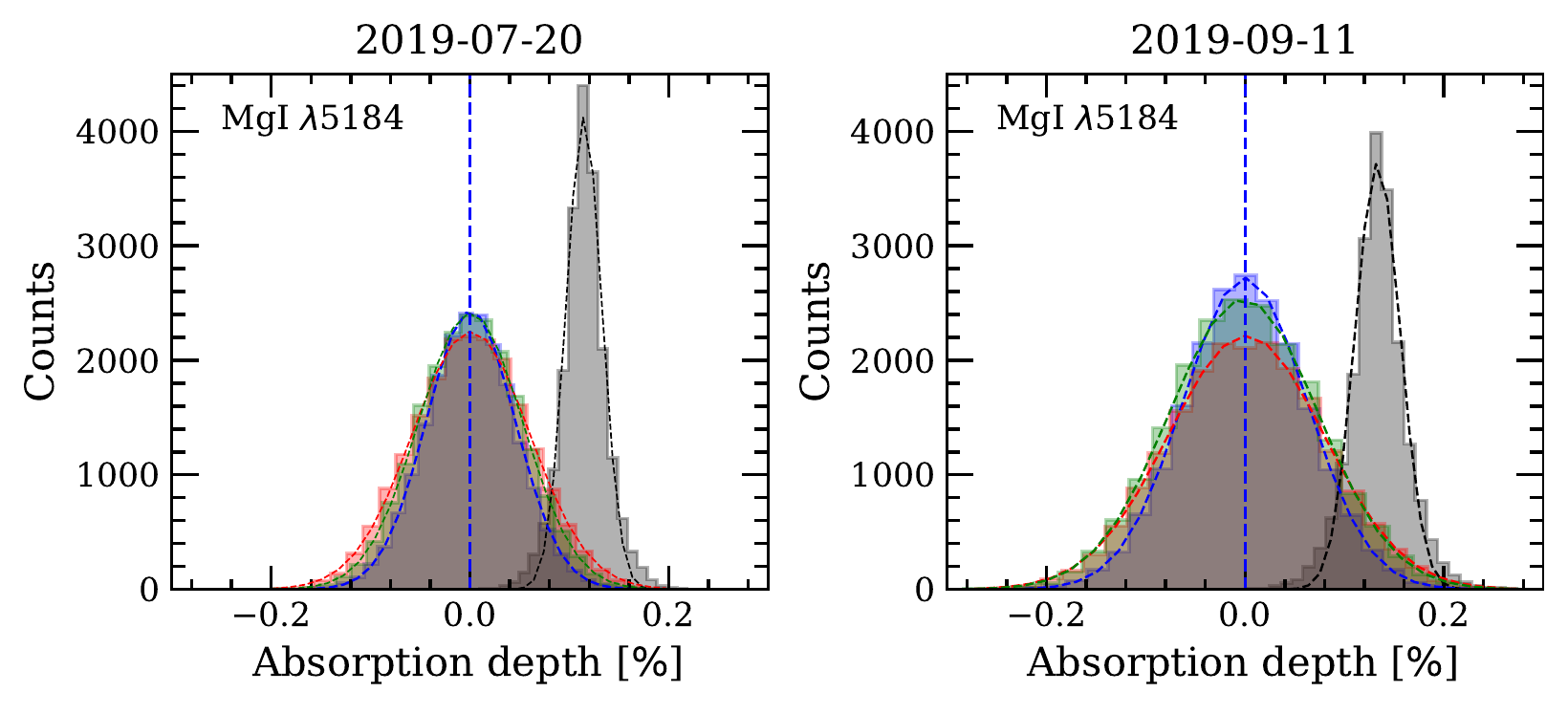}
\includegraphics[width=0.49\textwidth]{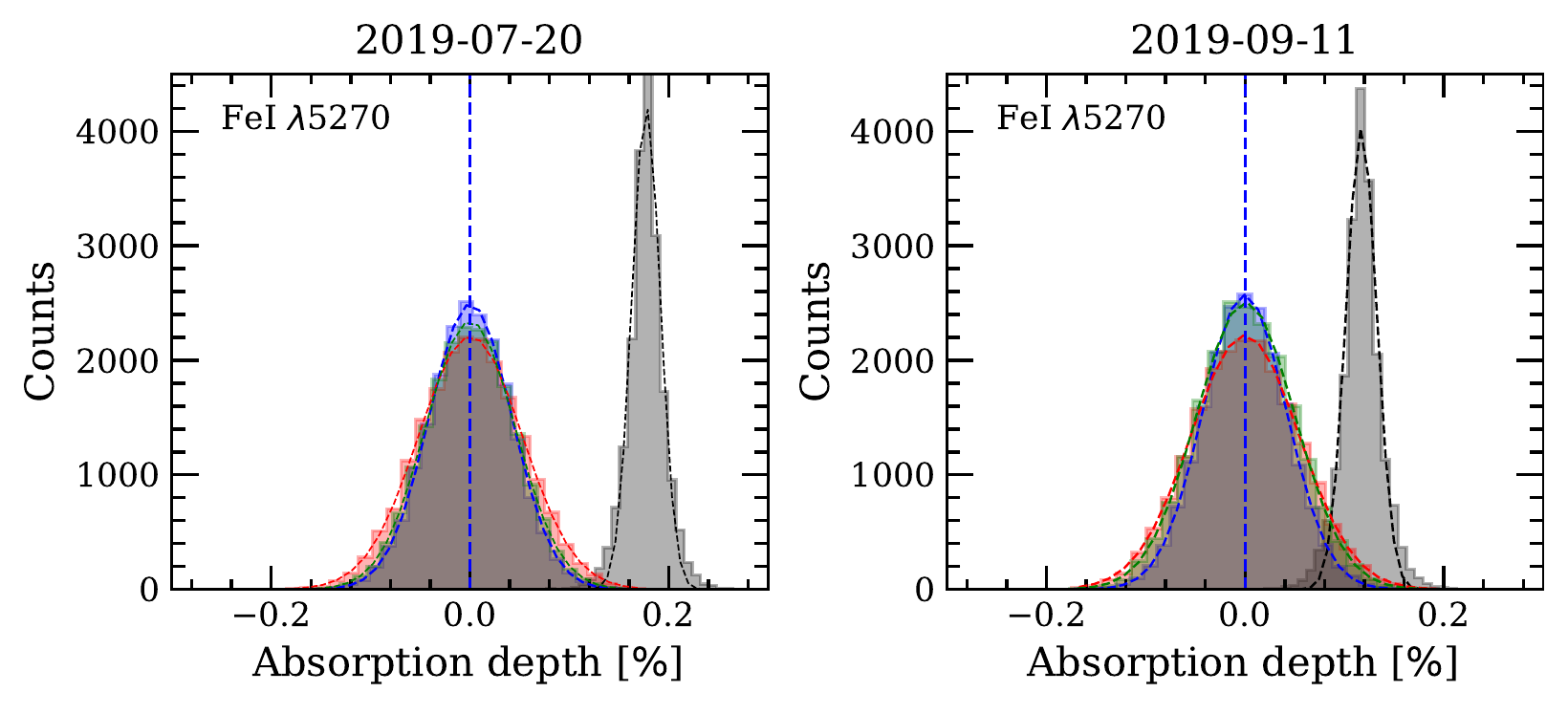}
\includegraphics[width=0.49\textwidth]{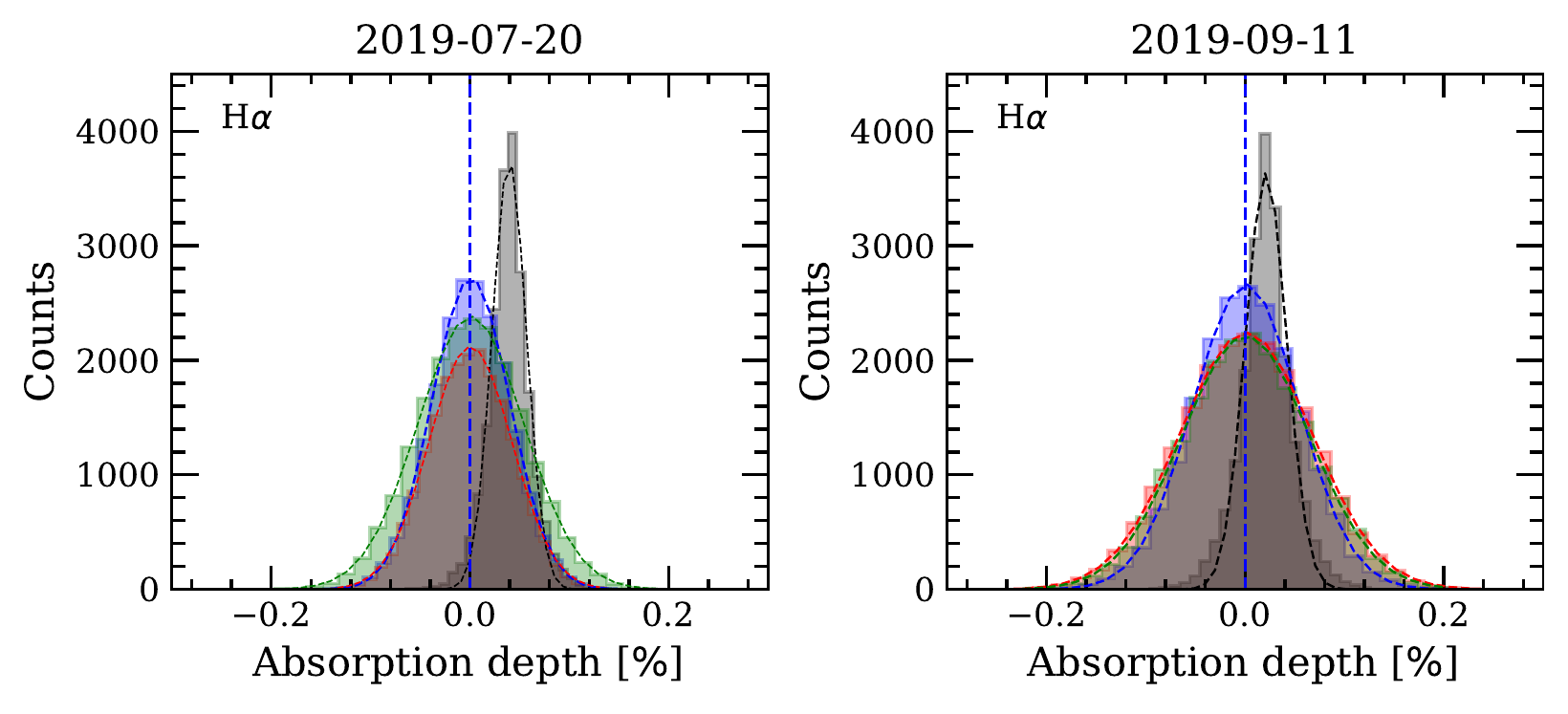}
\includegraphics[width=0.49\textwidth]{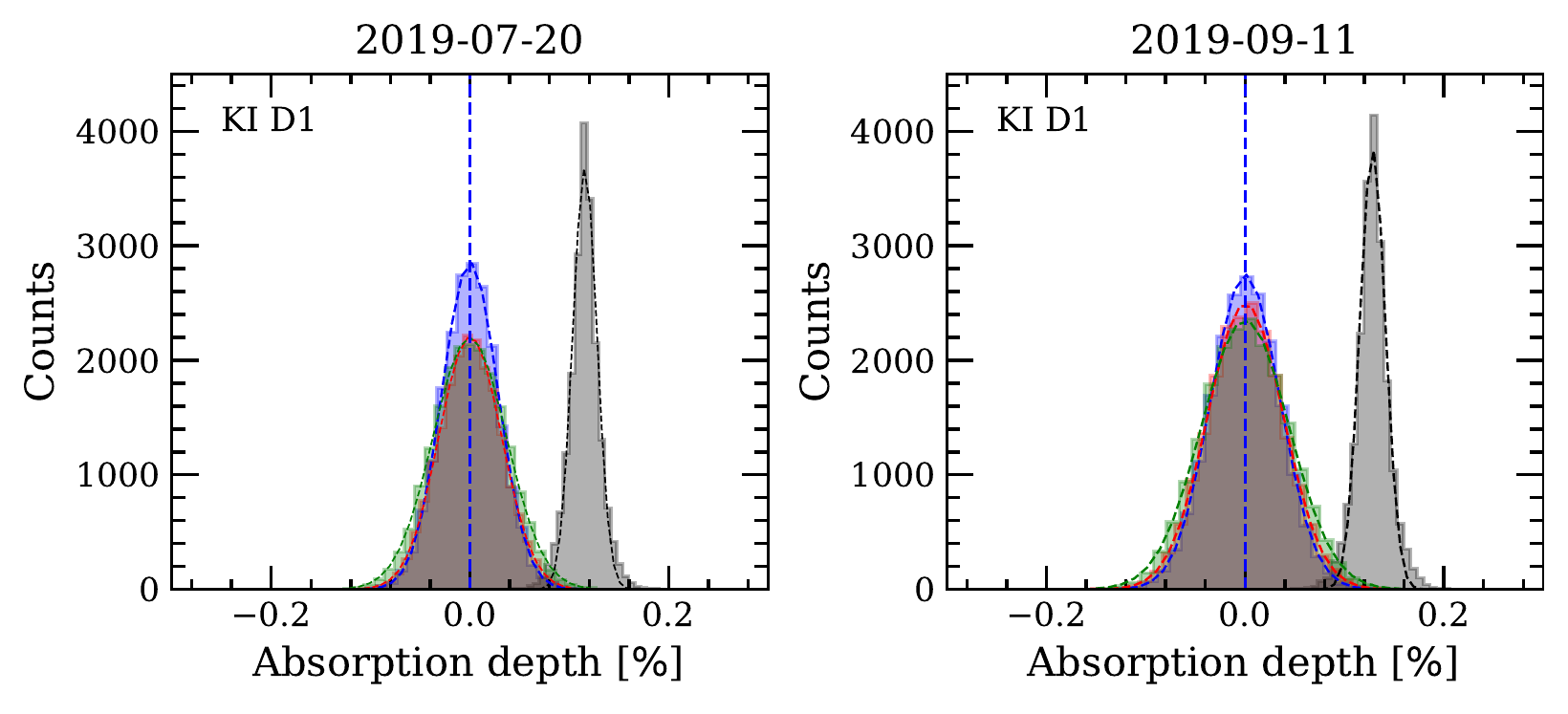}
\caption{Distributions of the EMC analysis of each individual night for different spectral lines: \ion{Na}{i} D1 and D2, \ion{Mg}{i} $\lambda5184$, \ion{Fe}{i} $\lambda5270$, H$\alpha$, and \ion{K}{i} D1. The absorption depth calculations are performed using $0.4~{\rm \AA}$ bandwidth except for the H$\alpha$ for which we use $0.5~{\rm \AA}$. In green we present the 'out-out' scenario, in red the 'in-in', in blue the 'mix-mix', and in grey the 'in-out'. The blue dashed vertical lines show the zero absorption level. In coloured dashed lines, we show the Gaussian fit to the histograms. }
\label{fig:EMC}
\end{figure*}

\clearpage

\section{TiO and VO cross-correlation maps}
\label{sec:ccfmolec}

\begin{figure*}[h]
\centering
\includegraphics[width=1\textwidth]{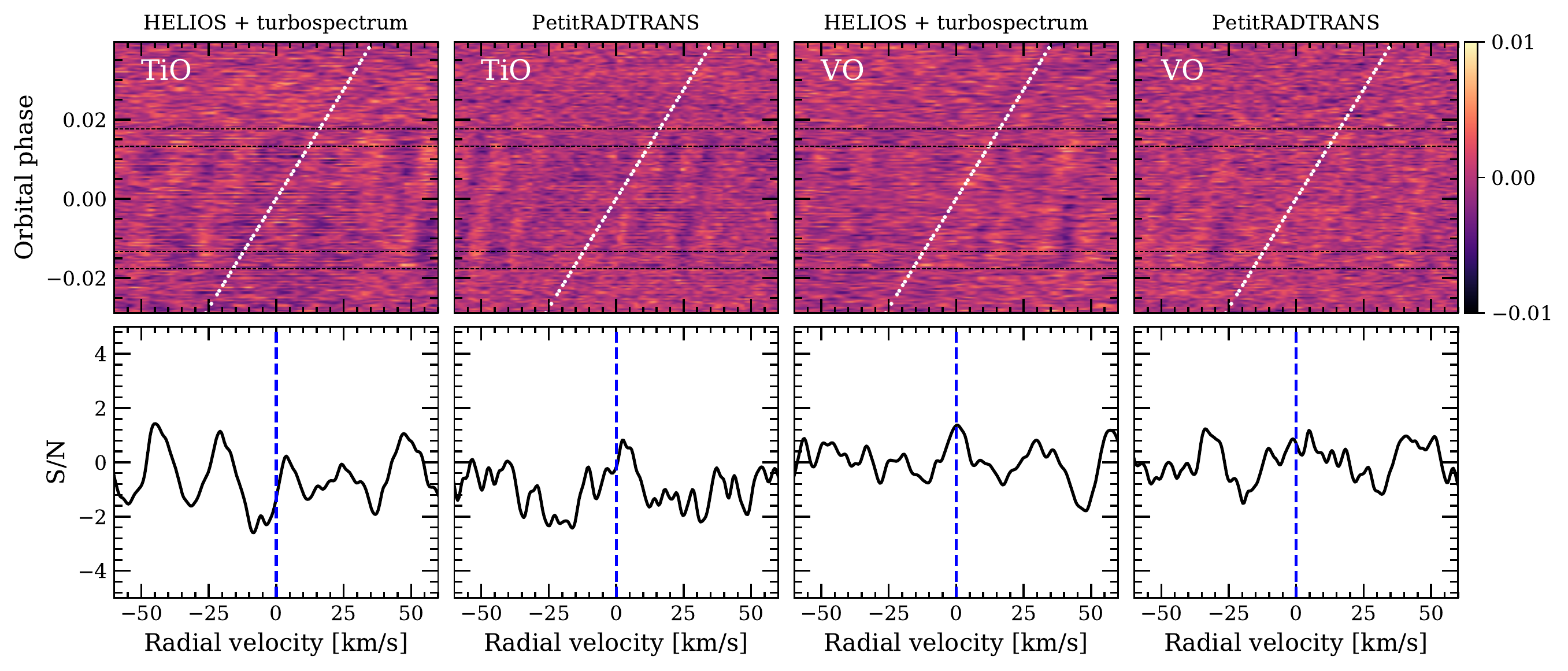}
\caption{Cross-correlation results of both nights combined for TiO and VO, obtained assuming different atmospheric models (HELIOS+{\tt turbospectrum} and {\tt PetitRADTRANS}). \textit{Top panel}: Cross-correlation maps of the TiO (first and second columns) and VO (third and fourth columns). \textit{Bottom panel:} Average of the in-transit cross-correlation values in the planet rest frame between the first and fourth contacts. The vertical axis is shown in S/N units for a better visualisation of the features strength, computed as described in Section~\ref{subsec:otherlines}. Positive S/N means correlation (see CCF values in the colour bar).}
\label{fig:molec}
\end{figure*}

\end{appendix}
\end{document}